\documentclass[12pt,doublespace]{JHEP3}
\JHEPspecialurl{http://jhep.sissa.it/JOURNAL/JHEP3.tar.gz}

\usepackage{mathrsfs}
\usepackage{amsmath,amssymb,amsfonts,xspace}
\usepackage{subfigure,graphicx}

\usepackage[all]{xy}

\allowdisplaybreaks

\numberwithin{equation}{section}

\DeclareSymbolFont{AMSa}{U}{msa}{m}{n}
\DeclareSymbolFont{AMSb}{U}{msb}{m}{n}
\DeclareMathSymbol{\fieldR}{\mathalpha}{AMSb}{"52}

\DeclareMathOperator{\Li}{Li}

\renewcommand{\Im}{\imag}

\newcommand{\be}{\begin{equation}}
\newcommand{\ee}{\end{equation}}
\newcommand{\beq}{\begin{eqnarray}}
\newcommand{\eeq}{\end{eqnarray}}
\newcommand{\bea}[2]{\be\label{#2}\begin{array}{#1}}
\newcommand{\eea}{\end{array}\ee}

\def\Tr{\,{\rm Tr}\, }

\def\Im{\,{\rm Im}\, }
\def\Re{\,{\rm Re}\, }

\def\({\left(}
\def\){\right)}
\def\[{\left[}
\def\]{\right]}
\def\p{\partial}

\def\11{1\!\! 1}

\def\hf{\frac{1}{2}}
\def\haf{\textstyle\frac{1}{2}}

\def\eps{\varepsilon}

\def\vrh{\varrho}

   \def\CX {{\cal X}}

\def\bX{\bar X}
\def\bZ{\bar Z}
\def\bY{ \bar Y }

\def\bW{ \bar W}

\def\bF{\bar F}

\def\bG{\bar G}

\def\ba{\bar a}

\def\bz{\bar z}
\def\bu{\bar u}
\def\bv{\bar v}

\def\tS{{\tilde S}}

\def\hcU{{\cal U}}

\def\vf{v^{\flat}}

\newcommand{\CP}{\IC P^1}

\renewcommand{\L}{\cal{L}}
\renewcommand{\d}{\mathrm{d}}
\newcommand{\de}{\mathrm{d}}

\newcommand{\I}{\mathrm{i}}
\newcommand{\cL}{\mathcal{L}}

\def\vrh{\varrho}

\newcommand{\half}{\frac{1}{2}}

\newcommand{\cA}{\mathcal{A}}
\newcommand{\cB}{\mathcal{B}}

\newcommand{\cS}{\mathcal{S}}

\newcommand{\cH}{\mathcal{H}}
\newcommand{\cK}{\mathcal{K}}
\newcommand{\cM}{\mathcal{M}}
\newcommand{\cW}{\mathcal{W}}
\newcommand{\cN}{\mathcal{N}}

\newcommand{\cX}{\mathcal{X}}
\newcommand{\cY}{\mathcal{Y}}
\newcommand{\cO}{\mathcal{O}}
\newcommand{\cP}{\mathcal{P}}
\newcommand{\cR}{\mathcal{R}}
\newcommand{\cT}{\mathcal{T}}
\newcommand{\cJ}{\mathcal{J}}
\newcommand{\cU}{\mathcal{U}}

\newcommand{\unit}{{\mathrm{1}}}

\newcommand{\cZ}{\mathcal{Z}}
\newcommand{\IR}{\mathbb{R}}
\newcommand{\IC}{\mathbb{C}}
\newcommand{\IZ}{\mathbb{Z}}
\newcommand{\IQ}{\mathbb{Q}}

\newcommand{\bfk}{{\bf k}}

\newcommand{\exm}{\mathcal{B}}

\newcommand{\fb}{\mathscr{F}}
\newcommand{\lb}{\mathscr{L}}

\newcommand{\stg}{{U}_\gamma}
\newcommand{\ctg}{{V}_\gamma}

\def\Hij#1{H^{[#1]}}

\def\hHij#1{H^{[#1]}_{\rm HK}}
\def\qHij#1{H^{[#1]}_{\rm QK}}

\def\etai#1{\eta_{[#1]}}

\def\xii#1{\xi_{[#1]}}
\def\mui#1{\mu^{[#1]}}
\def\txii#1{{\tilde\xi}^{[#1]}}
\def\ai#1{{\alpha}^{[#1]}}

\def\talp{\tilde\alpha}

\def\ui#1{^{[#1]}}

\def\ellg#1{\ell_{#1}}

\def\pa{\partial}

\newcommand{\tzeta}{{\tilde\zeta}}
\newcommand{\txi}{\tilde\xi}
\newcommand{\nn}{\nonumber}
\newcommand{\kahler}{{K\"ahler}\xspace}
\newcommand{\hk}{{hyperk\"ahler}\xspace}
\newcommand{\qk}{{quaternion-K\"ahler}\xspace}

\def\eps{\epsilon}
\def\bse{\begin{subequations}}
\def\ese{\end{subequations}}

\def\ellg#1{\ell_{#1}}

\def\qli2#1{\Psi_{#1}}

\def\hu{\hat u}
\def\hv{\hat v}
\def\ub{{\bf u}}
\def\vb{{\bf v}}

\def\zz{\cX}
\def\ind{s}
\def\varpiqk{t}
\def\varpihk{\zeta}
\def\sigp{{\theta}}
\def\phip{{\theta'}}
\def\kcon{p'_3}
\def\hcon{p'_+}
\def\ahcon{p'_-}
\def\pTheta{\Theta'}
\def\pcM{\cM'}
\def\pcZ{\cZ'}
\def\pomega{\omega'}

\def\pXi{\Xi'}
\def\pcX{\cX'}
\def\cff{\tau} 
\def\cfff{\nu} 
\def\rigZ{Z}
\def\brigZ{\bZ}
\def\rigX{X}

\def\Kc{K} 
\def\conhyper{\lambda}

\def\pzeta{\partial^{(\varpihk)}}
\def\bpzeta{\bar\partial^{(\varpihk)}}

\numberwithin{equation}{section}
\title{Wall-crossing,  Rogers dilogarithm,  \\ and the QK/HK correspondence}

\preprint{CERN-PH-TH/2011-239\\L2C:11-165\\arXiv:1110.0466v3}

\author{Sergei Alexandrov$^{1,2}$, Daniel Persson$^{3,4}$,
Boris Pioline$^{5,6}$\\

\vspace{2mm}
$^1$ {\it Universit\'e Montpellier 2, Laboratoire Charles Coulomb UMR 5221, F-34095,
Montpellier, France}

\vspace{2mm}
$^2$ {\it CNRS, Laboratoire Charles Coulomb UMR 5221, F-34095,
Montpellier, France}

\vspace{2mm}
$^3$  {\it Institut f\"ur Theoretische Physik, ETH Z\"urich, CH-8093 Z\"urich, Switzerland}

\vspace{2mm}
$^4$ {\it Fundamental Physics, Chalmers University of Technology, 412 96 Gothenburg, Sweden}

\vspace{2mm}
$^5$ {\it PH-TH Division, CERN, Geneva 23, Switzerland}

\vspace{2mm}
$^6$ {\it Laboratoire de Physique Th\'eorique et Hautes
Energies, CNRS UMR 7589, \\
Universit\'e Pierre et Marie Curie - Paris 6,
4 place Jussieu, 75252 Paris Cedex 05, France
} \\

\vspace*{-2mm} {\tt e-mail: \email{sergey.alexandrov@univ-montp2.fr},
\email{daniel.persson@itp.phys.ethz.ch},
\email{pioline@lpthe.jussieu.fr}
}}

\abstract{When formulated in twistor space, the D-instanton corrected hypermultiplet
moduli space  in $\cN=2$ string vacua and the  Coulomb branch  of
rigid $\cN=2$ gauge theories on $\IR^3\times S^1$ are strikingly similar and,
to a large extent, dictated by consistency with wall-crossing.
We elucidate  this similarity by showing that these two spaces
are related under  a general duality between, on one hand, \qk manifolds
with a quaternionic isometry  and, on the other hand,
 \hk manifolds  with a rotational isometry,
equipped with a canonical hyperholomorphic circle bundle  and a connection.
We show that the transition functions of the hyperholomorphic circle bundle relevant for
the hypermultiplet moduli space are given by the Rogers dilogarithm function, and that
consistency across walls of marginal stability is ensured by the motivic wall-crossing formula
of Kontsevich and Soibelman. We illustrate the construction on some simple examples
of wall-crossing related to cluster algebras for rank 2 Dynkin quivers. In an appendix we also provide a detailed
discussion on the general relation between wall-crossing and cluster algebras.}

\begin{document}

\tableofcontents
\setcounter{page}{2}
\newpage
\section{Introduction}
\label{intro}

In four-dimensional string vacua and gauge theories with $\cN=2$ supersymmetry,
BPS states, more often than not, tend to decay across certain real codimension-one walls
in the space of couplings or moduli $z^a$ \cite{Seiberg:1994rs,Ferrari:1996sv,Denef:2000nb,Denef:2007vg}.
A useful way to keep track of the resulting
dependence of the BPS index $\Omega(\gamma, z)$ on the moduli is to consider the
same theory on
$\IR^3\times S^1$: for large but finite radius of $S^1$, the low-energy effective action in
three dimensions receives  instanton corrections from four-dimensional BPS states
whose Euclidean worldline winds around the circle
\cite{Seiberg:1996nz,Seiberg:1996ns,Pioline:2005vi,Gunaydin:2005mx}.
Single-instanton corrections are
weighted by the BPS index, and are therefore discontinuous
across walls of marginal stability. Multi-instanton corrections ensure that
the low energy effective action is nonetheless regular in the vicinity
of the wall, both in the context of $\cN=2$ gauge theories \cite{Gaiotto:2008cd}
and $\cN=2$ string vacua \cite{Alexandrov:2008gh}.

In this work we elucidate
the geometric origin of the striking similarity between the twistorial constructions of the
moduli space in $\cN=2$ string and gauge theories on $\IR^3\times S^1$, and clarify
some aspects of our earlier computation of D-instanton corrections to the hypermultiplet metric
in $\cN=2$ string vacua  \cite{Alexandrov:2008gh}. In particular, we show that
both metrics are related by an instance of the QK/HK correspondence,  a general
duality between \qk (QK) spaces with a quaternionic isometry, and \hk (HK) spaces
with a rotational isometry, equipped with a canonical hyperholomorphic circle bundle. We show
that the BPS invariants determine a canonical hyperholomorphic circle bundle $\cP$ on
the HK space $\pcM$ dual to the hypermultiplet moduli space $\cM$,
which in turn determines the D-instanton corrected QK metric on $\cM$. The
functional equations obeyed by the Rogers dilogarithm play an essential role in
ensuring the consistency of the construction. To explain our results in more detail,
we start with a brief recap of \cite{Gaiotto:2008cd,Alexandrov:2008gh}.

\subsection*{$\cN=2$ field theory and symplectic geometry}

For gauge theories with rigid $\cN=2$ supersymmetry in four dimensions,
the discontinuity of the BPS index across walls of marginal stability is captured by the
Kontsevich-Soibelman (KS) wall-crossing formula \cite{ks} (see e.g. \cite{Pioline:2011gf} for a review
of the KS formula and equivalent versions thereof). This claim can be justified by
examining instanton corrections to the non-linear sigma model which describes the
effective low-energy dynamics on $\IR^3\times S^1$ \cite{Gaiotto:2008cd}.
Due to supersymmetry, the target space $\pcM$ of this sigma model
must  carry  a \hk (HK) metric of real dimension $4n$, where $n$ is the rank of the gauge group.
Instanton corrections to this metric are most conveniently formulated in terms
of the twistor space $\pcZ=\CP\times \pcM$ \cite{Hitchin:1986ea,Ivanov:1995cy,Alexandrov:2008ds}
(see \S\ref{sec_twi} for a summary of this approach).
Viewing $\pcZ$ as a fibration over $\CP$
with stereographic coordinate $\varpihk$, the fiber $\pcM(\varpihk)$ is isomorphic to
a (twisted) complex symplectic torus $(\IC^\times)^{2n}$ in complex structure $J(\varpihk)$.
Consequently, there exist  canonical complex Darboux coordinates
$\pXi=(\eta^\Lambda,\mu_\Lambda)$ (which are functions of
$(\varpihk,x'^\mu)\in \CP\times \pcM)$, such that the symplectic
form $\pomega(\varpihk)$ on $\pcM(\varpihk)$ is proportional to
\be
\label{omtintro}
 \langle \de \pXi, \de \pXi\rangle = 2\, \de \mu_\Lambda \wedge \de\eta^\Lambda \, ,
\ee
and such that the torus action corresponds to
integer translations of $(\eta^{\Lambda}, \mu_\Lambda)$ \cite{BPNeitzke}.
As argued in \cite{Gaiotto:2008cd}, due to
instanton corrections the Darboux coordinates $\pXi$ are discontinuous across
certain meridian lines on $\CP$, known as BPS rays, whose azimuthal angle is equal to the phase
of the central charge $Z_\gamma(z)$ of the corresponding BPS state.
It is natural to identify the symplectomorphism relating the Darboux coordinate
systems $\pXi$ across the BPS ray $\ell_\gamma$
with the abstract operator $U_\gamma$ appearing in the KS formula. Upon doing so, the KS
formula ensures the consistency of the construction across walls
of marginal stability, and hence the smoothness of the \hk metric.

\subsection*{$\cN=2$ supergravity and contact geometry}

For string vacua with local $\cN=2$ supersymmetry in four dimensions, a similar
relation holds between the BPS spectrum in $D=4$ and the target space metric of the
non-linear sigma model $\cM$  describing the vector multiplet moduli space
in $D=3$ (or, by T-duality, the hypermultiplet moduli space in $D=4$),
but with some important wrinkles \cite{Alexandrov:2008gh}. First,  unlike in field theory,
the index $\Omega(\gamma)$ of BPS states in string vacua tends to grow exponentially, and the resulting
instantonic series is divergent. This divergence is expected to be resolved by
gravitational instantons (or, in the T-dual set-up, NS5-brane instantons), which
have no counterpart in field theory \cite{Pioline:2009ia}.
These additional corrections are characterized by a non-trivial
dependence on the NUT potential (or Neveu-Schwarz axion)
$\sigma$, which parametrizes a certain compact
direction in $\cM$. However, for small string coupling
they are exponentially suppressed compared to the BPS-instantons.
In this paper we restrict to this weak coupling limit where gravitational (or NS5-) instantons are negligible,
so that the metric on $\cM$ has a Killing vector field $\partial_\sigma$, and we
ignore the divergence of the instantonic series caused by the exponential growth of $\Omega(\gamma)$.

Second, unlike the rigid case, the target space $\cM$ is no longer
\hk but rather \qk (see \S\ref{sec_twiqk} for a reminder of basis properties
of QK manifolds).  As a result, and in contrast to the hyperk\"ahler situation,
the twistor space $\cZ$ of a quaternion-K\"ahler
manifold $\cM$ is a non-trivial
fibration $\CP\to \cZ \to \cM$, and the fiber $\cM(\varpiqk)$ of the
opposite local fibration $\cM \to \cZ \to \CP$  is not a complex
manifold (we denote by $\varpiqk$ the stereographic coordinate on $\CP$,
at this point unrelated to the coordinate $\varpihk$ in the HK construction).
Nevertheless, the full twistor space $\cZ$ does admit a
canonical complex structure, in fact a complex {\it contact structure},
which serves as a replacement for the complex symplectic structure
in the \hk case \cite{MR664330}. In particular, there still exist
canonical `contact-Darboux' coordinates $(\Xi, \alpha)=(\xi^\Lambda,\txi_\Lambda,\alpha)$
(which are locally functions of $(\varpiqk,x^\mu)\in \CP\times \cM$),
such that the complex contact structure is given by the kernel of the holomorphic one-form
\cite{Alexandrov:2008nk}
\be
\label{c1f}
\cX=\de\alpha+\xi^\Lambda\de\txi_\Lambda.
\ee
The Killing vector field $\pa_\sigma$ on $\cM$ lifts to a holomorphic vector field on $\cZ$
which preserves the contact structure. The Darboux coordinates
$(\xi^{\Lambda}, \tilde\xi_\Lambda, \alpha)$ can be chosen such that
this vector field is  the Reeb vector\footnote{Recall that the Reeb vector of a contact structure
is the generator $R$ of the kernel of $\de\CX$, normalized such that $\CX(R)=1$.}
$\partial_\alpha$ of the contact structure \cite{Neitzke:2007ke}. For this property to hold
globally, the complex contact transformations $V\ui{ij}$ relating
the Darboux coordinate systems on the overlap
$\hcU_i\cap \hcU_j$ of two patches on $\cZ$, must descend to
complex symplectomorphisms on the {\it reduced twistor space}
$\cZ/\partial_\alpha$. Put differently, the  complex contact transformations  must
decompose into a symplectomorphism $U\ui{ij}$ acting on the Darboux coordinates
$\Xi=(\xi^\Lambda,\txi_\Lambda)$, and a shift of the Darboux
coordinate $\alpha$,
\be
\label{cxi}
V\ui{ij}\ :\quad
\Xi\ui{i}=U\ui{ij}\cdot \Xi\ui{j}\, ,
\qquad
\alpha\ui{i}=\alpha\ui{j}-\frac{1}{4\pi^2}\, S\ui{ij}(\Xi\ui{j})\, ,
\ee
such that the combined transformation preserves the contact one-form \eqref{c1f}.
The function $S\ui{ij}(\Xi\ui{j})$ is determined by the generating
function
of the symplectomorphism $U\ui{ij}$ (see \cite{Alexandrov:2008ds,Alexandrov:2008nk}).

As argued in \cite{Alexandrov:2008gh} on the basis of duality, mirror symmetry
and wall-crossing, it is again natural to identify the symplectomorphism $U\ui{ij}$  with
the KS operator $U_\gamma$, such that the KS wall-crossing formula ensures
the consistency of the construction of the reduced twistor space $\cZ/\partial_\alpha$
(if not of the full twistor space $\cZ$)
across walls of marginal stability.
In fact, upon trading the central charge function
(or stability data) $Z_\gamma(z)$ and the BPS invariants $\Omega(\gamma,z)$ of the $\cN=2$
gauge theory with those relevant for the string compactification,  identifying
the canonical Darboux coordinates $\pXi(\varpihk)$ and $\Xi(\varpiqk)$ and
ignoring the contact Darboux coordinate $\alpha(\varpiqk)$,
the construction of the twistor space $\cZ$ in \cite{Alexandrov:2008gh} is formally
isomorphic to the construction given in \cite{Gaiotto:2008cd} of the HK metric on the Coulomb branch $\pcM$ of
$\cN=2$ gauge theories. As a result,  the
reduced twistor space $\cZ/\partial_\alpha$
carries a HK metric, similar to the HK metric on $\pcM$. There are however two notable differences:
for prepotentials arising in Calabi-Yau compactifications, the HK metric on
$\cZ/\partial_\alpha$ has indefinite signature $(4,4n-4)$; in addition, since
the prepotential is homogeneous of degree 2, the HK metric
always admits a Killing vector field $\pa_{\phip}$ acting by R-symmetry
rotations. Keeping these differences in mind, we henceforth denote by
$\pcM$ the space $\cZ/\partial_\alpha$
equipped with the above Lorentzian-\hk metric.

\begin{center}
\begin{figure}[t]
\includegraphics[width=16cm]{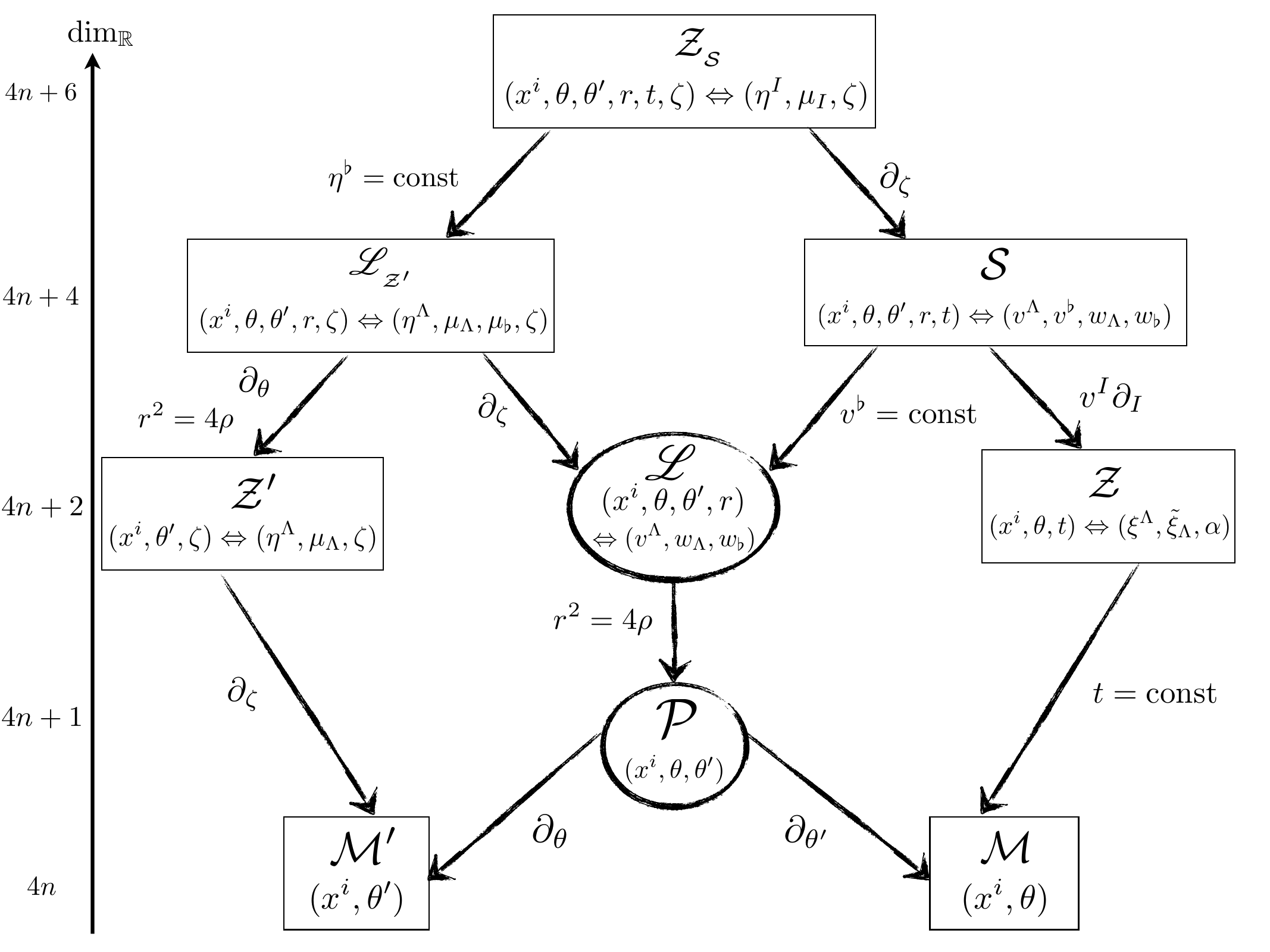}
\caption{Overview of the QK/HK correspondence.}
\label{QKHK}
\end{figure}
\end{center}

\subsection*{QK/HK correspondence, hyperholomorphic bundles and dilogarithm identities}

The fact that the D-instanton corrected QK metric on $\cM$ is captured
by a (Lorentzian) \hk metric on a different manifold $\pcM$ is a consequence of a general
geometric construction, which we dub the {\it QK/HK correspondence}.
We study this correspondence for its own sake in \S \ref{seq_qkhk}, and briefly
outline it here (see Figure \ref{QKHK} for orientation).

To any QK manifold $\cM$ with a quaternionic circle action $U(1)_A$,
generated by a Killing vector field $\partial_{\sigp}$,
the correspondence associates a HK manifold $\pcM$ of the same
dimension with an isometric  circle action $U(1)_R$,
generated by a rotational Killing  vector field $\partial_{\phip}$,
equipped with a canonical hyperholomorphic circle bundle $\cP$ and a connection
$\conhyper$.
This relation proceeds by lifting the  circle action $U(1)_A$
 on $\cM$ to a tri-holomorphic action on the Swann bundle $\cS$ (or \hk cone)
of $\cM$, and then constructing the \hk quotient
\be
\pcM\equiv \cS///U(1)_A
=\cS\cap \{\vec\mu=\vec m\}/U(1)_A\, ,
\ee
where $\vec\mu$ is the moment map of $U(1)_A$ action on $\cS$,
and $\vec m$ is a fixed unit norm vector (the direction of $\vec m$ is
immaterial, since it rotates under the $SU(2)_R$ isometric action on $\cS$).
The  result is a \hk
manifold $\pcM$ with positive signature if $\cM$ has positive scalar curvature, or
Lorentzian signature if $\cM$ has negative scalar curvature. Moreover, $\pcM$ admits
an isometric $U(1)_R$ action
which rotates the complex structures (here
$U(1)_R$ is the subgroup of  $SU(2)_R$
which preserves the vector $\vec m$). The level
set $\cP = \cS\cap\{\vec\mu=\vec m\}$ is
generically a circle bundle $\cP$ over $\cM$, equipped with a connection
$\conhyper$ given by the restriction of the Levi-Civita connection on $\cS$.
It is well known that this connection is hyperholomorphic, in the sense that
its curvature $\fb=\de\conhyper$ is of type (1,1) in all complex structures.
In a fixed complex structure, the circle bundle $\cP$ can
be extended to a complex line bundle $\lb$ with first Chern class $c_1(\lb)=\de\conhyper/(2\pi)$.
By the standard twistor construction,
it can be further lifted to  a holomorphic line bundle $\lb_{\pcZ}$ on the twistor
space $\pcZ$, with the property of being trivial along the $\CP$-fiber.

The QK/HK correspondence becomes particularly transparent at the level of
the twistor spaces $\cZ$ and $\pcZ$.
Each of them is determined by an open covering $\{\cU_i\}$ (respectively $\{\cU'_i\}$)
and a set of transition functions $\qHij{ij}(\Xi,\alpha)$ (respectively $\hHij{ij}(\pXi,\varpihk)$)
describing contact (respectively symplectic) transformations
between Darboux coordinates in different patches (see \S\ref{sec_twi}).
It turns out that the twistor spaces of a dual pair $(\cM,\pcM)$ can be described
by isomorphic coverings and  identical transition functions
\be
\hHij{ij}(\Xi)=\qHij{ij}(\Xi)\, .
\label{relHKQKintro}
\ee
This identification in \eqref{relHKQKintro}
is meaningful since the $U(1)_A$ isometry on $\cM$ requires
the transition functions $\qHij{ij}$
to be independent of $\alpha$, while the $U(1)_R$ isometry on $\pcM$
requires $\hHij{ij}$ to be independent of the $\CP$ coordinate $\varpihk$.
In such a scheme,  the Darboux coordinates
$(\Xi,\alpha)$ on $\cM$ and $(\Xi',\Upsilon)$ on $\lb_{\pcZ}\to \pcZ$
(where $\Upsilon$ is a holomorphic section of $\lb_{\pcZ}$)
are simply related
by\footnote{The last
relation in \eqref{identXiintro} must be altered in the presence of an anomalous dimension,
see \eqref{identW}.}
\be
\label{identXiintro}
\etai{i}^\Lambda(\varpihk)=\xii{i}^\Lambda(\varpiqk)\, ,
\qquad
\mui{i}_\Lambda(\varpihk )= \txii{i}_\Lambda(\varpiqk)\, ,
\qquad
\Upsilon\ui{i}(\varpihk) = e^{2\I\pi\ai{i}(\varpiqk)}\, ,
\ee
while the fiber coordinates $\varpiqk$ on $\cZ$ and $\varpihk$ on $\pcZ$ are
related by a phase rotation
\be
\varpiqk=\varpihk\, e^{-\I\phip}.
\label{relCP}
\ee
The transition functions for the line bundle $\lb_{\pcZ}$
are furthermore determined by the holomorphic section $\Upsilon$ via
\be
\label{fijLintro}
f_{ij} = \frac{\Upsilon\ui{i}}{\Upsilon\ui{j}}  = \exp\left( \frac{S\ui{ij}(\Xi)}{2\pi\I} \right),
\ee
where $S\ui{ij}$ is the same function which governs the shift of $\alpha$ under the contact transformation \eqref{cxi}
(the consistency conditions on $V\ui{ij}$  spelled
out in  \cite{Alexandrov:2008nk} ensure that $f_{ij}$ is indeed a Cech cocycle).
Thus, the geometry of the twistor space $\cZ$ of the QK manifold $\cM$ is completely
encoded in a suitable line bundle over the twistor space $\pcZ$
of the dual HK manifold $\pcM$.
As we shall see, one  advantage of this dual point of view is that it provides
a rigorous definition of the Darboux coordinates $\Xi(\varpiqk,x^\mu),\alpha(\varpiqk,x^\mu)$
on $\cZ$, which is difficult on the QK side due to the non-trivial fibration
of the twistor sphere $\CP$ over $\cM$.

On the basis of this correspondence, specifying  the D-instanton corrections to the QK metric
on $\cM$ is therefore equivalent to constructing
a specific hyperholomorphic line bundle $\lb\rightarrow\pcM$, defined by
transition functions of the form \eqref{fijLintro} such that the transformation \eqref{cxi}
leaves the contact one-form \eqref{c1f} invariant. Although not phrased in this way,
this problem was already addressed in \cite{Alexandrov:2008gh,Alexandrov:2009zh}, where
the generating function $S\ui{ij}$  of the symplectomorphism $U\ui{ij}$ was computed
in terms of the Spence dilogarithm function ${\rm Li}_2(z)$ and the BPS invariants
$\Omega(\gamma,z)$. While the formula obtained in \cite{Alexandrov:2009zh}
(reproduced below in \eqref{dirtySgamma}) was rather unilluminating,
one of the main new insights in the present work is to recognize that the
shifted, symplectic invariant coordinate
\be
\talp=-2\alpha-\xi^\Lambda\txi_\Lambda
\label{deftalp}
\ee
transforms in a much simpler fashion, namely $\talp\mapsto\talp + \Delta_\gamma\talp $
with\footnote{This relation holds when the quadratic refinement $\sigma(\gamma)$
is equal to $+1$, see \eqref{lga} for the general statement.}
\be
\label{lgaintro}
\Delta_\gamma\talp = \frac{1}{2\pi^2} \, \Omega(\gamma) \,
L\left(e^{-2\pi\I\langle\gamma,\Xi\rangle}\right),
\ee
where $L(z)$ is  known as the Rogers dilogarithm,
\be
\label{defrogers}
L(z) = \Li_2(z)+\frac12 \log z \log(1-z)\, .
\ee
As we shall see, the functional identities obeyed by the  Rogers dilogarithm, such as the five-term relation
\be
\label{Lreal5term2intro}
L(x) -L\left(\frac{x(1-y)}{1-xy}\right)- L\left(\frac{y(1-x)}{1-xy}\right) + L(y) - L(xy)=0\, ,
\quad 0<x,y<1 ,
\ee
are instrumental for ensuring
the consistency of the gluing conditions \eqref{cxi} across walls of marginal stability.
In \S \ref{secmot} we show that the semi-classical limit of
the motivic wall-crossing formula of Kontsevich and Soibelman \cite{ks}
(a much more powerful statement than the usual numerical
KS formula) implies a set of functional identities for the Rogers dilogarithm which
guarantee the existence of the line bundle $\lb_{\pcZ}$. We illustrate the construction
on some simple examples of wall-crossing involving a finite number of BPS states
on either side, where consistency is ensured by the famous pentagon identity
and perhaps less known hexagon and octagon identities for the Rogers dilogarithm.

\subsection*{Relation with cluster algebras}

Although this point of view is not essential for attaining our results, it should
be mentioned that the special wall-crossing identities mentioned above occur naturally  in the context
of cluster algebras (specifically, cluster algebras associated to rank 2 Dynkin quivers).
Indeed, cluster algebras provide a powerful method to generate wall-crossing identities
and dilogarithm identities\footnote{Such dilogarithm identities appeared first
in studies of two-dimensional integrable models \cite{Zamolodchikov:1991et,Ravanini:1992fi,Kuniba:1992ev,Gliozzi:1994cs}, see e.g. \cite{Kuniba:2010ir} for a recent review.
Cluster algebras can be regarded as an abstraction of the Y-systems appearing in
these models.}
 \cite{FG,Nakanishi:2009,Nakanishi:2010,Keller:2010,Keller:2010b}.
In fact, our derivation in \S \ref{secmot} is a simple generalization
of the analysis of \cite{Kashaev:2011} (in turn generalizing  \cite{Faddeev:1993rs}),
where it was pointed out
that these functional identities can be obtained as semi-classical limits
of quantum dilogarithm product identities for simply laced quivers
established in \cite{MR2470108,MR2402412,Keller:2011}. Our derivation
shows that the conjectural formulas of Nakanishi \cite{Nakanishi:2010}
for cluster algebras associated to non-simply laced quivers follow in the
same manner from the motivic wall-crossing formula of Kontsevich and
Soibelman \cite{ks}. In  Appendix \ref{app_cluster},  we summarize some basic facts
about cluster algebras, cluster transformations and their relation to wall-crossing,
and derive the pentagon, hexagon, and octagon identities from periods of
the cluster algebras associated to the $A_2$, $B_2$ and $G_2$ Dynkin quivers.

\subsection*{Historical remarks}

We close this introduction with some historical remarks and
pointers to related literature.
The basic tenets of the QK/HK correspondence were
noticed by A.~Neitzke and the third-named author in 2008
\cite{BPNeitzke}, in trying to understand the geometric meaning of the
`freezing procedure' used to extract the `rigid limit' of local $c$-map spaces
\cite{Rocek:2005ij,Saueressig:2007dr}.
After the main results in this article were
obtained, we learned from A. Neitzke that the QK/HK correspondence
had been independently discovered by A.~Haydys \cite{1143.53043}
in 2007 (cf. \cite{Hitchin-Newton2011,SwannTalk}
for further accounts of Haydys' construction). For completeness, we shall incorporate some
further insights gleaned from \cite{1143.53043}. Moreover, the fact that the local and
rigid $c$-map metrics are related  by the QK/HK-correspondence (see \S\ref{cmap})
appears to have been  noticed independently  by O.~Macia and A.~Swann \cite{SwannTalk}.
Our construction of the hyperholomorphic line bundle $\lb$
also seems closely related to work in progress by V.~Fock and A.~Goncharov in the
context of quantization of cluster varieties  \cite{FGgeometricquantization}.
Finally, A.~Neitzke has  independently
constructed a hyperholomorphic line bundle over the Coulomb
branch of $\cN=2$ gauge theories on $\IR^3\times S^1$ \cite{Neitzke:2011za}.\footnote{We
are grareful to A.~Neitzke for drawing our attention to \cite{1143.53043,Hitchin-Newton2011}
and sharing with us an advance draft of \cite{Neitzke:2011za}.}
His construction appears to match ours in superconformal cases where the prepotential is
homogeneous of degree 2.

\subsection*{Outline}

This paper is organized as follows. In \S\ref{seq_qkhk} we present the generalities
of the QK/HK correspondence, and illustrate it in the case of four-dimensional HK and
QK metrics with a rotational isometry (which can all be described in terms of solutions
to Toda equation) and in the case of $c$-map metrics. In particular, we show
that the `local' and rigid' $c$-map metrics, for the same choice of prepotential, are related
by the QK/HK correspondence. In \S \ref{contactwallcrossing}, we specialize to
the hypermultiplet moduli space of $\cN=2$ string vacua, and express
the contact transformations \eqref{cxi} across BPS rays in terms of the Rogers
dilogarithm. We show that the consistency of the construction across walls of marginal
stability is ensured by the classical limit of the motivic wall-crossing formula. We illustrate
this mechanism on simple examples of wall-crossing, related to cluster algebras
of type $A_2, B_2$ and $G_2$. Appendix \ref{sec_enhrog} recalls the definition and main properties of the Rogers
dilogarithm and some of its variants. Finally, for completeness we review in Appendix \ref{app_cluster} some basic
aspects of cluster algebras and their relation to wall crossing.

\section{The QK/HK correspondence}
\label{seq_qkhk}

In this section we present a general geometric correspondence between
QK and HK manifolds with a rotational
Killing vector field. More precisely, to any real $4n$-dimensional QK manifold
$\cM$ with a quaternionic isometry, we associate a
HK manifold $\pcM$ of the same dimension, equipped with a rotational Killing vector field and
 a canonical hyperholomorphic circle bundle $\cP\rightarrow \pcM$ with
connection $\conhyper$.
In \S \ref{sec_twi} we start with a brief reminder on the twistorial description
of HK and QK manifolds.
In \S \ref{quotient} we construct $(\pcM,\cP)$ by lifting the quaternionic
isometry on $\cM$ to a  tri-holomorphic action
on the Swann bundle $\IR^4/\IZ_2\rightarrow \cS\rightarrow \cM$, and
then taking the \hk quotient.
In \S \ref{oneQKdimension},
we study the correspondence in detail for QK/HK manifolds of real dimension 4, where
both sides can be described in terms of the same solution to Toda equation.
Finally, in \S \ref{cmap} we work out the QK/HK correspondence for $c$-map spaces,
and show that the  rigid and the local $c$-map (including the one-loop deformation
parameter) are related by the QK/HK correspondence.

\subsection{Twistorial descriptions of HK and QK manifolds: a reminder\label{sec_twi}}

In this section we briefly recall some basic features and relevant formulae for the twistorial description of HK
and QK manifolds (see \cite{Alexandrov:2008ds,Alexandrov:2008nk} for more details).

\subsubsection{Hyperk\"ahler manifolds and symplectic geometry \label{sec_twihk}}

A dimension $4n$ Riemannian manifold $\pcM$ is \hk (HK) if it has restricted holonomy group
$USp(n)\subset SO(4n)$. We shall also allow for Lorentzian-HK manifolds, whose holonomy
group lies in $USp(1,n-1)\subset SO(4,4n-4)$. In either case,
$\pcM$ can be described analytically in terms of its twistor space $\pcZ=
\pcM\times \CP$.  $\pcZ$ admits a complex symplectic structure, more precisely a
closed two-form
\be
\pomega=\pomega_+-\I\varpihk \pomega_3+\varpihk^2 \pomega_-\, ,
\qquad
\pomega_\pm=-\frac12\,(\pomega_1\mp\I\pomega_2)
\ee
valued in the $\cO(2)$ line bundle over $\CP$, and
non-degenerate along the fibers of the projection $\pcZ\to\pcM$.
Here $\pomega_i$ are the symplectic forms on $\pcM$ associated to the complex
structures $J_i$, $i=1,2,3$, satisfying the quaternion algebra and $\varpihk$
is a complex coordinate on $\CP\cong SU(2)/U(1)$ parametrizing the complex structure
\be
J(\varpihk) =\I \, \frac{\bar\varpihk-\varpihk}{1+|\varpihk|^2}\, J_1
+\frac{\varpihk+\bar\varpihk}{1+|\varpihk|^2} \, J_2
+\frac{1-|\varpihk|^2}{1+|\varpihk|^2}\, J_3\, .
\label{complexstr}
\ee
In particular, the pull-back of $\pomega$ to $\pcM$
is the \kahler form with respect to $J(\varpihk)$.

Locally, on a patch
$\hcU'_i$ of an open covering $\cup\, \hcU'_i$ of $\pcZ$,
there exist complex  Darboux coordinates
$(\etai{i}^\Lambda,\mui{i}_\Lambda)$ such that\footnote{We use a patch-dependent normalization, such that in the patch $\cU'_+$
around the north pole of $\CP$ ($\varpihk=0$) it is given by $\kappa_+=\I\varpihk/4$.}
\be
\pomega=\kappa_i\,\de\etai{i}^\Lambda\wedge \de\mui{i}_\Lambda \, .
\ee
The complex symplectic structure is conveniently  encoded in  a set of
holomorphic functions $\hHij{ij}(\etai{i},\mui{j},\varpihk)$ which generate
symplectomorphisms between the Darboux coordinates
$(\etai{i}^\Lambda,\mui{i}_\Lambda)$ on the overlap $\hcU'_i\cap\hcU'_j$ \cite{Alexandrov:2008ds}.
The functions $\hHij{ij}$ are subject to certain cocycle and reality conditions. Moreover,
since $\pomega$
is defined only up to closed two-forms which vanish on the fiber of the projection $\pcZ\to\pcM$,
$\hHij{ij}$ may in general depend explicitly on $\varpihk$.
To obtain the metric, one needs to `parametrize the twistor lines', i.e.
to find the Darboux coordinates $(\eta^{\Lambda}, \mu_\Lambda)$
as functions of $\varpihk$ and of the coordinates on $\pcM$.
The Darboux coordinates are determined by the following integral equations
\cite{Alexandrov:2009zh}
\be
\begin{split}
\etai{i}^\Lambda(\varpihk)& = x^\Lambda +
\varpihk^{-1} v^\Lambda - \varpihk \bv^\Lambda
-\frac12 \sum_j \oint_{C_j}\frac{\de\varpihk'}{2\pi\I \varpihk'}\,
\frac{\varpihk'+\varpihk}{\varpihk'-\varpihk}\, \p_{\mui{j}_\Lambda }\hHij{ij}(\varpihk') ,
\\
\mui{i}_\Lambda(\varpihk)& = \vrh_\Lambda +
\half  \sum_j \oint_{C_j} \frac{\de \varpihk'}{2 \pi \I \varpihk'} \,
\frac{\varpihk' + \varpihk}{\varpihk' - \varpihk}
\, \p_{\etai{i}^\Lambda } \hHij{ij}(\varpihk'),
\end{split}
\label{txiqline}
\ee
where $C_j$ is the contour surrounding the projection of $\hcU'_j$ on $\CP$
in the counterclockwise direction, while the complex variables $v^\Lambda$ and
real variables $x^\Lambda, \vrh_\Lambda$ serve as coordinates on $\pcM$.
The sums in \eqref{txiqline} run over all patches including those
which do not intersect with $\cU'_i$ --- the corresponding transition functions
are obtained by analytic continuation and by applying the
cocycle condition. Once the Darboux coordinates are known,
the K\"ahler potential in complex structure $J_3\equiv J(\varpihk=0)$
is given by the contour integral \cite{Lindstrom:2008gs,Alexandrov:2009zh}
\be
\label{Kdefrew}
K_{\pcM}= \frac{1}{8\pi}\sum_j\oint_{C_j} \frac{\de\varpihk}{\varpihk}
\[\hHij{ij}-\etai{i}^\Lambda \p_{\etai{i}^\Lambda}\hHij{ij}
+\(\varpihk^{-1} v^{\Lambda}-\varpihk \bv^{\Lambda} \)\p_{\etai{i}^\Lambda} \hHij{ij} \].
\ee
A set of complex coordinates on $\pcM$ in this complex structure is given by the
leading Laurent coefficients in the expansion of $\etai{+}^\Lambda$ and $\mui{+}_\Lambda$
at small $\varpihk$, namely $v_\Lambda$ and
\be
w_\Lambda\equiv \frac{\I}{2}\,\mui{+}_\Lambda|_{\varpihk=0}=\frac{\I}{2}\,\vrh_\Lambda +
\frac{1}{8\pi}  \sum_j \oint_{C_j} \frac{\de \varpihk}{ \varpihk}
\, \p_{\etai{i}^\Lambda } \hHij{+j}.
\label{defholcoor}
\ee
For HK manifolds with a rotational isometry, the transition functions $\hHij{ij}$
must have no explicit dependence on the fiber coordinate $\varpihk$.

\subsubsection{Quaternion-\kahler manifolds and contact geometry \label{sec_twiqk}}

A $4n$-dimensional Riemannian manifold $\cM$ is \qk (QK) if it has restricted
holonomy group $USp(n)\times SU(2) \subset SO(4n)$. The Ricci scalar $R$ is then
constant, and the curvature of the $SU(2)$ part of the Levi-Civita connection, rescaled
by $1/R$, provides a  triplet of quaternionic 2-forms $\vec \omega$. While $R$ can
take either sign, hypermultiplet moduli spaces  in $\cN=2$ supergravity or string theory
models have $R<0$. The degenerate limit $R\to 0$ recovers the case of HK manifolds
discussed in \S\ref{sec_twihk}.

A QK manifold $\cM$
 can be described analytically in (at least) two equivalent ways,
either in terms of  its
twistor space $\cZ$, or in terms of the Swann bundle $\cS$ (and its twistor space $\cZ_\cS$).
The Swann bundle, or HK cone, is
the total space $\cS$ of the $\IR^4/\IZ_2$ bundle over $\cM$ twisted with the $SU(2)$
part of the Levi-Civita connection on $\cM$ \cite{MR1096180,deWit:2001dj}.
It is a HK manifold\footnote{More precisely, $\cS$ carries a \hk metric if $\cM$
has positive scalar curvature, or a pseudo-\hk
metric with signature $(4,4n)$ if $\cM$ has negative scalar curvature.}
with a homothetic action of $\IR^+$ and an isometric
action of $SU(2)_R$ (here $R$ stands for R-symmetry).
The twistor space $\cZ_{\cS}$ of the Swann bundle
provides an analytic description of the  QK space $\cM$ in terms of a homogeneous
complex symplectic structure. On the other hand, the twistor space $\cZ$
is the total space  of the $\CP$ bundle over
$\cM$ twisted with the \emph{projectivized} $SU(2)_R$ connection on $\cM$.
$\cZ$ is a \kahler-Einstein space equipped with a canonical complex contact structure,
given by the kernel of the one-form
\be
\label{defDt}
D\varpiqk= \de \varpiqk+p_+ - \I p_3  \varpiqk+p_- \varpiqk^2,
\ee
where $\varpiqk$ is a complex coordinate on $\CP$, and $p_\pm,p_3$
is the $SU(2)$ part of the Levi-Civita connection on $\cM$. The curvature
of the latter is related to the triplet of covariantly constant two-forms by
(we set the cosmological constant $\Lambda=-6$)
\be
\label{ptoom}
\omega_+=-\hf\(\de p_+ + \I p_+\wedge p_3\) ,
\qquad
\omega_3=-\hf\(\de p_3-2\I p_+\wedge p_-\).
\ee
Note that $D\varpiqk$ is defined only projectively, as it rescales under $SU(2)$
rotations. More precisely, it is valued in the $\cO(2)$ line bundle on $\CP$ \cite{MR664330}.

The two descriptions of $\cM$ outlined above are closely related, since the twistor space $\cZ$ is the quotient
of the Swann bundle $\cS$ by the $\IC^\times$ action which combines the dilation
and $U(1)_R\subset SU(2)_R$ rotation. The complex contact structure
on $\cZ$ is then simply the projectivization of the homogeneous  complex symplectic
structure on $\cS$. The HKC metric on $\cS$ and the \kahler-Einstein metric
on $\cZ$ are related to the \qk metric on $\cM$ by \cite{Alexandrov:2008nk}
\be
\de s^2_\cS=\de r^2+r^2\[\frac{1}{4}\(D \phip\)^2+ \de s^2_\cZ\] ,
\qquad
\de s^2_\cZ =
\frac{|D \varpiqk|^2}{(1+\varpiqk\bar\varpiqk)^2}-\hf\,\de s^2_\cM\, ,
\label{metricS}
\ee
where $(r,\phip)$ parametrizes the $\IC^\times$ fiber
and
\be
\label{defDphip}
D\phip = \de \phip+\frac{\I}{1+\varpiqk\bar\varpiqk}\[ \varpiqk\de\bar\varpiqk-\bar\varpiqk\de\varpiqk
-\I(1-\varpiqk\bar\varpiqk)p_3+2\varpiqk p_--2\bar\varpiqk p_+\].
\ee
Moreover, the \kahler forms on $\cS$ and $\cZ$ are given by
\be
\label{om3SZ}
\omega^3_{\cS} = \hf\,r\, \de r \wedge D\phip + r^2 \, \omega_{\cZ}\, ,
\qquad
\omega_{\cZ}=\frac{\I}{2}\,\frac{D \varpiqk \wedge D\bar\varpiqk}{(1+\varpiqk\bar\varpiqk)^2}
-\frac{(1-\varpiqk\bar\varpiqk)\omega_3+2\I\varpiqk \omega_- -2\I \bar\varpiqk \omega_+ }{2(1+\varpiqk\bar\varpiqk)}
\ee
while the complex symplectic form on $\cS$ is
\be
\omega^+_{\cS} = - \frac{r^2 e^{\I\phip}}{2(1+\varpiqk\bar\varpiqk)}\biggl[\omega_+-\I\varpiqk \omega_3+\varpiqk^2 \omega_-
-\(\frac{\I}{2}\,D\phip+\frac{\de r}{r}\)\wedge D\varpiqk \biggr].
\label{holom-om}
\ee

Locally, on a patch
of an open covering $\{\hcU_i\}$ of $\cZ$,
one can always find complex  Darboux coordinates  $(\xii{i}^\Lambda,\txii{i}_\Lambda,\ai{i})$
such that the contact one-form \eqref{defDt}, suitably rescaled by a function $e^{\Phi\ui{i}}$, takes the form
\be
\cX^{[i]}\equiv 4 \, e^{\Phi\ui{i}}\, \frac{D\varpiqk}{\I\varpiqk} =  \de\ai{i}+ \xi_{[i]}^\Lambda \de \tilde{\xi}^{[i]}_\Lambda\, .
\label{contform}
\ee
The function $\Phi\ui{i}$, which we refer to as the `contact potential', is holomorphic along
the $\CP$ fiber, and defined up to an additive holomorphic function on
$\hcU_i$.  It provides, among other things,
a K\"ahler potential for the \kahler-Einstein metric on $\cZ$ \cite{Alexandrov:2008nk}:
\be
\label{Knuflat}
K\ui{i}_{\cZ} = \log\frac{4(1+\varpiqk\bar \varpiqk)}{|\varpiqk|}+\Re\Phi\ui{i}\, .
\ee

Globally, the complex contact structure on $\cZ$ can be specified by a set of generating
functions $\qHij{ij}(\xii{i}^\Lambda,\txii{j}_\Lambda,\ai{j})$ for complex contact transformations
between Darboux coordinates on overlaps
$\hcU_i\cap \hcU_j$,
subject to cocycle and reality conditions \cite{Alexandrov:2008nk}. Unlike the HK case,
the transition functions $\qHij{ij}$ are independent of the coordinate $\varpiqk$ on the
$\CP$ fiber.
In the case when the QK-manifold $\cM$ has a quaternionic isometry
$\partial_\sigp$, one may choose the Darboux coordinates such that the Killing
vector lifts to the holomorphic action $\pa_\alpha$. As a result, the
transition functions $\qHij{ij}$ become independent of the coordinate $\alpha\ui{j}$,
and the contact potential $\Phi^{[i]}$ becomes constant on
$\CP$ \cite{Alexandrov:2008gh}.\footnote{As we shall see, $e^\Phi$ is related to the
norm of the moment map of $\partial_\sigp$ on $\cM$, or to the moment map of
$\pa_\phip$ on the dual HK manifold $\pcM$.}
In this case the Darboux coordinates are determined by the following system of integral equations:
\beq
\xii{i}^\Lambda(\varpiqk)& =& A^\Lambda +
\varpiqk^{-1} Y^\Lambda - \varpiqk \, \bY^\Lambda
-\frac12 \sum_j \oint_{C_j}\frac{\de\varpiqk'}{2\pi\I \varpiqk'}\,
\frac{\varpiqk'+\varpiqk}{\varpiqk'-\varpiqk}
\,\p_{\txii{j}_\Lambda }\qHij{ij},
\nonumber \\
\txi_\Lambda^{[i]}(\varpiqk)& = &  B_\Lambda +
\half  \sum_j \oint_{C_j} \frac{\de \varpiqk'}{2 \pi \I \varpiqk'} \,
\frac{\varpiqk' + \varpiqk}{\varpiqk' - \varpiqk}
\, \p_{\xii{i}^\Lambda } \qHij{ij},
\label{txiqlineQK}
\\
\ai{i}(\varpiqk)& = & B_\alpha +
\half  \sum_j \oint_{C_j} \frac{\de \varpiqk'}{2 \pi \I \varpiqk'} \,
\frac{\varpiqk' + \varpiqk}{\varpiqk' - \varpiqk}
\( \qHij{ij}- \xii{i}^\Lambda \p_{\xii{i}^\Lambda}\qHij{ij}\)
+4\I c \log\varpiqk .
\nonumber
\eeq
Here the complex variables $Y^\Lambda$, up to an overall phase rotation
which can be absorbed into a phase rotation of $\varpiqk$,
and the real variables $A^\Lambda,B_\Lambda,B_\alpha$
serve as coordinates on $\cM$. It is convenient to fix the phase freedom in  $Y^\Lambda$
by requiring $Y^0\equiv \cR$ to be real.
Moreover, $B_\alpha$ is related to the coordinate $\sigp$ along the isometric direction
by $\pa_{B_\alpha}=\frac14 \pa_\sigp$. Finally,
$c$ is a real constant known as an anomalous dimension \cite{Alexandrov:2008nk},
which characterizes the singular behavior of the Darboux coordinate $\alpha$
at the north and south poles of $\CP$. It plays an important physical role in describing
 the one-loop correction to the hypermultiplet moduli space metric in type II
 string compactifications.\footnote{In fact, there are
 other anomalous dimensions, called $c_\Lambda$,
which introduce logarithmic singularities in $\txi_\Lambda$ and further affect $\alpha$ \cite{Alexandrov:2008nk}.
In this work we restrict ourselves to the case of vanishing $c_\Lambda$ because they do not seem to play
any role in physical applications. From the point of view of the QK/HK correspondence, it appears
that their inclusion does not affect the dual HK metric,
but does affect the hyperholomorphic connection.}

The procedure to extract the metric from the solutions of \eqref{txiqlineQK} was outlined
in \cite{Alexandrov:2008nk,Alexandrov:2008gh}.
Similarly to $K_{\pcM}$, the contact potential can be computed from the transition
functions $\qHij{ij}$ and the solutions of  \eqref{txiqlineQK} using
\be
e^{\Phi}=\frac{1}{16\pi} \sum_j\oint_{C_j}\frac{\de\varpiqk}{\varpiqk}
\(\varpiqk^{-1} Y^{\Lambda}-\varpiqk \, \bY^{\Lambda} \)
\p_{\xii{i}^\Lambda } \qHij{ij}
-c.
\label{contpotconst}
\ee

\subsection{The QK/HK correspondence}
\label{quotient}

We start from a QK manifold $\cM$ of real dimension $4n$, with a quaternionic Killing vector
field\footnote{Recall that a quaternionic  vector field is a vector field which preserves
the canonical closed 4-form $\vec \omega \wedge \vec \omega$ on the QK manifold $\cM$.}
 which we denote by $\partial_{\sigp}$. We assume that the action of $\partial_{\sigp}$
exponentiates to a circle action of a group which we denote by $U(1)_A$ ($A$ stands
for axion). By the moment map construction \cite{MR872143,MR960830}, the vector
field $\partial_\sigp$ lifts to a tri-holomorphic Killing vector field on the Swann
bundle $\cS$.
We abuse notation and denote by the same symbol $\partial_\sigp$ the vector
field on $\cM$ and its tri-holomorphic lift to $\cS$, and by
$\vec \mu$ its moment map.

Let us now perform the \hk quotient of $\cS$ by $U(1)_A$.
This proceeds by first restricting to a fixed level set of
the moment map,
\be
\label{defPm}
\cP(\vec m) = \cS\, \cap\{\vec\mu=\vec m \}
\ee
and then performing the usual Riemannian quotient by $U(1)_A$. For $\vec m\neq 0$,
the action of $U(1)_A$ on $\cP(\vec m)$ has at most a finite stabilizer, and the quotient
\be
\pcM = \cP(\vec m) / U(1)_A
\ee
is a \hk orbifold. Due to the $SU(2)_R$ and dilation symmetries on $\cS$,
the spaces $\cP(\vec m)$ (respectively, $\pcM(\vec m)$) for varying $\vec m\neq 0$ are canonically
isomorphic, and the induced (respectively, quotient)
metric depends only on the norm of $\vec m$, by an overall factor.
We shall set  $|\vec m|=1$ in the following and omit the dependence on the vector $\vec m$.

As usual, the \hk quotient may be decomposed
in three steps: (i) impose $\mu_+=0$, where $\mu_+$ is the complex valued
projection of the moment map $\vec\mu$ on the plane orthogonal to $\vec m$,
(ii) impose $\mu_3=1$, where $\mu_3=\vec\mu\cdot\vec m$,
and (iii) mod out by $U(1)_A$. Step (i) defines a complex
submanifold of $\cS$ in complex structure $\vec m\cdot \vec J$. Steps (ii) and (iii)
are equivalent to modding out by the complexification $\IC^\times_A$ of $U(1)_A$,
which equips $\pcM$ with a complex structure which we continue to denote by
$ \vec m\cdot\vec J$. Since $\pcM$ is independent of the direction $\vec m$,
it admits an $S^2$ worth of complex structures, and is indeed \hk, with positive
signature if $\cM$ has positive scalar curvature, or Lorentzian signature is $\cM$ has
negative scalar curvature.  Moreover, since $\cP$
is invariant under the $U(1)_R$ subgroup of $SU(2)_R$ which leaves invariant the
direction of the vector $\vec m$, and since $U(1)_A$ commutes with $SU(2)_R$,
the action of $U(1)_R$ descends to an isometric action on the quotient $\pcM$.
We denote by $\pa_\phip$ the corresponding vector field. Denoting by $J_3$
the projection of $\vec J$ along $\vec m$ and by $J_+$
the orthogonal projection, the $U(1)_R$ action
rotates the complex structure according to
\be
\label{U1RJ}
J_3\mapsto J_3, \qquad J_+\mapsto e^{\I\phip} J_+\, .
\ee

By construction, the level set $\cP$ is the total space of a circle bundle over $\pcM$. It is equipped
with a canonical connection one-form $\conhyper$, namely the restriction of the Levi-Civita
connection on $\cS$. It is a well-known fact  that this connection is hyperholomorphic, i.e.
that the curvature $\fb=\de\conhyper$ is of type (1,1) in all complex structures
on $\pcM$ \cite{MR967469,MR1139657,MR1486984}. The complex line bundle $\lb$
associated to the circle bundle $\cP$ in complex structure $J_3$ is isomorphic to the locus
$\{\mu_+=0\}\cap \cS$ referred to above. The connection $\lambda$ on $\cP$
defines a unitary connection on $\lb$ which we continue to denote by the same symbol.

\subsubsection{From QK to HK\label{sub_qktohk}}

Let us now perform the quotient procedure discussed above
using the explicit formula for the metric on the Swann bundle \eqref{metricS}.
To this end, note that any QK metric with a quaternionic Killing vector $\pa_\sigp$ can be written as
\be
\label{dsqkgen}
\de s^2_{\cM} =  \cff \(\de\sigp+\Theta\)^2 + \de s^2_{\cM/\pa_\sigp},
\ee
where $\Theta$ is a connection one-form and $\cff$ is a function on $\cM$ invariant
under $\pa_\sigp$.\footnote{Note that $\sigp$ and $\Theta$ depend on the choice of coordinate but
$\cff$ and $D\sigp\equiv \de\sigp+\Theta$ are defined unambiguously.}
We choose an $SU(2)$ frame such that the Lie derivative of $\vec p$
with respect to $\pa_\sigp$ vanishes, and such that the QK moment map $\vec \mu_{\cM}$
for $\pa_\sigp$ is aligned along the third axis. Denoting  by
$1/(4\rho^2)$ its squared norm, so that $\vec \mu_{\cM} = (0,0,1)/(2\rho)$,
from \eqref{ptoom} it follows that the component
$p_3$ of the $SU(2)$ connection on $\cM$ must take the form
\be
p_3 = - \frac{1}{\rho} \( \de \sigp + \Theta\) + \pTheta
\label{genformp3}
\ee
for a certain one-form connection $\pTheta$ on $\cM/\p_\sigp$. For later convenience,
we further trade the function $\tau$ in \eqref{dsqkgen} for a function $\cfff$ such that
\be
\label{defcfff}
\cff=\frac{\cfff+\rho}{2\rho^2\cfff}.
\ee
From \eqref{metricS}, one then finds the following metric on the Swann bundle
\be
\de s^2_\cS=\de r^2+r^2\[\frac{1}{4}\(D \phip\)^2
+ \frac{|D \varpiqk|^2}{(1+\varpiqk\bar\varpiqk)^2}-\frac{\cff}{2} \(\de\sigp+\Theta\)^2
-\frac12 \,\de s^2_{\cM/\pa_\sigp}  \].
\ee

We now perform the \hk quotient of $\cS$ with respect to the tri-holomorphic
action of  $\pa_\sigp$. From \eqref{om3SZ} and \eqref{holom-om},
one finds that the components of the moment map $\vec \mu$ on the Swann bundle
are given by
\be
\mu_3=-\frac{1-\varpiqk\bar\varpiqk}{1+\varpiqk\bar\varpiqk}\,\frac{r^2}{4\rho}\ ,
\qquad
\mu_+=\frac{\I\varpiqk\, e^{\I\phip}}{1+\varpiqk\bar\varpiqk}\,\frac{r^2 }{4\rho}\ .
\ee
Therefore, the level set $\cP(\vec m)$ in \eqref{defPm}, with $\vec m$ a fixed
unit norm vector, is obtained by setting
\be
r=2\sqrt{\rho},
\qquad
t=\zeta e^{-\I\phip},
\label{fixmom}
\ee
and holding $\varpihk$ constant. As we shall see momentarily, $\varpihk$
parametrizes the twistor fiber of $\pcM$, and the last equality in \eqref{fixmom}
establishes the relation \eqref{relCP} between
the $\CP$ coordinates on $\cZ$ and $\pcZ$.
After completing squares, the restriction to $\cP(\vec m)$ of the metric on $\cS$
can be written
\be
\label{metricP}
\de s^2_{\cP}=\de s^2_{\pcM} - \frac1{\cfff} \left(
\de\sigp+\conhyper \right)^2\, ,
\ee
where $\de s^2_{\pcM}$ is a metric which is degenerate along $\pa_\sigp$ and
invariant under $\pa_\phip$,
\be
\label{dsPhk}
\de s^2_{\pcM} =\frac{\de\rho^2}{\rho} + 4 \rho \, |p_+|^2   +
(\cfff+\rho) \(\de\phip+\pTheta\)^2 -2\rho \, \de s^2_{\cM/\pa_\sigp}
\ee
and $\conhyper$ is the one-form
\be
\conhyper =\cfff \( \de\phip+\pTheta\) + \Theta\, .
\label{lambdlbTheta}
\ee
Performing the quotient with respect to $\pa_\phip$, the metric \eqref{dsPhk}
gives the metric on the HK space $\pcM$ dual to the QK space $\cM$,
while the one-form $\conhyper$ on the circle bundle $\cP(\vec m)$
is  the hyperholomorphic connection afforded by the QK/HK correspondence.
It is noteworthy that both the metric \eqref{dsPhk} and connection \eqref{lambdlbTheta}
are independent of the parameter $\varpihk$.

To check that the metric \eqref{dsPhk} is \hk, one may construct the
\kahler form $\pomega_3(\varpihk)$ and complex symplectic form
$\pomega_+(\varpihk)$ on  $\pcM$ by restricting the corresponding forms
\eqref{om3SZ}, \eqref{holom-om} on $\cS$ to $\cP(\vec m)$. With some further efforts, one finds that
$\pomega_3(\varpihk),\pomega_+(\varpihk)$ can be integrated
to the one-forms
\be
\kcon(\varpihk)=
\frac{(1-\varpihk\bar\varpihk) \kcon+2\I\varpihk\ahcon-2\I\bar\varpihk\hcon}{1+\varpihk\bar\varpihk} ,
\qquad
\hcon(\varpihk)=
\frac{\hcon-\I\varpihk\kcon+\varpihk^2\ahcon}{1+\varpihk\bar\varpihk} ,
\label{rot-con}
\ee
where $\kcon,\hcon$ are related to the $SU(2)$ connection on $\cM$ by
\be
\label{pqktohk}
\kcon=\rho \left( \de\phip+p_3\right) +\de\theta ,
\qquad
\hcon=\rho\, e^{\I\phip} p_+ .
\ee
In the first of these equations, the last term $\de\theta$ was chosen so as to cancel
the contraction $\pa_{\sigp}\cdot \kcon$.
In particular, these formulae identify the parameter $\varpihk$ as the standard stereographic
coordinate on the twistor space of $\pcM$, as anticipated below \eqref{fixmom}.
They also show that the Killing vector $\pa_\phip$ on $\pcM$ leaves the
complex structure $J_3=J(\varpihk=0)$ on $\pcM$ invariant, and rotate
$J_\pm=J_\pm(\varpihk=0)$ according to \eqref{U1RJ}. Furthermore, they identify the
coordinate $\rho$ as the moment map of the Killing vector $\pa_\phip$
with respect to $\pomega_3=\pomega_3(\varpihk=0)$. By the usual argument, this implies
that the coefficient of $\de\rho^2$ in the metric \eqref{dsPhk} must be inversely related to the
coefficient of $(\de\phip+\pTheta)^2$, namely
\be
\label{hkgenrho}
\de s^2_{\pcM} =\frac{\de\rho^2}{\cfff+\rho}+(\cfff+\rho) \(\de\phip+\pTheta\)^2 +\de s^2_{\pcM//\pa_\phip}
\ee
where $\de s^2_{\pcM//\pa_\phip}$ is the metric on the \kahler quotient of $\pcM$ by
$U(1)_R$.  In particular, the complex structure $J_3$ maps the one-form $\de\rho$
to $(\cfff+\rho)\( \de\phip+\pTheta\)$, and as a result
\be
\label{dmdbarrho}
\I (\pa-\bar\pa)\rho = -(\cfff+\rho)  \( \de\phip+\pTheta\) ,
\ee
where $\pa$ is the Dolbeault derivative in complex structure $J_3$.
Combining \eqref{genformp3}, \eqref{lambdlbTheta}, \eqref{pqktohk}  and \eqref{dmdbarrho},
we see that the hyperholomorphic one-form $\conhyper$ can be rewritten as
\be
\label{contohyp}
\conhyper = -\I (\pa-\bar\pa)\rho - \kcon\, ,
\ee
with curvature
\be
\label{FLgen}
\fb = \de \conhyper  = 2\I \, \pa\bar\pa \rho - \pomega_3 \, ,
\ee
in agreement with Eq. (14) in \cite{1143.53043}. In particular,
the hyperholomorphic curvature $\fb$ can be derived from the \kahler potential
\be
\label{KltoKc}
K_{\lb} = 2\rho - \Kc_{\pcM}\, ,
\ee
where $ \Kc_{\pcM}$ is a \kahler potential for $\pomega_3$ in complex structure $J_3$.

\subsubsection{From HK to QK}

The above construction can be inverted as follows.
Let $\pcM$ be a HK manifold with a rotational Killing vector $\pa_\phip$,
which lifts to a $U(1)_R$ circle action, and
acts on the \kahler form $\pomega_3$ and complex symplectic form $\pomega_+$
(in a fixed complex structure $J_3$) via
\be
\cL_{\pa_\phip}\pomega_3=0\, ,
\qquad
\cL_{\pa_\phip}\omega_\pm'=\pm\I\, \omega_\pm'\, .
\ee
We assume that $\pomega_3$ lies in an integer cohomology class.
Let $\rho$ be the moment map of $\pa_\phip$ with respect to $\pomega_3$.
As explained in \cite{Hitchin-Newton2011}, the two-form $\fb\equiv
2\I \, \pa\bar\pa \rho - \pomega_3$, where $\pa$ is the Darboux derivative
in complex structure $J_3$, is  of type (1,1) in all complex structures, hence
it defines a hyperholomorphic circle bundle $\cP$ on $\pcM$ with first
Chern class $c_1(\cP)=\fb/(2\pi)$. Let $\conhyper$ be a connection on $\pcM$,
such that $\de\lambda=\fb$.

Since $\rho$ is the moment map for the circle action, the \hk metric on $\pcM$ can always be written in  the form \eqref{hkgenrho}
for some function $\cfff$ and connection 1-form $\pTheta$ with
$\pa_\phip\cdot \cfff=\pa_\phip\cdot \pTheta=0$. Here, $\rho$ and $\cfff$
are defined up to an additive constant, but their sum $\rho+\cfff$ is unambiguous.
In addition, $\phip$ and $\pTheta$ depend on the choice of coordinate but
the combination $\de\phip+\pTheta$ is unambiguous.
Since $\lambda$ satisfies \eqref{FLgen}, it  can be written
as \eqref{contohyp} for some one-form $\kcon$ such that $\de \kcon=\pomega_3$.
Using \eqref{dmdbarrho}, this can be further decomposed into
\be
\label{defkcon}
\kcon = \rho \left( \de \phip+\pTheta \right) -\Theta ,
\qquad
\conhyper =\cfff \( \de\phip+\pTheta\) + \Theta
\ee
for some connection $\Theta$ with $\pa_\phip\cdot \Theta=0$.

We now equip the circle bundle $\cP$ with the metric \eqref{metricP}, invariant
under the $U(1)_A$ action generated by the Killing vector $\pa_\sigp$.
The $U(1)_R$ isometric action on $\pcM$ lifts to a $U(1)_R$ isometric
action on $\cP$ generated by the Killing vector $\pa_\phip$
(indeed, any other lift of the form $\pa_\phip+ \cfff_0 \pa_{\sigp}$ can be brought to this
form  by tuning the additive constant in $\cfff$). Using the
formulae \eqref{dsqkgen} and \eqref{dsPhk},
the metric on $\cP$ can be rewritten as
\be
\label{dsPqk}
\de s^2_{\cP} = -2\rho \, \de s^2_{\cM} + \frac{\de\rho^2}{\rho} + 4\rho \, |p_+|^2
+ \rho \, ( \de\phip + p_3)^2\ ,
\ee
where the metric element $\de s^2_{\cM}$ is degenerate along the direction $\pa_\phip$,
and expressed in terms of the HK metric on $\pcM$  and connection $\lambda$ via
\be
\label{qkgenrho}
\de s^2_{\cM} = \frac{(\de\rho)^2}{4\cff \rho^4}+ \cff \(\de\sigp+\Theta\)^2
+2\frac{|p'_+|^2}{\rho^2} -\frac{1}{2\rho}  \de s^2_{\pcM//\pa_\phip}\, ,
\ee
where the function $\cff$ is defined in terms of $\cfff$ and $\rho$ by \eqref{defcfff}.
The metric element \eqref{qkgenrho} defines a non-degenerate QK metric on the quotient
$\cP/U(1)_R$, whose  $SU(2)$ connection $\vec p$  is obtained in terms of the
\kahler and complex symplectic connections $\kcon$, $\hcon$ on $\pcM$
by inverting \eqref{pqktohk}. Note that the metric \eqref{qkgenrho} on $\cP/U(1)_R$
differs from the standard Riemanniann quotient metric $\de s^2_{\cP} -\rho \, ( \de\phip + p_3)^2$
due the second and third terms in \eqref{dsPqk}, as well as the conformal rescaling by
$-1/(2\rho)$. The fact that the metric \eqref{qkgenrho} is \qk follows from the
Swann bundle construction in \S\ref{sub_qktohk}.

It is also important to emphasize that the QK metric \eqref{qkgenrho} depends on the
hyperholomorphic connection $\conhyper$, and not only on its curvature. In particular,
shifting $ \conhyper\mapsto  \conhyper+c\,\de\theta'$ where $c$ is an arbitrary constant
does not affect the curvature $\fb=\de\conhyper$, but  does in general affect
the functions $\cfff$ and $\cff$ and therefore
the dual QK metric, leading to a one-parameter family of
inequivalent QK metrics.
On the other hand, a shift of $\conhyper$ (and, simultaneously, an opposite
shift of $\kcon$ such that
\eqref{contohyp} is preserved) by a closed one-form $\de\phi$ with $\partial_\phip\cdot  \de\phi=0$
can be reabsorbed by a redefinition of the coordinate $\sigp$
and one-form $\Theta$, such that the dual QK space is unaffected.

Finally, to make contact with  \cite{1143.53043}
we observe that \eqref{qkgenrho} can alternatively be written as
\be
\label{eqhayd}
\de s^2_{\cM} =-\frac{1}{2\rho}\left[ \de s^2_{\pcM} - \frac1{\cfff}
\left(\de\sigp+ \conhyper \right)^2 \right]
+\frac{1}{2\rho^2}\left[ \left( \de\sigp+\conhyper -K_0\right)^2
+ \vec K^2 \right],
\ee
where $K_0$ and $\vec K=\pa_{\phip}\cdot \vec\pomega$ the one-forms
obtained by contracting  $\pa_{\phip}$ with the HK metric and the triplet
of symplectic forms $\vec\pomega$, respectively:
\be
K_0=(\cfff+\rho)  \( \de\phip+\pTheta\) ,
\qquad
K_3=\de\rho,
\qquad
K_+=\I \hcon .
\ee
Eq. \eqref{eqhayd} reproduces Eq. (18) in \cite{1143.53043} (after correcting
a misprint in this reference, namely $\psi^2$
should appear outside the bracket).

\subsubsection{Darboux coordinates and transition functions}
\label{subsubsec-darboux}

In this subsection we shall relate the twistorial descriptions of the QK manifold $\cM$ and
of  the HK manifold $\pcM$ together with the hyperholomorphic line bundle $\lb$.
The relation is provided by the formulae  \eqref{relHKQKintro} and
\eqref{identXiintro} previewed in the introduction. Here we establish them from
the quotient procedure underlying the QK/HK correspondence.

Let us choose Darboux coordinates $(\xi^\Lambda,\txi_\Lambda,\alpha)$ on $\cZ$
such that the isometry $\pa_\sigp$ lifts to the holomorphic action $\pa_{\alpha}$ on $\cZ$.
Denoting by  $\sqrt{\vf}$ the complex coordinate on the fiber of the Swann
bundle $\IC^\times\to\cS\to\cZ$ (following the notations
of \cite{Neitzke:2007ke,Alexandrov:2008nk}), one can choose the following combinations
\be
v^\flat,
\qquad
v^\Lambda=v^\flat\xi^\Lambda,
\qquad
w_\Lambda=\frac{\I}{2}\,\txi_\Lambda,
\qquad
w_\flat=\frac{\I}{2}\,\alpha+2c\log\vf
\label{holcoorS}
\ee
as complex Darboux coordinates on $\cS$
in complex structure $J_3$,
such that  $\omega_\cS^+=\de v^I\wedge\de w_I$ with
$I=\flat,0,1,\dots$.
The coordinate $\vf$ is related to the moment map $\vec\mu$ by $\mu_+=-\I\vf$.

We now consider the \hk quotient $\pcM=\cS///\pa_\sigp$ at level $\vec m$.
A set of complex coordinates on $\pcM$ in complex structure $J_3$ can be obtained by restricting the
complex coordinates $v^\Lambda,w_\Lambda$ to the locus $\mu_+={\rm const}$.
Thus, we can identify the complex Darboux coordinates $\Xi'=(\eta^\Lambda,\mu_\Lambda)$
on $\pcM$ with the complex Darboux coordinates  $\Xi=(\xi^\Lambda,\txi_\Lambda)$ on $\cZ$.
On the other hand, it was shown in \cite{Alexandrov:2008nk} that the coordinate $\varpiqk$
on the $\CP$ fiber of $\cZ$  is related to the coordinate $\varpihk$
on the $\CP$ fiber of $\cZ_\cS$ by $\varpiqk=(\bar\pi_2\varpihk+\pi^1)/(-\bar\pi_1\varpihk+\pi^2)$,
where $(\pi_1,\pi_2)$ are coordinates on the $\IC^2/\IZ_2$ fiber of the bundle $\cS\rightarrow \cM$. Since we
work in complex structure $J_3$ on $\cS$, we have $\pi^1=0$ and therefore
$\varpiqk=(\bar\pi_2/\pi^2) \varpihk$. Using (3.43) in  \cite{Alexandrov:2008nk},
we arrive at the following identifications of the Darboux coordinates on $\pcZ$ and $\cZ$,
\be
\label{identXi}
\etai{i}^\Lambda(\varpihk)=\xii{i}^\Lambda(\varpiqk)\, ,
\qquad
\mui{i}_\Lambda(\varpihk )= \txii{i}_\Lambda(\varpiqk)\, ,
\qquad
\varpiqk=\varpihk\, e^{-\I\phip}\, .
\ee
This identification implies that the patches on $\cZ$ and $\pcZ$ are in one-to-one correspondence
and allows to conclude that the same transition functions and covering which
define the complex contact structure
on $\cZ$ also define the complex symplectic structure
on $\pcZ$, i.e.
\be
\hHij{ij}(\Xi)=\qHij{ij}(\Xi).
\label{relHKQK}
\ee
In particular, the fact that
$\pcM$ admits  a rotational isometry follows from the
independence of $\hHij{ij}$ on the $\CP$ coordinate $\varpihk$.
Moreover, the coordinates $(v^\Lambda,x^\Lambda,\vrh_\Lambda)$ on $\pcM$
which appear in the integral equations \eqref{txiqline} for $\pcM$ are naturally
identified with the coordinates $(Y^\Lambda, A^\Lambda,B_\Lambda)$ appearing
in the integral equations \eqref{txiqlineQK} for $\cM$ through
\be
v^\Lambda=Y^\Lambda e^{\I\phip},
\qquad
x^\Lambda=A^\Lambda,
\qquad
\vrh_\Lambda=B_\Lambda.
\label{identrealcoor}
\ee
Together with the identification of the $\CP$ variables in \eqref{identXi}, this allows
to relate the coverings $\cU_i$ and $\cU'_i$ of the two dual twistor spaces.

Having identified the Darboux coordinates $\xi^\Lambda=\eta^\Lambda$, $\txi_\Lambda=\mu_\Lambda$ on
$\cM$ and $\pcM$ via \eqref{identrealcoor}, it is natural to apply the same identifications to
the contact Darboux coordinate $\ai{i}$ in \eqref{txiqlineQK}, which gives
\be
\ai{i}(\varpihk) =  B_\alpha +
\half  \sum_j \oint_{C_j} \frac{\de \varpihk'}{2 \pi \I \varpihk'} \,
\frac{\varpihk' + \varpihk}{\varpihk' - \varpihk}
\( \hHij{ij}- \etai{i}^\Lambda \p_{\etai{i}^\Lambda}\hHij{ij}\)
+4c\, (\phip + \I \log\varpihk),
\ee
and ask about its meaning on the HK side. We first restrict to the case with no
anomalous dimension, $c=0$. By construction, $\ai{i}$ is holomorphic
 in complex structure $J(\varpihk)$ in the patch $\cU'_i$, and on the
 overlap of two patches $\cU'_i\cap \cU'_j$, satisfies
\be
S\ui{ij} \equiv \frac{1}{(2\pi)^2} \left( \ai{j}-\ai{i} \right)
= \frac{1}{(2\pi)^2} \left( \hHij{ij}- \etai{i}^\Lambda \p_{\etai{i}^\Lambda}\hHij{ij} \right) .
\ee
Thus, $\Upsilon\ui{i}\equiv e^{2\I\pi\ai{i}}$ can be viewed as a holomorphic section
of a line bundle $\lb_{\pcZ}$ over the twistor space $\pcZ$, with transition functions
given by $S\ui{ij}$. The restriction of the line bundle $\lb_{\pcZ}$
to the fibers of the fibration $\pcZ\to \pcM$ is trivial, since it admits a
nowhere-vanishing  section $\Upsilon$. By the usual twistor correspondence
(see e.g. \cite{ward-wells})
this descends to a hyperholomorphic line bundle $\lb$ on $\pcM$, and therefore
to a hyperholomorphic circle bundle $\cP$, whose fiber is parametrized by
$B_\alpha$. The connection
$\lambda$ on $\lb$ can be obtained by requiring that the covariant derivative
$D\Upsilon\equiv (\de -8\pi\I\conhyper)\Upsilon$ be of type (1,0) in complex
structure $J(\varpihk)$ \cite{Gaiotto:2011tf}. Equivalently,
\be
\conhyper = \frac14 \( \bpzeta \ai{i} + \pzeta \bar\alpha\ui{i} \),
\label{lambda}
\ee
where $\pzeta$ is the Dolbeault derivative in complex structure $J(\varpihk)$.
In particular, the r.h.s. of \eqref{lambda} is independent of $\varpihk$,
since the connection $\conhyper$ is.
Moreover, a \kahler potential\footnote{It should be noted that
$\frac{1}{2\i}(\pzeta-\bpzeta)K\ui{i}_{\lb}(\varpihk)$ in general differs from
$\conhyper$ by a closed, $\varpihk$-dependent one-form.}
for $\fb=\de\lambda$
in complex structure $J(\varpihk)$  is given by the log-norm of $\Upsilon$,
\be
\fb=\I\pzeta\bpzeta K\ui{i}_{\lb}(\varpihk),
\qquad
K\ui{i}_{\lb}(\varpihk)= \hf\Im \ai{i}.
\label{Khhgent0}
\ee

Let us now discuss the effect of the anomalous dimension. In this case $\ai{+}$ and
$\ai{-}$ are no longer regular in their respective patches, but have a logarithmic
singularity $4\I c\log\varpihk$ at $\varpihk=0$ and $\varpihk=\infty$, respectively.
This singularity can however be cancelled by singling out one of the Darboux
coordinates, say $\eta^0$, and defining
\be
\label{identW}
\Upsilon\ui{i}\equiv (\etai{i}^0)^{-8\pi \, c\ui{i}} e^{2\I\pi\ai{i}}\, ,
\ee
where $c\ui{+}=-c\ui{-}=c$ and $c\ui{i}=0$ otherwise.
This defines a section of a holomorphic line bundle $\lb_{\pcZ}$ on $\pcZ$
which is trivial along the real twistor lines, and therefore again
a hyperholomorphic curvature $\fb=\de\lambda$ on $\pcM$
with \kahler potential
\be
K\ui{i}_{\lb}(\varpihk)= \hf \Im \left[ \ai{i} + 4 \I  c\ui{i} \log \etai{i}^0\right] .
\ee
On the other hand, comparing \eqref{genformp3} with the expression following from \eqref{contform},
one finds that the moment map $\rho$ of the $U(1)_R$ action on $\pcM$
 coincides with the contact potential on $\cM$, $\rho=e^\Phi$.
Plugging these results into \eqref{KltoKc}, one obtains, in particular, that a \kahler
potential for the HK metric on $\pcM$ in complex structure $J_3$ is given by
\be
K_{\pcM}(0)=2 e^\Phi-\hf\lim\limits_{\varpihk\to0} \, \Im \[\alpha-4\I c\log\varpihk\]
-c  \log v^0\bv^0 \, .
\label{relpottt}
\ee
This indeed agrees with \eqref{Kdefrew}, \eqref{txiqlineQK} and \eqref{contpotconst}
up to a K\"ahler transformation given by the last term.

Thus, we see that the twistorial description of the QK manifold $\cM$
is completely equivalent to the twistorial description of the HK manifold $\pcM$ endowed
with a  hyperholomorphic circle bundle $\cP\rightarrow\pcM$ with connection \eqref{contohyp}.
The advantage of the
description in terms of $(\pcM,\cP)$ is that the twistor space $\pcZ$
is trivially fibered over $\CP$, so that the Darboux coordinates \eqref{txiqline} are valid
globally on $\pcZ$, whereas $\cZ$ is a non-trivial fibration by $\CP$'s, and therefore
does not admit such coordinates  globally.
In the rest of this paper, we shall use $\varpiqk$ and $\varpihk$
interchangeably for the twistor coordinate on $\pcZ$, and  similarly $\Xi$ and $\pXi$ for the
Darboux coordinates on $\pcM$, keeping in mind the identifications \eqref{identXi}.

\subsection{The QK/HK correspondence in one quaternionic dimension}
\label{oneQKdimension}

In this subsection we consider the QK/HK correspondence for one-dimensional
quaternionic manifolds. Recall that in one quaternionic dimension, HK and QK
manifolds correspond to self-dual Einstein spaces with zero and
non-zero cosmological constant, respectively. Moreover, the triplet of
\hk forms is self-dual, while hyperholomorphic
connections are connections with  anti-self dual curvature. In addition,
self-dual Einstein metrics with one rotational Killing vector field
are classified by solutions of the continual Toda equation
\be
\p_{z} \p_{\bar z}  T+\p_ \rho^2 \, {\rm e}^T = 0\, .
\label{Toda}
\ee
We shall see that the QK/HK correspondence relates QK and HK
manifolds associated to the same solution of \eqref{Toda}.

\subsubsection{Tod Ansatz for one-dimensional QK manifolds with one isometry}
\label{sec_TodaQK}

On the QK side, self-dual Einstein metrics with one Killing vector field can be
cast locally into the form of Tod's Ansatz \cite{MR1423177}
\be
\label{dstodaqk}
\de s_\cM^2 =\frac12\left[
\frac{P}{\rho^2} \left( \de \rho^2 + 4 {\rm e}^T \de z \de\bar z \right)
+ \frac{1}{P\rho^2}\,(\de \sigp+ \Theta )^2\right] ,
\ee
where $(\rho,z,\bar z, \sigp)$ are local coordinates, with $\pa_\sigp$ coresponding
to the Killing vector field.  Here $T$ is a solution of the Toda equation \eqref{Toda},
$P\equiv 1- \haf\,\rho \pa_ \rho T $, and $\Theta$ is a connection one-form such that
\be
\label{dth}
\de \Theta = \I (\pa_z P \de z - \pa_{\bar z} P \de \bar z)\wedge \de  \rho
- 2\I\, \partial_\rho(P {\rm e}^T)\de z\wedge \de\bar z\, .
\ee
This condition is integrable by virtue of \eqref{Toda} and gauge transformations of the one-form
$\Theta$ can be reabsorbed into redefinitions of the coordinate $\sigp$.
The self-dual part of the Levi-Civita connection can be chosen as
\be
\label{qkconn}
p_3 = - \frac{1}{\rho} \( \de \sigp + \Theta\) + \pTheta \, ,
\qquad
p_+ = \frac{e^{T/2}}{\rho}\, \de z = (p_-)^*\, ,
\ee
where we introduced another one-form
\be
\label{defThetap}
\pTheta\equiv \frac{\I}{2} \left( \pa_z T \de z - \pa_{\bar z} T \de \bar z \right).
\ee
The triplet of quaternionic two-forms \eqref{ptoom}
is then covariantly constant, verifying the \qk property of the metric.

It will be important to note that  the Toda equation \eqref{Toda} is invariant under
the symmetry
\be
\label{Todasym}
T(\rho,z,\bar z)\mapsto \tilde T(\rho,z,\bar z)=T(\rho+c,g(z),\bar g(\bz)) + \log | \de g/\de z|^2,
\ee
where $g(z)$ is any holomorphic function of $z$ and $c$ any real constant.
The effect of the function $g(z)$ can be absorbed by a holomorphic change of
coordinates, but this is not so for the constant $c$. Thus, QK metrics with one
Killing vector come (at least locally) in one-parameter families. For later reference
we note that under the symmetry \eqref{Todasym}, the one-forms $\Theta$ and $\pTheta$ vary by
\be
\Theta\mapsto \Theta-c\,\pTheta,
\qquad
\pTheta\mapsto\pTheta-\Im\de \log \frac{\de g}{\de z},
\label{trans1form}
\ee
where all quantities on the right hand side are understood as functions of $\rho+c$ and $g(z)$.
As a result of \eqref{trans1form}, the curvature of the circle bundle generated by $\pa_{\sigp}$
receives a contribution proportional to $\de\pTheta$.

\subsubsection{Toda Ansatz for one-dimensional HK manifolds with one rotational isometry}
\label{sec_TodaHK}

On the HK side, self-dual Ricci-flat metrics with one rotational Killing vector field
can be cast into the Boyer-Finley Ansatz \cite{MR660020,MR759978,Bakas:1989yj,
Bakas:1996gf},
\be
\label{dstodahk}
\de s_{\pcM}^2 = \frac12\left[ \pa_\rho T \left( \de \rho^2 + 4 {\rm e}^T \de z \de\bar z \right)
+ \frac{4}{\pa_\rho T}\(\de \phip+\pTheta \)^2\right],
\ee
where $T(\rho,z,\bar z)$ is again a solution of the Toda equation \eqref{Toda}, and
$\pTheta$ is the connection one-form \eqref{defThetap}. We choose
\be
\ub=\frac{\sqrt{\partial_\rho T}}{2}\, \de \rho + \frac{\I}{\sqrt{\partial_\rho T}}
\left(\de \phip+\pTheta \right),
\qquad
\vb=e^{\frac12\,T+\I\phip} \sqrt{\partial_\rho T} \de z
\ee
 as a basis of the space of $(1,0)$ forms in complex structure $J_3$.
The  self-dual two-forms
\beq
\label{om3hk}
\pomega_3&=&
\I \( \ub\wedge \bar \ub+ \vb\wedge \bar \vb\) =
\de\rho \wedge\left(\de \phip+\pTheta \right)
+\I\, e^T \partial_\rho T\, \de z\wedge \de\bz
\\
\pomega_+&=&\ub\wedge \vb= \de\( e^{T/2+\I\phip}\) \wedge \de z
\, ,\qquad \qquad
\pomega_-=(\pomega_+)^*\, ,
\eeq
are closed by virtue of the Toda equation, and the corresponding complex structures
satisfy the quaternion algebra, verifying the \hk property. They can be integrated
to one-forms
\be
\label{kahlercon1d}
\kcon= \rho \left( \de \phip+\pTheta \right)-\Theta ,
\qquad
\hcon= e^{T/2+\I\phip}\,  \de z\, ,
\ee
where $\Theta$ is the same connection which features in the QK metric \eqref{dstodaqk}.
In fact, the one-forms $\kcon, \hcon$ are related to the $SU(2)$ connection \eqref{qkconn}
of the QK metric by exactly the same equations as \eqref{pqktohk}.

The \kahler connection $\kcon$ can be further integrated to a \kahler potential \cite{MR660020}.
For this purpose, one must first integrate the Toda
potential $T(\rho,z,\bz)$ to a function $\cL(\rho,z,\bz)$ such that
\be
\p_\rho\cL=T \, ,
\qquad
\p_{z} \p_{\bar z}  \cL+\p_ \rho \, {\rm e}^T = 0\, .
\label{TodaL}
\ee
This determines the function $\cL$ up to the addition of the real part of a holomorphic
function of $z$. We fix this ambiguity by requiring that
\be
\label{ThetaL}
\Theta=\rho\,\pTheta - \frac{\I}{2}\,\(\pa_z\cL\, \de z - \pa_{\bz}\cL \, \de\bar z\) .
\ee
Indeed, any solution of \eqref{dth} can be put in this form.
Then the Legendre transform of $\cL$ with respect to $\rho$
\be
\label{KToda}
\Kc_{\pcM}(z,\bz,u,\bu)= \langle \rho\, \log(u \bu) - \cL(\rho,z,\bz)\rangle_{\rho}\, ,
\ee
provides a K\"ahler potential for the HK metric in the complex structure $J_3$
\cite{MR660020}, with complex coordinates $z,u$.
Using \eqref{TodaL}, one verifies that
the \kahler potential \eqref{KToda} satisfies the Monge-Amp\`ere equation
\be
\pa_{z\bar z}^2 \Kc_{\pcM}\, \pa_{u\bu}^2 \Kc_{\pcM} - \pa_{z\bar u}^2 \Kc_{\pcM}\, \pa_{u\bar z}^2 \Kc_{\pcM}=1
\ee
and reproduces the metric \eqref{dstodahk} provided one identifies $u=e^{T/2+\I\phip}$.
Moreover, using \eqref{ThetaL} one may check that it also reproduces
the connection \eqref{kahlercon1d},
\be
\kcon = \frac{1}{2\I} (\pa-\bar\pa)\Kc_{\pcM} = \rho\, \de\phip
+\frac{\I}{2}\,\(\pa_z\cL\, \de z - \pa_{\bz}\cL \, \de\bar z\) .
\ee

\subsubsection{Hyperholomorphic connection and QK/HK correspondence}

Using the general prescription in \S\ref{sub_qktohk} with
\be
\cff=\frac{1}{2P\rho^2}\, ,
\qquad
\cfff=\frac{2P}{\pa_\rho T}\, ,
\qquad
\cfff+\rho=\frac{2}{\pa_\rho T}
\ee
it is immediate to check that  the HK metric \eqref{dstodahk}
is related to the QK metric \eqref{dstodaqk} with the same Toda
potential under the QK/HK correspondence. The hyperholomorphic
connection afforded by this correspondence is given by \eqref{lambdlbTheta},
\be
\label{hyperholcon1d}
\conhyper
=\frac{2P}{\pa_\rho T}  \left( \de \phip+\pTheta \right) + \Theta .
\ee
Indeed, one may check that the curvature $\fb=\de \conhyper$  is
a linear combination of anti-self-dual forms,
\be
\begin{split}
\fb=
 & \,\frac{2\I}{(\pa_\rho T)^2}\biggl[e^{-\hf\, T-\I\phip} \pa^2_{\rho z} T\, \bar\ub\wedge \vb
-  e^{-\hf\, T+\I\phip} \pa^2_{\rho\bar z} T \, \ub \wedge \bar\vb \biggr.
\\
&\biggl. \qquad\qquad
+\left( \hf\, (\pa_\rho T)^2+e^{-T}\pa^2_{z\bar z}T\right) (\ub\wedge \bar \ub- \vb\wedge \bar \vb)\biggr] ,
\end{split}
\ee
and derives from the \kahler potential \eqref{KltoKc} where $\Kc_{\pcM}$
is given in \eqref{KToda}.

Under the symmetry \eqref{Todasym}, we note that the HK metric \eqref{dstodahk}
is invariant, up to a change of coordinates
$(\rho,z,\phip)\mapsto (\tilde\rho=\rho+c,\tilde z=g(z),\tilde\phip=\phip-\Im\log g'(z))$.
However, the hyperholomorphic connection, \kahler connection and
K\"ahler potential  do transform,
\be
\label{lambdac}
\conhyper \mapsto \conhyper +c \de\tilde\phip\, ,
\qquad
\kcon \mapsto  \kcon- c \de\tilde\phip\, ,
\qquad
K_{\pcM}\mapsto K_{\pcM}-c \log(u\bu)\, .
\ee
The parameter $c$ determines how the  $U(1)_R$ action
on $\pcM$ lifts to an action on the total space of the circle bundle $\cP$,
and leads to a one-parameter family of dual QK metrics.
It is also worth noting that in one quaternionic dimension, unlike in higher
dimensions (cf. \eqref{dsPhk}),
the quotients of the QK and HK manifold by their respective $U(1)$ action
are related by a conformal rescaling,
\be
\de s^2_{\cM/\partial_\sigp} \propto \de s^2_{\pcM//\partial_\phip} \propto \de\rho^2 +
4 e^T \de z\, \de \bar z\, .
\ee

The QK/HK correspondence also provides a relation between the above description
based on the Toda equation and the twistor framework.
This can be done using expressions of the Toda coordinates $\rho,z$ and potential
$T$ in terms of the data on the twistor space $\cZ$ found in \cite{Alexandrov:2009vj} and
the dictionary \eqref{identrealcoor}
between the coordinates on the dual QK and HK spaces.
As a result, one finds that
\be
T=\log(v\bv/4),
\qquad
\rho=e^\Phi,
\qquad
z=\frac{\I}{2}\,\vrh
+\frac{1}{8\pi}\sum_j\oint_{C_j}\frac{\de \varpiqk}{\varpiqk}\,\p_{\xii{i}}\Hij{ij},
\label{HKident}
\ee
where $\Phi$ given in \eqref{contpotconst} is understood as a function of $v$ and $\vrh$.
Note that here $z$ coincides with $w$ defined in \eqref{defholcoor}.
Using these identifications, one can also show that
the K\"ahler potential obtained by the Legendre transform \eqref{KToda}
differs from the one given in \eqref{Kdefrew} by a K\"ahler transformation
proportional to the anomalous dimension $c\log(v\bv/4)$.

\subsubsection{Examples\label{sec_todaex}}

Let us now illustrate the general formulae obtained in this subsection on three
simple examples of QK manifolds, the sphere $S^4$, hyperbolic space $H^4$,
and the `perturbative universal hypermultiplet moduli space' (a deformation of
a non-compact version of $\IC P^2$). The latter is a special case of the
local $c$-map spaces discussed in \S\ref{cmap}. As it turns out, these three
QK manifolds are dual to the same HK manifold, namely $\IR^4$ with its
flat metric (with negative definite signature when $\cM=H^4$), but equipped
with a different hyperholomorphic connection.

\subsubsection*{\bf $S^4$ and $H^4$ vs. flat space}

The standard round metric on $S^4$ (respectively, the standard metric on the
four-dimensional hyperbolic space $H^4$) can be cast into the Tod Ansatz
\eqref{dstodaqk} by choosing
\be
\label{potTflat}
T=2\log \frac{\epsilon(4\rho-1)}{4\cosh(z+\bz)}\, ,
\qquad
\pTheta=4\Theta=-\I\tanh(z+\bz)(\de z-\de\bz)\, ,
\ee
where $\epsilon=1$ for $\cM=S^4$ and $\eps=-1$ for $\cM=H^4$, and $\rho$
lies in the range where $\eps(4\rho-1)>0$  \cite{Alexandrov:2009vj}.
By changing coordinates to
\be
\rho=\frac14\,(1+\epsilon R^2),
\qquad
z=\hf\(\log\tan \delta+\I(\beta-\gamma)\),
\qquad
\phip=\beta+\gamma,
\ee
the metric on the dual HK space \eqref{dstodahk} can be written as
\be
\de s_{\pcM}^2=\epsilon\[\de R^2+R^2\(\de\delta^2+\sin^2\delta\, \de\beta^2
+\cos^2\delta\, \de\gamma^2\)\],
\ee
which is recognized as the flat metric on $\IR^4$ in Hopf coordinates,
with positive signature for $\eps=1$, or negative signature for $\eps=-1$.
The hyperholomorphic connection \eqref{hyperholcon1d} evaluates to the flat connection
\be
\conhyper = -\frac14\,\de\phip=-\frac14\, \de(\beta+\gamma) ,
\qquad
\fb=0.
\label{flathypercon}
\ee
It may be checked explicitly that it derives from the K\"ahler potential \eqref{KltoKc},
where
\be
\Kc_{\pcM}=2\epsilon|u|\cosh(z+\bz)+\frac14\,\log u\bu = 2\rho +\frac14\,\log u\bu - \frac12
\label{flatKM}
\ee
follows by Legendre transform from the potential
\be
\cL=\hf\, (4\rho-1)\(\log\frac{\epsilon(4\rho-1)}{4\cosh(z+\bz)}-1\) .
\ee

The twistor space associated to $\cM$ is $\cZ=\IC P^3$ for $\eps=1$, or $\IC P^{2,1}$
for $\eps=-1$. In either cases $\cZ$ can be described in the language of \S\ref{sec_twi}
by three open patches $\cU_+,\cU_-,\cU_0$ where $\cU_\pm$ covers a neighborhood of
$\varpiqk=\mp \eps e^{\pm(z+\bar z)}$ and $\cU_0$ covers the rest of $\CP$, with
 transition functions and anomalous dimension  \cite{Alexandrov:2009vj}
\be
\Hij{0\pm}=\pm\hf\,\xi\log\xi\, ,
\qquad
c=-\frac14\, .
\ee
The Darboux coordinates \eqref{txiqlineQK} in the patch $\cU_0$  are then given by
\be
\begin{split}
\xi=&\, \epsilon(4\rho-1)\(\frac{\varpiqk^{-1}-\varpiqk}{2\cosh(z+\bz)}-\epsilon\tanh(z+\bz)\) ,
\\
\txi=&\, -\I\(2z+\log\frac{1+\epsilon\varpiqk e^{-z-\bz}}{1-\epsilon\varpiqk e^{z+\bz}}\) ,
\\
\alpha=&\, 4\sigp-\I\log\varpiqk .
\end{split}
\ee
After performing the replacement \eqref{identXi}, the same formulae
provide Darboux coordinates $\eta,\mu$ on $\pcZ$ and a holomorphic section
$\Upsilon=e^{2\pi\I\alpha}$ of $\lb_{\pcZ}$. Indeed, one may check
that the one-forms $\de\xi, \de\txi$ and $\de\alpha-4\lambda$ are of type (1,0) in complex
structure $J(\varpihk)$, consistently with \eqref{lambda}.

\subsubsection*{\bf Perturbative universal hypermultiplet}

We now consider the family of QK metrics  \cite{Antoniadis:2003sw}
\be
\label{dsuh}
\de s^2_\cM = \frac{\rho+2c}{4\rho^2(\rho+c)}\,\de \rho^2
+\frac{\rho+2c}{16 \rho^2} \( (\d\zeta^0)^2+4 (\de\tzeta_0)^2\)
+ \frac{\rho+c}{64 \rho^2(\rho+2c)} (\de\sigma+\tzeta_0\,\de\zeta^0- \zeta^0\, \de\tzeta_0)^2 ,
\ee
where $c$ is a real parameter, and $\rho$ lies in the range $\rho>\max(0,-2c)$.
The metric \eqref{dsuh} describes the weak coupling limit of the hypermultiplet moduli
space in type IIA string theory compactified on a rigid Calabi-Yau three-fold $X$,
where $c$ is determined by the Euler number of $X$
(see \S\ref{cmap} and \S\ref{sec_lifting} for further details on this set-up). For $c=0$,
the metric \eqref{dsuh} reduces to the $SU(2,1)$-invariant metric
on $\IC P^{1,1}$ (a non-compact version of $\IC P^2$).
The metric \eqref{dsuh} may be cast into Tod's Ansatz \eqref{dstodaqk}
by choosing
\be
\label{TodaUH}
z=-\frac14\,(\zeta^0-2\I\tzeta_0),
\qquad
\sigp=-\frac18\, \sigma ,
\qquad
T=\log(\rho +c) .
\ee
Using
\be
P=\frac{\rho+2c}{2(\rho+c)}\, ,
\qquad
\Theta=-\frac{\I}{2} ( z \de \bar z - \bar z \de z)=\frac18(\zeta^0\de\tzeta_0-\tzeta_0\de\zeta^0)\, ,
\qquad
\pTheta=0\, ,
\ee
we find the dual HK metric \eqref{dstodahk} and hyperholomorphic connection
\eqref{hyperholcon1d}
\be
\label{dsuhhk}
\de s_{\pcM}^2 = \frac{\de\rho^2}{2(\rho+c)}+2\de z\,\de\bz+2(\rho+c)(\de\phip)^2,
\qquad
\conhyper = (\rho+2c) \de \phip   -\frac{\I}{2}\, ( z \de \bar z - \bar z \de z) .
\ee
By changing coordinates to $(\rho,z)=(\hf\, R^2-c,\tilde R e^{\I\tilde\phip}/\sqrt{2})$, we recognize
\eqref{dsuhhk} as the flat metric on $\IR^4$ in bi-polar coordinates, equipped
with a constant anti-self dual field,\footnote{Since the metric \eqref{dsuh} has a curvature
singularity at $\rho=-2c$ when $c<0$, this example shows that the HK dual of a smooth
QK manifold need not be geodesically complete.}
\be
\label{UHflat}
\de s^2_{\pcM}=\de R^2+R^2(\de\phip)^2+\de \tilde R^2+\tilde R^2(\de \tilde\phip)^2,
\qquad
\conhyper=\hf\( R^2\, \de\phip-\tilde R^2\, \de\tilde\phip\)+c \de\phip .
\ee
One may check that $\lambda$ derives from the K\"ahler potential
\be
K_{\lb}=-z\bar z+u\bu +c\log u\bu\, ,
\ee
related via  \eqref{KltoKc} and \eqref{KToda} to the function $\cL$
and \kahler potential $\Kc_{\pcM}$
\be
\cL=(\rho+c)\, \log(\rho+c)-\rho-z\bz\, ,
\qquad
\Kc_{\pcM}
=z\bz+u\bu-c\log (u\bu)
\, .
\ee
Although the logarithmic term in $\Kc_{\pcM}$ and $K_\lb$ can be removed by a K\"ahler transformation,
it is needed in order to correctly reproduce the hyperholomorphic connection and, as a consequence,
the dual QK metric \eqref{dsuh}.

The twistor space $\cZ$ can be covered by three patches $\cU_+,\cU_-$ and $\cU_0$,
covering the north pole ($\varpiqk=0$), south pole ($\varpiqk=\infty$), and equator in $\CP$,
respectively, with transition
functions \cite{Alexandrov:2008nk,Alexandrov:2009vj}
\be
\Hij{+0}=-\frac{\I}{4}\,\xi^2\, ,
\qquad
\Hij{-0}=\frac{\I}{4}\,\xi^2\, ,
\ee
and anomalous dimension $c$. The Darboux
coordinates \eqref{txiqlineQK} in the patch $\cU_0$ read
\be
\begin{split}
\xi/2 =& -(z+\bz)+\sqrt{\rho+c} \, \(\varpiqk^{-1}  -  \varpiqk \)\\
\I\txi =& z-\bz+\sqrt{\rho+c} \, \( \varpiqk^{-1}  +  \varpiqk \) \\
-2\alpha-\xi\txi=&-8\sigp + 4\I \sqrt{\rho+c} \, \( z \, \varpiqk^{-1}  +  \bz \, \varpiqk \)
- 8\I c  \log t\, .
\end{split}
\ee
Upon performing the replacement \eqref{identXi}, the same formulae
provide Darboux coordinates $\eta,\mu$ on $\pcZ$ and a holomorphic section
$\Upsilon=e^{2\pi\I\alpha}$ of $\lb_{\pcZ}$. As in the previous example, one may check
that the one-forms $\de\xi, \de\txi$ and $\de\alpha-4\lambda$ are of type (1,0) in complex
structure $J(\varpihk)$.

\subsection{Local c-map vs. rigid c-map}
\label{cmap}

In this subsection we demonstrate that the $4n$-dimensional QK space $\cM$,
 obtained by the local $c$-map
procedure \cite{Ferrara:1989ik} from a $(2n-2)$-dimensional projective special \kahler manifold
$\cS\cK$
with homogeneous prepotential $F(X^\Lambda)$, is dual via the QK/HK correspondence
to the HK manifold $\pcM$ obtained by the rigid $c$-map \cite{Cecotti:1988qn} from
the $2n$-dimensional rigid special \kahler manifold with the same (homogeneous)
prepotential $F(X^\Lambda)$. The correspondence continues to hold for the
one-loop deformed local $c$-map \cite{Alexandrov:2007ec,Alexandrov:2008nk},
which is dual to the same HK manifold but with a different hyperholomorphic connection.
The universal hypermultiplet manifold considered above is a particular
example with quadratic prepotential. The main results in this subsection
were obtained in \cite{BPNeitzke}.

\subsubsection{Local c-map}
\label{localcmap}

We start by describing a one-parameter
family of QK metrics associated to any special \kahler space $\cS\cK$.
Its relevance to physics comes from the fact that this family
describes the perturbative hypermultiplet moduli space in type II
string theory compactified on a Calabi-Yau threefold (see \S\ref{sec_lifting}).
Let $F(X^\Lambda)$ be a holomorphic function of $n$ coordinates $X^\Lambda$, $\Lambda=0,\dots,n-1$,
homogeneous of degree 2, which encodes the geometry
of  $\cS\cK$. The family of QK metrics, parametrized by a real constant $c$ is given in coordinates
$\rho,z^a,\zeta^\Lambda,\tzeta_\Lambda,\sigma$ by
\cite{Robles-Llana:2006ez,Alexandrov:2007ec}
\be
\de s_{\cM}^2=\frac{\rho+2c}{4\rho^2(\rho+c)}\,\de \rho^2
+\frac{\rho+c}{2\rho}\, \de s^2_{\cS\cK}
+\frac{\de s^2_\cT }{4\rho}
+ \frac{c}{2\rho^2}\,e^{\cK}\, | X^\Lambda \de \tzeta_\Lambda - F_{\Lambda} \de \zeta^\Lambda|^2
+ \frac{\rho+c}{64 \rho^2(\rho+2c)} D\sigma^2\,  ,
\label{hypmetone}
\ee
where $\de s^2_{\cS\cK}=2\cK_{a\bar b}\de z^a \de \bz^b$ is the projective special \kahler metric with \kahler potential
and \kahler connection
\be
\cK=-\log[ \I ( \bX^\Lambda F_\Lambda - X^\Lambda \bF_\Lambda)] ,
\qquad
\cA_K=\frac{\I}{2}\,( \cK_a \de z^a -  \cK_{\ba} \de \bz^{\ba}) ,
\ee
$\de s^2_\cT$ is the \kahler metric on the torus $\cT$
\be
\label{dsT}
\de s^2_\cT = -\frac12\,
(\de \tzeta_\Lambda - \bar\cN_{\Lambda\Lambda'} \de \zeta^{\Lambda'})
 \Im \cN^{\Lambda\Sigma}
(\de \tzeta_\Sigma - \cN_{\Sigma\Sigma'} \de \zeta^{\Sigma'}) ,
\ee
$\cN_{\Lambda\Lambda'}$ is the `Weil period matrix'\footnote{This terminology
is borrowed from the context of Type IIA string theory compactified on a Calabi-Yau,
where $\cN_{\Lambda\Lambda'}$ corresponds to the period matrix of the Weil intermediate Jacobian,
while $\tau_{\Lambda\Lambda'}$ is the period matrix of the Griffiths intermediate Jacobian.}
\be
\label{defcN}
\cN_{\Lambda\Lambda'} =\bar \tau_{\Lambda\Lambda'} +2\I \frac{ [\Im\tau \cdot X]_\Lambda
[\Im \tau \cdot X]_{\Lambda'}}
{X^\Sigma \, \Im\tau_{\Sigma\Sigma'}X^{\Sigma'}}\, ,
\qquad
\tau_{\Lambda\Sigma}=\pa_{X^\Lambda} \pa_{X^\Sigma} F\, ,
\ee
and
\be
D\sigma \equiv
\de\sigma+ \tzeta_\Lambda \de \zeta^\Lambda -  \zeta^\Lambda \de \tzeta_\Lambda
+ 8 c \, \cA_K .
\label{sigmaconnection}
\ee
For prepotentials $F$ arising in Calabi-Yau compactifications, the quadratic forms
$\Im\cN$ and $\Im\tau$ have signature $(0,n)$ and $(n-1,1)$, respectively.
In the one-modulus case with $F=-\frac{\I}{4}\, (X^0)^2$, $\cS\cK$ is trivial
and the metric \eqref{hypmetone} reduces to \eqref{dsuh}.

The twistor space $\cZ$ can be read off from the Legendre transform construction
of \eqref{hypmetone} \cite{Rocek:2005ij,Robles-Llana:2006ez}.
It can be covered by three patches $\cU_+,\cU_-,\cU_0$ around the north pole  ($\varpiqk=0$),
south pole  ($\varpiqk=\infty$) and equator, respectively,
with transition functions
\be
\label{Hcmap}
\Hij{+ 0}=F(\xi^\Lambda), \qquad \Hij{- 0}=\bF(\xi^\Lambda) \, ,
\ee
and anomalous dimension $c$.
The canonical Darboux coordinates on $\cZ$ in the patch $\cU_0$
are given by \cite{Neitzke:2007ke}
 \be
\label{gentwi}
\begin{split}
\xi^\Lambda &= \zeta^\Lambda + 2e^{\cK/2} \sqrt{\rho+c}
\left( \varpiqk^{-1} X^{\Lambda} -\varpiqk \,\bX^{\Lambda}  \right),
\\
\txi_\Lambda &= \tzeta_\Lambda +2 e^{\cK/2} \sqrt{\rho+c}
\left( \varpiqk^{-1} F_\Lambda-\varpiqk \,\bF_\Lambda \right),
\\
\talp&= \sigma + 2 e^{\cK/2}\sqrt{\rho+c}
\left[\varpiqk^{-1} (F_\Lambda \zeta^\Lambda - X^\Lambda \tzeta_\Lambda)
-\varpiqk \, (\bar F_\Lambda \zeta^\Lambda - \bar X^\Lambda \tzeta_\Lambda )\right]
- 8 \I c\log \varpiqk \, ,
\end{split}
\ee
where $\talp$ is related to $\alpha$ by \eqref{deftalp}. Unlike the latter, the former
is invariant under simultaneous symplectic transformations
of the vectors $(X^\Lambda,F_\Lambda)$ and $(\zeta^\Lambda,\tzeta_\Lambda)$.
The metric \eqref{hypmetone} is invariant under the Killing vector field $\pa_\sigma$
(as well as translations along $\zeta^\Lambda,\tzeta_\Lambda$, which we ignore in
this section). The vector field $\pa_\sigma$ lifts to the holomorphic vector field
$\pa_\alpha$ on $\cZ$.

\subsubsection{Rigid c-map}
\label{rigidcmap}

On the other hand, the same prepotential $F$,
via the rigid $c$-map construction\footnote{The rigid $c$-map construction
does not require that $F$ be homogeneous, but we restrict to this case as it is the
one relevant for the QK/HK correspondence.}
produces the following \hk metric  \cite{Cecotti:1988qn}
\be
\label{rcmap}
\de s^2_{\pcM} = \frac14\,\de s^2_{\cR\cS\cK}
+  \frac{1}{2}\,\de s^2_{\cT_{\rm rig}},
\ee
where
\be
\de s^2_{\cR\cS\cK} =-4\Im\tau_{\Lambda\Sigma} \de \rigZ^\Lambda  \de \brigZ^\Sigma
\ee
is the metric on the rigid special \kahler manifold $\cR\cS\cK$ with prepotential $F(Z^\Lambda)$,
and
\be
\de s^2_{\cT_{\rm rig}}=-\frac12 \left(\de \tilde\zeta_\Lambda
-\bar{\tau}_{\Lambda\Sigma}\de \zeta^{\Sigma}\right)\Im \tau^{\Lambda\Lambda'}
\left(\de \tilde\zeta_{\Lambda'}-\tau_{\Lambda'\Sigma'}\de \zeta^{\Sigma'}\right)
\label{rigidtorus}
\ee
with $\tau_{\Lambda\Sigma} = \partial_{\rigZ^{\Lambda}}\partial_{\rigZ^{\Sigma}}F(\rigZ)$
is the flat metric on its cotangent space.

For prepotentials $F$ arising in Calabi-Yau compactifications, the metric \eqref{rcmap}
has signature $(4,4n-4)$.
In complex structure $J_3$ where  $\rigZ^\Lambda$ and
\be
W_\Lambda=\frac{\I}{2}(\tzeta_\Lambda-\tau_{\Lambda\Sigma}\zeta^{\Sigma})
\ee
are complex coordinates, the \kahler form and holomorphic symplectic form
are
\be
\label{omKCrcmap}
\pomega_3 = \frac{1}{2\I}\, \Im\tau_{\Lambda\Sigma} \de \rigZ^\Lambda \wedge \de \brigZ^\Sigma
+\frac{1}{2\I}\, [\Im\tau^{-1}]^{\Lambda \Sigma} \, DW_\Lambda \wedge D \bar W_\Sigma ,
\qquad
\pomega_+ = \hf\, \de \rigZ^\Lambda\wedge \de W_\Lambda  ,
\ee
where $DW_\Lambda$ are the (1,0)-forms
\be
DW_\Lambda \equiv
\de W_\Lambda - \frac{1}{2\I}\,\p_{Z^\Lambda}\tau_{KL} [\Im\tau^{-1}]^{KK'}
(W_{K'}+\bar W_{K'}) \de \rigZ^L \, .
\ee
It is straightforward to check that $\pomega_3,\pomega_\pm$ are closed and
that the associated complex structures $J_3=\pomega_3\, g^{-1}, J_\pm=\pomega_\pm\, g^{-1}$
satisfy the quaternion algebra.
The \kahler form $\pomega_3=\I\pa\bar\pa K_{\pcM} $ derives from the
\kahler potential
\be
\label{kcecotti}
K_{\pcM} = \frac{1}{4\I} \( \rigZ^\Lambda \bG_\Lambda- \brigZ^\Lambda G_\Lambda \)
- \frac{1}{4}\,(W_\Lambda + \bW_\Lambda) [\Im\tau^{-1}]^{\Lambda\Sigma}
(W_\Sigma + \bW_\Sigma) ,
\ee
where $G_\Lambda\equiv \pa F(\rigZ^\Lambda)/\pa \rigZ^\Lambda$.
The metric \eqref{rcmap} is invariant under $(\rigZ^\Lambda,W_\Lambda)\mapsto
 (e^{\I\phip}\, \rigZ^\Lambda,W_\Lambda)$, while the complex structures transform as in \eqref{U1RJ}.

The twistor
space $\pcZ$ can again be read off from the Legendre transform construction
of \eqref{rcmap} \cite{Rocek:2006xb}. It involves three patches $\cU'_+, \cU'_-,
\cU'_0$ around $\varpihk=0$, $\varpihk=\infty$ and around the equator, and the
same transition  functions as in \eqref{Hcmap},
\be
\label{Hrcmap}
\Hij{+ 0}=F(\eta^\Lambda), \qquad \Hij{- 0}=\bF(\eta^\Lambda) \, .
\ee
In fact, the Legendre construction of the HK metric \eqref{rcmap} is closely similar
to that of the Swann bundle $\cS$ of the QK metric \eqref{hypmetone}. To wit,
 the Legendre construction of $\cS$ involves  one additional $\cO(2)$ multiplet $\eta^\flat$,
known as the superconformal compensator, and the generalized prepotential for $\cS$
is just obtained by rescaling the  generalized prepotential for $\pcM$ by a factor
of $1/\eta^\flat$, ensuring the proper homogeneity degree for superconformal invariance.
Thus, in this Legendre construction procedure, the rigid $c$-map $\cM$ is obtained from
$\cS$ by freezing the $\cO(2)$ multiplet  to a fixed value,  which determines
the complex structure in which $\pcM$ is obtained.
Mathematically, this freezing corresponds to performing the HK quotient with respect to the $U(1)_A$ isometry.
As a result, the Darboux coordinates \eqref{txiqline}
obtained from \eqref{Hrcmap} are closely similar to \eqref{gentwi},
\be
\label{gentwihk}
\begin{split}
\eta^\Lambda &=\zeta^\Lambda+ \varpihk^{-1} \rigZ^\Lambda - \varpihk\brigZ^\Lambda\, ,
\\
\mu_\Lambda&=\tzeta_\Lambda+  \varpihk^{-1} G_\Lambda - \varpihk\bG_\Lambda\, .
\end{split}
\ee
The use of the same variables $\zeta^\Lambda,\tzeta_\Lambda$ as in the local $c$-map metric \eqref{hypmetone} will
be justified in \eqref{rcmapid} below.

\subsubsection{QK/HK-correspondence for c-map spaces}
\label{QKHKcmap}

The fact that the local $c$-map $\cM$ and rigid $c$-map $\pcM$ are described by
identical transition functions \eqref{Hcmap}, \eqref{Hrcmap}
shows, by itself, that $\cM$ and $\pcM$ are dual under
the QK/HK correspondence. To confirm this, we
note that the Darboux coordinate systems \eqref{gentwi}
and \eqref{gentwihk} are related under the general identification \eqref{identXi},
provided
the coordinates $\rigZ^\Lambda,W_\Lambda$ on $\pcM$ are related to
$\rho,z^a,\zeta^\Lambda,\tzeta_\Lambda$ on $\cM/\pa_{\sigma}$ via
\be
\label{rcmapid}
\rigZ^\Lambda = \sqrt{2} \, R \, e^{\I\phip+\cK/2} X^\Lambda(z^a) ,
\qquad
W_\Lambda
= \frac{\I}{2} \left( \tzeta_\Lambda - \tau_{\Lambda\Sigma} \zeta^\Lambda \right) ,
\ee
where, as in \eqref{UHflat}, we define $R=\sqrt{2(\rho+c)}$.

Moreover, by applying the
procedure of \S\ref{sub_qktohk} to the  QK metric \eqref{hypmetone} with
\be
\cff=\frac{\rho+c}{\rho^2(\rho+2c)}\, ,
\qquad
\cfff=\rho+2c\, ,
\qquad
\de\sigp+\Theta=-\frac18 D\sigma\, ,
\qquad
\pTheta=\cA_K ,
\ee
and using  \eqref{dsPhk}, \eqref{rcmapid} one finds that the HK metric dual
to \eqref{hypmetone} is given by
\be
\de s^2_{\pcM} = \de R^2-\frac12 R^2 \, \de s^2_{\cS\cK}+ R^2(\de\phip+\cA_K)^2
-\frac12\de s^2_\cT
+ \frac{1}{2R^2} | Z^\Lambda \de \tzeta_\Lambda - G_{\Lambda} \de \zeta^\Lambda|^2\, .
\ee
It is straightforward to check that this  agrees with  the rigid $c$-map metric \eqref{rcmap}.
The hyperholomorphic connection afforded by the QK/HK
correspondence is easily obtained from \eqref{lambdlbTheta},
\be
\label{hyperholgen}
\conhyper =(\rho+c) (\de \phip+\cA_K)+
\frac18 \(\zeta^\Lambda \de \tzeta_\Lambda-\tzeta_\Lambda \de \zeta^\Lambda \)
+ c \de\phip\, .
\ee
Its curvature can be expressed as
\be
\label{hyperholrigid}
\fb = \frac{1}{2\I}\, \Im\tau_{\Lambda\Sigma} \de \rigZ^\Lambda \wedge \de \brigZ^\Sigma
-\frac{1}{2\I} \, [\Im\tau^{-1}]^{\Lambda \Sigma} \, DW_\Lambda \wedge D \bar W_\Sigma
 \, ,
\ee
and is indeed of type (1,1) in all complex structures. It is worthwhile to note
that it differs from  the \kahler form in \eqref{omKCrcmap} by a flip of sign (and a rescaling).
Indeed, it can be derived from the \kahler potential
\be
\label{kcecottih}
K_{\lb} = \frac{1}{4\I} \( \rigZ^\Lambda \bG_\Lambda- \brigZ^\Lambda G_\Lambda \)
+ \frac14\, (W_\Lambda + \bW_\Lambda) [\Im\tau^{-1}]^{\Lambda\Sigma}
(W_\Sigma + \bW_\Sigma)
\ee
in complex structure $J_3$, which differs also from \eqref{kcecotti} by a flip of sign.
Furthermore, it can be checked that the
Darboux coordinate
\be
\talp= \sigma +2\I \left( \varpihk^{-1} \rigZ^\Lambda W_\Lambda + \varpihk \bar \rigZ^\Lambda\bar
W_\Lambda \right) - 8  c\( \phip + \I \log \varpihk\) \, ,
\ee
obtained by replacing $\varpiqk$ by $\varpihk e^{-\I\phip}$ in the third
equation of \eqref{gentwi}, satisfies
\be
\conhyper = -\frac12 \(\bpzeta \talp + \pzeta \bar\talp\)
\ee
in any complex structure. This of course agrees with \eqref{lambda}, since
$\alpha$ and $-\frac12\talp$ differ by a holomorphic function. Thus,
$\tilde\Upsilon\equiv e^{-\I\pi\talp}$ provides
a holomorphic section of the bundle $\lb_{\pcZ}$, related to $\Upsilon=e^{2\I\pi\alpha}$
by a (complexified) gauge transformation.

\section{D-instantons, wall-crossing and contact geometry}
\label{contactwallcrossing}

In this section we apply the general correspondence discussed in \S\ref{seq_qkhk}
to the geometry of the hypermultiplet moduli space $\cM$ in $\cN=2$ string vacua.
As indicated in the introduction, the twistorial construction of the QK
manifold $\cM$ presented in \cite{Alexandrov:2008gh}  is closely similar to the twistorial construction
of the HK moduli space $\pcM$ of the Coulomb branch of $\cN=2$ rigid field theories
 in \cite{Gaiotto:2008cd}. In this section, we shall show that  $\cM$
arises by applying the QK/HK
correspondence to the HK manifold $\pcM$ constructed in  \cite{Gaiotto:2008cd} (using
the central charge function and BPS invariants relevant for the string vacuum at hand),
equipped with a suitable hyperholomorphic line bundle $\lb$. After reviewing the
construction of $\pcM,\cM$ in \S \ref{GMNreview} and  \S\ref{sec_lifting},
we show in \S\ref{sec_rogers}
that the requisite line bundle $\lb$ can be constructed by lifting the
KS-symplectomorphisms $\stg$ across BPS rays on $\pcZ$ to
contact transformations $\ctg$ on $\cZ$ (or to gauge transformations
of the hyperholomorphic line bundle $\lb$ on $\pcM$), using the Rogers dilogarithm function.
Generalizing the techniques of Kashaev and Nakanishi
\cite{Kashaev:2011}, building on earlier work by Faddeev and Kashaev \cite{Faddeev:1993rs},
we show in \S \ref{secmot} that the classical limit of
the motivic KS wall crossing formula implies a set of functional identities
for the Rogers dilogarithm which ensure the consistency of the construction.
In \S \ref{pentagonalcontact}
we give a detailed illustration of the general contact wall crossing formula for the
so called pentagon, hexagon and octagon relations for the Rogers dilogarithm.

\subsection{Wall-crossing in $\cN=2$ gauge theories and symplectic geometry}
\label{GMNreview}

In this section we briefly review the construction of the Coulomb branch of
$\cN=2$ gauge theories on $\IR^3\times S^1$ with emphasis on the phenomenon of wall-crossing \cite{Gaiotto:2008cd}.

\subsubsection{The Coulomb branch of four-dimensional $\cN=2$ gauge theories}

Let us first focus on the Coulomb branch of an $\cN=2$ gauge theory
on $\mathbb{R}^4$ with rank $n$ gauge group $G$. For simplicity
we restrict to the case where the flavor symmetry is trivial and, as in \S\ref{cmap},
assume that the theory is superconformal, as it is the case relevant for the
QK/HK correspondence. The
moduli space $\cB$ is an $n$-dimensional rigid special K\"ahler manifold
parametrized
by $n$ complex valued scalar fields
$z^{i}, i=1, \dots, n$, specifying the vevs of the vector multiplet scalars.
At a generic point on $\cB$ the gauge group $G$ is broken to $U(1)^n$
and there are correspondingly $n$ massless gauge fields. The electric
and magnetic charges $\gamma=(p^{\Lambda}, q_\Lambda)$ are
sections of a local system $\Gamma\rightarrow \cB$ of rank $2n$
lattices $\Gamma_z\cong \IZ^{2n}$ fibered over each point $z^i\in\cB$.
The lattice $\Gamma_z$ is even and self-dual  with respect to the
symplectic inner product $2\langle\cdot,\cdot\rangle$,
\be
\left<\gamma, \gamma'\right>= q_\Lambda p'^{\Lambda}-q'_\Lambda p^{\Lambda} \in\IZ\, .
\label{SymplecticPairing}
\ee
In what follows we identify $\Gamma$
with its dual $\Gamma^{\star}$.

The geometry of $\cB$ can be encoded in a holomorphic  Lagrangian
section
\be
\Omega(z)=(Z^{\Lambda}(z), G_\Lambda(z)),
\ee
$\Lambda=0, 1, \dots, n-1,$ of the
symplectic vector bundle $\Gamma\otimes\IC$ over $\cB$. The Lagrangian
property $\langle \de\Omega,\de\Omega \rangle=0$ implies that $G_\Lambda$
 is locally given by  $\partial_{Z^{\Lambda}}F(Z)$ for some holomorphic
 function $F(Z^\Lambda)$ known as the prepotential. Superconformal
 invariance implies that $F$ is homogeneous of degree 2.
 The \kahler metric on $\cB$ derives from
 the K\"ahler potential
\be
K_\cB= \I \langle \Omega, \bar\Omega \rangle
=\I(\bar{\rigZ}^{\Lambda}G_\Lambda-\rigZ^{\Lambda}\bar{G}_\Lambda) ,
\label{KahlerpotB}
\ee
while the complexified gauge coupling (or period matrix)
is given by the second derivative of the prepotential,
$\tau_{\Lambda\Sigma}$.
A key object in the study of $\cN=2$ theories is the central charge function (or stability data)
$Z' \, :\, \Gamma \to \mathbb{C}$, defined as the inner product:
\be
Z'_\gamma(z) = \left< \gamma, \Omega(z)\right> = q_\Lambda \rigZ^\Lambda(z)
- p^\Lambda G_\Lambda(z) .
\label{centralcharge}
\ee

\subsubsection{Compactification to three dimensions and the semi-flat metric}

Upon compactification on a circle, the low-energy dynamics is described
by an $\cN=4$ supersymmetric sigma model on $\IR^3$, with complex $2n$-dimensional
\hk target space $\pcM$. Topologically, $\pcM$ is a twisted torus bundle $\cT_z\rightarrow
\pcM\rightarrow \cB$ over the four-dimensional Coulomb branch $\cB$. The torus
fiber $\cT_z=\Gamma\otimes_{\mathbb{Z}}\mathbb{R}/\mathbb{Z}$
over each point $z^i\in \cB$ parametrizes the holonomies $C=(\zeta^{\Lambda}, \tilde\zeta_\Lambda)$ of
the electric and magnetic Abelian gauge fields around the circle. Invariance
under large gauge transformations requires that the holonomies are valued in
$\mathbb{R}/\mathbb{Z}$, i.e. are periodic under integer translations
\be
\zeta^{\Lambda} \, \mapsto \, \zeta^{\Lambda}+n^\Lambda,
\qquad
\tilde\zeta_\Lambda \, \mapsto \, \tilde\zeta_\Lambda +m_\Lambda,
\qquad
H=(n^{\Lambda}, m_\Lambda)\in \mathbb{Z}^{2n} .
\label{abeliantransl}
\ee
In the infinite radius limit\footnote{We set
the radius of the circle to $r=2$ using superconformal invariance. The infinite
radius limit then corresponds to the boundary of $\cB$ where $Z^\Lambda$ is scaled
to infinity.}, the HK metric on $\pcM$ is given by the rigid $c$-map metric \eqref{rcmap},
with $\cB$ playing the role of the rigid special \kahler manifold.
In this context, the  rigid $c$-map metric \eqref{rcmap} is also known as the
`semi-flat' metric on $\pcM$.

The corresponding twistor space
$\pcZ$ admits a canonical set of complex Darboux coordinates
$\pXi=(\eta^{\Lambda}, \mu_\Lambda)$ given in \eqref{gentwihk}. Defining, for any
$\gamma\in\Gamma$,
\be
\pXi_\gamma\equiv \langle \gamma,\pXi\rangle,
\qquad
\Theta_\gamma= \langle \gamma,C\rangle,
\ee
these Darboux coordinates can then be written as
\be
\pXi^{\text{sf}}_\gamma\equiv\Theta_\gamma +\varpihk^{-1}Z'_\gamma-\varpihk\bZ'_\gamma
\ .
\label{semiflatDarboux}
\ee
The translations \eqref{abeliantransl}
of $(\zeta^{\Lambda}, \tilde\zeta_\Lambda)$ then lift to a holomorphic action on $\pcZ$:
\be
\eta^{\Lambda}\, \mapsto\, \eta^{\Lambda}+n^{\Lambda},
\qquad
\mu_\Lambda \, \mapsto \, \mu_\Lambda +m_\Lambda.
\label{holabeliantranl}
\ee
This twistorial description of the semi-flat metric will play an important role in what follows.

While the metric \eqref{rcmap} is correct in the strict $R=\infty$ limit, at finite radius
it fails to take into account instanton effects arising from $D=4$ BPS states whose
Euclidean worldline  winds around the circle. These effects are particularly important
near singularities in $\cB$ where these BPS states become massless, and are expected
to resolve the singularity of the semi-flat metric \eqref{rcmap}. As shown in \cite{Gaiotto:2008cd},
the corresponding quantum corrections to the HK metric on $\pcM$
are largely dictated by consistency with wall-crossing, to which we now turn.

\subsubsection{BPS-instantons and wall-crossing}

The  Hilbert space $\cH(z)$ of single-particle states in a four-dimensional gauge theory
depends on the values of the scalar fields $z^i$ and is graded by the charge lattice $\Gamma$,
\be
\cH(z)=\bigoplus_{\gamma\in \Gamma}\cH_\gamma(z).
\ee
 The index (or second helicity supertrace)
\be
\Omega(\gamma,z)=-\frac{1}{2}\text{Tr}_{\cH_\gamma(z)}(2J_3)^2 (-1)^{2J_3} \in \IZ\, ,
\label{defindex}
\ee
where $J_3$ is generator of the little group in $D=4$, is sensitive only to
BPS states, i.e. to single-particle states whose mass $M$ saturates the bound
$M\geq |Z'_\gamma(z)|$, which is determined by the central charge function \eqref{centralcharge}.
The index $\Omega(\gamma,z)$ is a locally constant
function of $z\in \cB$ but may jump at co-dimension one subspaces corresponding
to walls of marginal stability:
\be
\label{defwall}
W(\gamma_1,\gamma_2) = \{ z\in \cB \, : \, \arg[Z'_{\gamma_1}(z)]=\arg[Z'_{\gamma_2}(z)]\}\, ,
\ee
 where $(\gamma_1, \gamma_2)$ are two primitive charge vectors. Across the wall
 $W(\gamma_1,\gamma_2)$, BPS bound states of particles with charges $\gamma^{(m)}$
 lying in the two-dimensional sublattice spanned by $\gamma_1$ and $\gamma_2$
 become unstable, leading to a jump in the index $\Omega(\gamma,z)$ for
 $\gamma=\sum_m \gamma^{(m)}$. The jump of $\Omega(\gamma,z)$ across the wall is
 determined by the Kontsevich-Soibelman (KS) wall-crossing formula \cite{ks}.

 The KS wall-crossing formula holds the key to the construction of the twistor space $\pcZ$ for
 the instanton-corrected metric on $\pcM$ as follows \cite{ks,Gaiotto:2008cd}. At a fixed
 point $z^i$ on the 4D Coulomb branch, the instanton correction from a BPS state
 with total charge $\gamma$ induces a discontinuity in the canonical
 Darboux coordinates $\pXi$
 across a meridian line $\ell_\gamma$ on the twistor fiber, also known as
 a BPS ray, which extends from the north ($\varpihk=0$) to the south ($\varpihk=\infty$) pole at a
 longitude determined by the phase of the central charge $Z'_\gamma(z)$:
\be
\ell_\gamma=\{ \varpihk \in \mathbb{C}^{\times}:\  Z'_\gamma(z)/\varpihk \in \I\IR^-\}\, .
\label{lgam}
\ee
The discontinuity in $\pXi$ is given by the action of a (twisted) complex symplectomorphism $\stg$
which is most  conveniently represented in terms of its action on (twisted) holomorphic
Fourier modes $\pcX_\gamma$ (with respect to the abelian translation group
\eqref{holabeliantranl}), defined by
\be
\label{defpcXg}
\pcX_\gamma\equiv \sigma(\gamma)\, e^{-2\pi\I \langle \gamma,\pXi\rangle}\, .
\ee
Here $\sigma(\gamma)$
is a quadratic refinement
of the intersection form on $\Gamma$, i.e. a homomorphism
$\sigma: \Gamma\to U(1)$ satisfying the cocycle relation
\be
\label{qrifprop}
\sigma(H+H') = (-1)^{\langle H, H' \rangle}\, \sigma(H)\, \sigma(H') \, .
\ee
In the basis where $H=(n^{\Lambda}, m_\Lambda)$
the quadratic refinement can be parametrized by characteristics
$\Theta=(\theta^\Lambda,\phi_\Lambda)\in(\Gamma\otimes \IR)/\Gamma$
such that \cite{Belov:2006jd}
\be
\sigma_\Theta(H) = e^{-\pi \I  m_\Lambda n^\Lambda + 2\pi \I\left(m_\Lambda \theta^\Lambda
- n^\Lambda \phi_\Lambda\right)}\, .
\label{quadraticrefinement}
\ee
We shall restrict ourselves to the case where $\sigma_\Theta(H)$ is real, i.e. where
the characteristics are half integer.
The action of the symplectomorphism $\stg$ on $\pcX_\gamma$ is then
\be
\label{ugamma}
\stg(z)\, \quad :\, \quad  \pcX_{\gamma'}\, \longmapsto\,
\pcX_{\gamma'} (1- \pcX_\gamma)^{\Omega(\gamma,z) \langle \gamma, \gamma' \rangle}\, .
\ee

The symplectomorphisms $\stg$ are naturally identified with the abstract operators
featuring in the KS wall-crossing formula, which we are now ready to state: as $z\in \cB$
crosses the wall $W(\gamma_1, \gamma_2)$, the jump in the index $\Omega(\gamma,z)$
for $\gamma$ lying in the two-dimensional sublattice spanned by $\gamma_1, \gamma_2$
should be such that the following product of symplectomorphisms stays constant:
\be
A(\gamma_1, \gamma_2; z)=
\prod_{\gamma=m_1\gamma_1+m_2\gamma_2\atop m_1, m_2\geq 0} \stg(z),
\ee
where the factors are ordered so that $\text{arg}(Z'_\gamma)$ decreases from left to right
(corresponding to a clockwise ordering of the BPS rays $\ell_\gamma$).
Equivalently, this may be rewritten as
\be
\label{ewall}
\prod_{\substack{m_1\geq 0,m_2\geq 0, \\ m_1/m_2\downarrow}}
U_{m_1\gamma_1+m_2\gamma_2}(z_+)
=
\prod_{\substack{m_1\geq 0,m_2\geq 0, \\ m_1/m_2\uparrow}}
U_{m_1\gamma_1+m_2\gamma_2}(z_-),
\ee
where $z_{\pm}$ denote points infinitesimally close on opposite sides of the wall.
By applying the Baker-Campbell-Hausdorff formula repeatedly, one may rewrite
the product of factors appearing on the l.h.s. in the opposite order and express
the BPS index $\Omega(\gamma)$ on one side of the wall in terms of its
value on the other side (see e.g. \cite{Pioline:2011gf} for more details).
With this twistorial interpretation of the operators $\stg$, it is now clear
that the KS formula ensures
that the complex symplectic structure on $\pcZ$ defined by the
collection of symplectomorphisms $\{\stg(z), \gamma\in\Gamma\}$ is unchanged as
$z\in\cB$ crosses the wall.

According to a standard procedure, the HK metric can be obtained by `parametrizing the twistor lines',
i.e. determining the Darboux coordinates $\pXi_\gamma$ in terms of the
coordinates $\rigX^\Lambda,\zeta^\Lambda,\tzeta_\Lambda$ on $\pcM$ and
of the coordinate $\varpihk$ on the twistor fiber, and plugging them into the complex
symplectic two-form \eqref{omtintro}. The gluing conditions \eqref{ugamma}
for the Darboux coordinates
across the BPS rays, as well as the boundary conditions at $\varpihk=0$ and $\varpihk=\infty$,
can be summarized in the following system of integral equations
\cite{Gaiotto:2008cd,Alexandrov:2009zh}:
\be
\label{xiint}
\pcX_\gamma= \pcX^{\text{sf}}_\gamma\,
\exp\left[\frac{1}{4\pi \I}\sum_{\gamma'} \Omega(\gamma')\, \langle \gamma,\gamma'\rangle
\int_{\ell_{\gamma'} }\frac{\de \varpihk'}{\varpihk'} \frac{\varpihk+\varpihk'}{\varpihk-\varpihk'}
\,\log\left(1-
\pcX_{\gamma'}(\varpihk')\right)\right].
\ee
Solving \eqref{xiint} iteratively with respect to increasing number of instantons
(by first plugging in $\pcX^{\text{sf}}_\gamma$ on the r.h.s. and successively repeating this procedure)
generates an infinite series of multi-instanton corrections to the semi-flat metric.
The one-instanton corrections
correspond to the first correction in this iterative scheme and are weighted by a
factor $\Omega(\gamma,z)e^{-4\pi |Z'_\gamma(z)|-2\pi\I \langle \gamma,C\rangle}$,
which is discontinuous across walls
of marginal stability. Nevertheless, multi-instanton corrections conspire so as to produce
a smooth metric across the walls. In addition, these instanton corrections
resolve the codimension 2 singularities arising from BPS states becoming massless,
at least when the BPS state has primitive charge \cite{Gaiotto:2008cd}.

\subsection{Wall-crossing in $\cN=2$ supergravity and contact geometry}
\label{sec_lifting}

We now turn to wall-crossing in the context of four-dimensional $\cN=2$ supergravities
arising as low-energy limits of type II string compactifications on compact Calabi-Yau threefolds.

\subsubsection{The hypermultiplet sector of $\cN=2$ string vacua}

In string theory the analogue of the hyperk\"ahler Coulomb branch $\pcM$
is the vector multiplet moduli space $\cM$ in type IIA/B on $X\times \mathbb{R}^{3}\times S^{1}$,
where $X$ is a compact Calabi-Yau threefold. By T-duality along the compactification
circle, the same space $\cM$ also describes the hypermultiplet moduli space
in type IIB/A on $X\times \mathbb{R}^4$. In either case local $\cN=2$ supersymmetry
requires that $\cM$ be \qk \cite{Bagger:1983tt}. For definiteness we shall use
a terminology adapted to the hypermultiplet sector of type IIA string theory.
The dictionary for translating to other set-ups can be found
in \cite{Alexandrov:2008gh,Alexandrov:2010ca}.

In this language, $\cM$
is parametrized by the expectation value $\rho$ of the dilaton, the NS-axion $\sigma$
(the 4D dual of the $B$-field), the periods $(\zeta^{\Lambda}, \tilde\zeta_\Lambda)$
of the RR 3-form $C$ along a symplectic basis $(\cA^{\Lambda}, \cB_\Lambda)$ of
the (D-brane) charge lattice $\Gamma\equiv H_{3}(X, \mathbb{Z})$, together with
$n-1$ complex scalars $z^{a}$, $a=1, \dots, n-1$, corresponding to coordinates
on the complex structure moduli space $\cM_X$ of the Calabi-Yau.
In the weak coupling limit, to all orders in $1/\rho$, the metric on $\cM$
is given by the $c$-map metric \eqref{hypmetone}
after identifying the projective special K\"ahler manifold $\cS\cK$ with
the complex structure moduli space $\cM_X$, and fixing the parameter $c$ to
\be
\label{cval}
c= -\chi(X)/(192 \pi)\, ,
\ee
where $\chi(X)$ is the Euler number of $X$.
The twisted torus $\cT$ is then identified
with the intermediate Jacobian $H^{3}(X, \mathbb{R})/H^3(X, \mathbb{Z})$
in the Weil complex structure, with metric given in \eqref{dsT}. This torus is in turn
fibered over $\cM_X$ with total space $\cJ_X\rightarrow \cM_X$ known as the relative
intermediate Jacobian.
Similarly as in field theory the torus coordinates $C\equiv (\zeta^{\Lambda}, \tilde\zeta_\Lambda)$
are periodic with integer periods, due to  large gauge transformations of the RR 3-form.
However now the large gauge transformations involve an
additional  shift of the NS-axion $\sigma$:
\be
\zeta^{\Lambda}\,\mapsto\,\zeta^\Lambda+n^{\Lambda},
\qquad
\tilde\zeta_\Lambda\,\mapsto\, \tilde\zeta_\Lambda+m_\Lambda,
\qquad
\sigma \,\mapsto\,\sigma +2\kappa +\langle C-2\Theta,H\rangle -n^\Lambda m_\Lambda,
\label{Heisenberg}
\ee
where $(n^\Lambda,m_\Lambda, \kappa)\in \mathbb{Z}^{2n}\times \mathbb{Z}$.
The characteristics $\Theta=(\theta^{\Lambda}, \phi_\Lambda)\in
H^{3}(X, \mathbb{R})/H^3(X, \mathbb{Z})$
appearing in \eqref{Heisenberg} are conjecturally
identified with those appearing in the quadratic refinement \eqref{quadraticrefinement}
\cite{Alexandrov:2010np}.
Due to the periodicity under $\sigma \rightarrow \sigma +2$, the NS-axion $\sigma$
parametrizes the fiber of a
circle bundle over $\cJ_X$. Hence, at fixed large value of $\rho$ (weak-coupling limit)
the manifold $\cM_\rho$ is the total space of the fibration \cite{Alexandrov:2010np,Alexandrov:2010ca}:
\be
S^{1}_\sigma\,   \longrightarrow\,   \cM_\rho  \, \longrightarrow\, \cJ_X\, ,
\label{sigmabundle}
\ee
equipped with a connection $D\sigma$ given by \eqref{sigmaconnection}.
The fibration \eqref{sigmabundle} is the stringy generalization of the twisted
torus bundle $\cT_z \rightarrow \pcM\rightarrow \cB$ in $\cN=2$ field theories
on $\mathbb{R}^3\times S^1$.

As explained in \S\ref{localcmap}, the perturbative metric \eqref{hypmetone}
on $\cM$ is most conveniently described in terms of its
twistor space,
with Darboux coordinates $\Xi=(\xi^{\Lambda}, \tilde\xi_\Lambda)$ and  $\talp$ given
in \eqref{gentwi}. In particular, the large gauge transformations
\eqref{Heisenberg} lift to the holomorphic action on $\cZ$,
\be
\label{heisalgz}
\xi^\Lambda \mapsto \xi+n^{\Lambda},
\qquad
\tilde\xi_\Lambda\mapsto \tilde\xi_\Lambda+m_\Lambda,
\qquad
\talp\mapsto\talp +2\kappa+ \langle \Xi-2\Theta,H\rangle - n^\Lambda m_\Lambda  ,
\ee
leaving invariant the holomorphic Fourier modes
\be
\cX_\gamma=\sigma(\gamma)\, e^{-2\pi\I\langle\gamma,\Xi\rangle}\, ,
\ee
which are the direct analogues of the holomorphic Fourier modes \eqref{defpcXg} in the
field theory case.

\subsubsection{D-brane instantons and wall-crossing}

Similarly to the semi-flat metric on $\pcM$, the perturbative metric
on $\cM$ is only physically valid in the strict weak-coupling limit
where $\rho=\infty$. At finite values of the coupling there are
additional effects arising from D-brane instantons, i.e. Euclidean
D-branes wrapping supersymmetric cycles in the internal
manifold. For type IIA compactified on a  Calabi-Yau $X$
these correspond to D2-branes wrapping special Lagrangian
3-cycles (sLags) in $X$, with homology class
$\gamma=[q_\Lambda\cA^{\Lambda}-p^{\Lambda}\cB_\Lambda]\in H_3(X,
\mathbb{Z})$. In the weak coupling limit $\rho\to\infty$ and in the
one-instanton approximation,
corrections to the metric on $\cM$ are of the form
$\Omega(\gamma,z)e^{-8\pi |Z_\gamma|/g_s-2\pi \I\langle \gamma,C\rangle}$,
where $g_s\equiv \rho^{-1/2}$ is the
string coupling, $Z_\gamma(z)$ is the central charge function given by a period integral
of the holomorphic 3-form $\Omega^{3,0}\in H^{3,0}(X,\mathbb{C})$:
\be
\label{Zsugra}
Z_\gamma(z)=e^{\cK/2} \int_\gamma \Omega^{3,0}=
e^{\cK/2}\left( q_\Lambda X^\Lambda(z) - p^\Lambda F_\Lambda(z) \right)
\, \in \, \mathbb{C} \, ,
\ee
 and $\Omega(\gamma,z)$ is the
generalized Donaldson-Thomas (DT) invariant, counting the number
of stable sLags in homology class $\gamma$.
Just like the BPS indices in
rigid $\cN=2$ field theories, the DT invariants $\Omega(\gamma)$ are locally constant functions
of $z^{a}\in \cM_X$ but may jump on codimension 2 subspaces
$W(\gamma_1,\gamma_2) $ defined as in \eqref{defwall} (with $\cB$ replaced by $\cM_X$),
with a jump determined by the  KS wall-crossing formula \eqref{ewall}.
Importantly, D-brane instanton corrections (unlike NS5-instanton corrections)
are independent of the NS-axion
$\sigma$, and therefore preserve the Killing vector $\pa_\sigma$.

The D-instanton corrected metric on $\cM$, or rather its
twistor space $\cZ$, was constructed in \cite{Alexandrov:2008gh,Alexandrov:2009zh},
based on consistency with S-duality and
mirror symmetry. The
construction is formally identical to the construction of the twistor space $\pcZ$ of
instanton corrected Coulomb branch $\pcM$ in $\cN=2$ gauge theories, in particular
the holomorphic Fourier modes $\cX_\gamma$
satisfy the same discontinuities \eqref{ugamma} across
BPS rays and integral equations \eqref{xiint} as in the field theory
case.\footnote{The quadratic refinement $\sigma(\gamma)$ was ignored in
\cite{Alexandrov:2008gh,Alexandrov:2009zh}, but has been  included later
in \cite{Alexandrov:2010ca}.} In addition, the
discontinuity of the contact coordinate $\alpha$ was specified in
\cite{Alexandrov:2008gh,Alexandrov:2009zh} in terms of the Spence dilogarithm function, ensuring
that the combined transformation of $(\xi^\Lambda,\txi_\Lambda,\talp)$ preserves the contact one-form.
However, this construction was unsatisfactory on two counts:
i) requiring that the change of Darboux coordinates is a contact transformation
determined the shift of $\talp$  only up to an additive constant, and there could
have been a global
obstruction in choosing these constants, and ii) the notion of Darboux coordinates being
discontinuous across BPS rays ignored the fact that, unlike the field theory twistor space
$\pcZ$,  the stringy twistor space $\cZ$ is {\it not} trivially fibered over $\CP$, but rather
is a non-trivial $\CP$ bundle over $\cM$. In the rest of this section, we shall use the
QK/HK correspondence to put the construction of \cite{Alexandrov:2008gh,Alexandrov:2009zh}
on a more rigorous basis.

\subsection{D-brane instantons, QK/HK correspondence and Rogers dilogarithm
\label{sec_rogers}}

As outlined in \S\ref{intro}, our approach is based on the
fact that D-instanton corrections preserve the continuous isometry $\pa_\sigma$
corresponding to shifts along the NS-axion $\sigma$. Therefore, the D-instanton corrected
\qk metric on $\cM$ can be  equivalently described in terms of the dual HK metric on $\pcM$ and
hyperholomorphic line bundle $\lb$. At the level of twistor spaces, the complex
contact geometry on the twistor space $ \cZ$ over  $\cM$ is then equivalently described by
the complex symplectic geometry of $ \pcZ$, equipped
with the holomorphic line bundle $\lb_{\pcZ}$.

The results of \S\ref{subsubsec-darboux} imply that
 the Darboux coordinates $\Xi'=(\eta^\Lambda, \mu_\Lambda)$ on $\pcZ$ are identified
 with the Darboux coordinates $\Xi=(\xi^{\Lambda}, \tilde\xi_\Lambda)$
on the reduced twistor space  $\cZ/\partial_\alpha $, while
the additional contact coordinate $\Upsilon=e^{2\I \pi \alpha}$ on $\cZ$
parametrizes the $\mathbb{C}^{\times}$-fiber of $\lb_{\pcZ}\to \pcZ$.
The construction of the twistor space $\cZ$ obtained in
\cite{Alexandrov:2008gh}  can now be rephrased as follows:
\begin{itemize}
\item[i)] The dual twistor space  $\pcZ$
is given by the same construction as in $\cN=2$ field theory, with
the appropriate central charge function \eqref{Zsugra} and BPS invariants
$\Omega(\gamma, z)$. In other words,
$\pcZ$ is described by complex symplectomorphisms $U_\gamma$Ê \eqref{ugamma}
relating the Darboux coordinates $\Xi'$
across the BPS rays \eqref{lgam}.
\item[ii)] The holomorphic line bundle $\lb_{{\pcZ}}$ over $\pcZ$
is defined by transition functions \eqref{fijLintro} evaluated on the
dual coordinates $\Xi'=(\eta^{\Lambda}, \mu_\Lambda)$:
\be
\label{fijL}
f_{\gamma} = \frac{\Upsilon_+}{\Upsilon_-}=   \exp\left( \frac{S_\gamma(\Xi')}{2\pi\I} \right),
\ee
where $S_\gamma(\Xi')$ can be computed from Eq. (3.32) in \cite{Alexandrov:2009zh} and reads
\be
\label{dirtySgamma}
S_\gamma(\Xi') =  \Omega(\gamma) \[ {\rm Li}_2\big(\cX'_\gamma\big)
-2\pi \I \, q_\Lambda \eta^{\Lambda}\,  \log\left(1-\cX'_\gamma\right)
+\frac12\, {\Omega(\gamma)} \, p^{\Lambda}q_\Lambda
\left[\log\left(1-\cX'_\gamma\right)\right]^2 \] .
\ee
\end{itemize}

To elucidate the transition function \eqref{dirtySgamma}, it is useful to change
coordinates and use the symplectic invariant coordinate $\talp$ \eqref{deftalp}
in place of $\alpha$. Eq. \eqref{dirtySgamma} can then be rewritten as
\be
\label{cleanSgamma}
\Delta_\gamma\talp = \talp_+ - \talp_-= \frac{1}{2\pi^2}\, \tS_\gamma(\Xi'),
\ee
where the function $\tS_\gamma$ is defined by
\be
\label{lga}
\tS_\gamma(\Xi') =  \Omega(\gamma) \, L_{\sigma(\gamma)}(\cX'_\gamma)\,  .
\ee
Here, $L_\eps(z) $ is a variant of the Rogers dilogarithm $L(z)$ \eqref{defrogers} defined by
\be
\label{defrogersv}
L_\eps(z) \equiv {\rm Li}_2(z)+\frac12\, \log (\eps^{-1} z) \log(1-z)\, .
\ee
To check that the combination of the shift \eqref{cleanSgamma} and the symplectomorphism
\eqref{ugamma} preserves the contact
one-form
\be
\cX=-\frac{1}{2}\Big(\de\tilde\alpha +\left<\Xi', \de\Xi'\right>\Big) ,
\label{holconnection}
\ee
we note that under the complex symplectomorphism $\stg$,
\be
\label{varconn}
\stg\, \,  :\,\,  \left<\Xi', \de\Xi'\right>\  \mapsto\
\left<\Xi', \de\Xi'\right>
- \frac{\Omega(\gamma)}{2\pi \I} \left[ \Xi'_\gamma \, \de \log(1-\cX'_\gamma)
-  \log(1-\cX'_\gamma) \, \de \Xi'_\gamma\right] ,
\ee
and use the following properties\footnote{See \eqref{dlm1uv}, \eqref{Lm1vsL} for a precise statement
of these properties.} of the function $L_\eps(z)$:
\be
L_\eps(z) = L(z) - \frac12\, \log\eps\, \log(1-z)\, ,
\ee
\be
\label{dlm1z}
\de L_\eps (z) = -\frac{1}{2} \left( \frac{\log(1-z)}{z} + \frac{\log (\eps^{-1} z)}{1-z} \right)\de z\, .
\ee
It is also important to note that the shift \eqref{cleanSgamma} is consistent with the
invariance under the large gauge transformations \eqref{heisalgz}
thanks to the monodromy of the Rogers dilogarithm around $z=0$.
Indeed, consider the action \eqref{heisalgz} on
the Darboux coordinates $(\Xi'_-,\talp_-)$  on one side of a BPS ray.
Under this action, the Fourier mode
$\cX'_{\gamma}$ rotates by $e^{-2\pi\I\langle\gamma,H\rangle}$, where $H=(n^{\Lambda}, m_\Lambda)\in \mathbb{Z}^{2n}$.
On the other side, the Darboux coordinates $(\Xi'_+,\talp_+)$ determined
by \eqref{vxia} will transform the same way as in \eqref{heisalgz} provided
$\tS_\gamma$ transforms as
\be
\tS_\gamma \mapsto \tS_\gamma -\I\pi\,  \Omega(\gamma) \,
\langle\gamma,H\rangle\, \log\left( 1- \cX'_\gamma \right)  .
\ee
This property is indeed ensured by the last term in \eqref{defrogersv}.

Postponing the issue of consistency with wall-crossing to the next subsection,
we conclude that the geometry of the D-instanton corrected hypermultiplet
moduli space $\cM$ is obtained via the QK/HK correspondence from the HK
manifold $\cM'$ and hyperholomorphic connection $\conhyper$, whose
twistor space $\pcZ$ is governed by the
discontinuity conditions
\beq
\label{vxia}
\ctg  \quad & : & \quad   (\Xi', \talp)\, \,   \longmapsto\,\,
\Big( \stg \cdot \Xi', \, \talp+ \frac{1}{2\pi^2}\, \tS_\gamma(\Xi') \Big)
\eeq
across the BPS ray \eqref{lgam}. These discontinuities can be viewed either as
contact transformations on $\cZ$, or as a combination of a
complex symplectomorphism on $\pcZ$ and a gauge transformation
on $\lb_{\pcZ}$. These gluing conditions together with the regularity
conditions at $\varpihk=0$ and $\varpihk=\infty$ can be summarized
by the same integral equations \eqref{xiint} for the holomorphic Fourier modes
$\cX'_\gamma$, supplemented by an integral formula for the
holomorphic section $\tilde\Upsilon\equiv e^{-\I\pi\talp}$ of $\lb_\pcZ$
\beq
\tilde\Upsilon &=& \exp\left[ -\I\pi \left( \sigma
+\varpihk^{-1} \cW-\varpihk \bar \cW \right)
+\frac{\chi_X}{24} \left( \log \varpihk -\I\phip\right)
+\frac{1}{8\pi^2}\sum_\gamma
\int_{\ellg{\gamma}}\frac{\d \varpihk'}{\varpihk'}\,
\frac{\varpihk+\varpihk'}{\varpihk-\varpihk'}\,
\tS_\gamma\(\Xi\) \right],
\label{manyalpha}
\eeq
where we defined
\be
\cW=G_\Lambda \zeta^\Lambda-\rigZ^\Lambda\tzeta_\Lambda
+\frac{1}{8\pi^2}\sum_\gamma \Omega(\gamma) \, Z'_\gamma(z)\,
\int_{\ellg{\gamma}}\frac{\d \varpihk'}{\varpihk'}\,
\log\(1-\cX'_\gamma\) .
\ee
This result can be easily translated into an expression for the contact coordinate $\talp$
on the QK side using \eqref{identXiintro} and \eqref{rcmapid}. This expression is equivalent
to the one obtained in  \cite{Alexandrov:2009zh}, Eq. (3.32), but is considerably simpler thanks
to the use of the Rogers dilogarithm. By plugging the solution
of \eqref{xiint} and \eqref{manyalpha} (after implementing
the identifications \eqref{identXi}) into the contact one-form \eqref{holconnection}
and matching to \eqref{defDt}, one may extract the D-instanton corrections
to the perturbative metric \eqref{hypmetone} in a systematic fashion.

\subsection{Dilogarithm identities and wall-crossing}
\label{secmot}

We now return to the issue of the consistency of the  set of discontinuities
 \eqref{vxia} with wall-crossing. For the
construction to be independent of the value of the moduli $z\in \cM_X$, the
transformations $\ctg$ should satisfy  the obvious generalization of the KS wall-crossing identity,
\be
\label{ewallV}
\prod_{\substack{m_1\geq 0,m_2\geq 0, \\ m_1/m_2\downarrow}}
V_{m_1\gamma_1+m_2\gamma_2}(z_+)
=
\prod_{\substack{m_1\geq 0,m_2\geq 0, \\ m_1/m_2\uparrow}}
V_{m_1\gamma_1+m_2\gamma_2}(z_-)\ .
\ee
By construction this formula reduces to \eqref{ewall} when projecting onto the base of the fibration $\lb_{{\pcZ}}\to \pcZ$.
This implies in particular that the left
and right-hand sides of \eqref{ewallV} differ at most by a translation $\talp\to\talp+\Delta\talp$ along the
$\mathbb{C}^{\times}$-fiber, which is given by the cumulative effect of the translations in
\eqref{vxia}. To present the explicit form of this shift it is useful to first rewrite the wall-crossing
formula \eqref{ewallV} by assembling all the operators on one side:
\be
\label{ewallV2}
\prod_{\ind} V_{\gamma_\ind }^{\eps_\ind} = \unit \, ,
\ee
where the product runs over all charge vectors appearing in \eqref{ewallV}, and $\eps_\ind$ is a sign which
changes from $+1$ on the right of the product (corresponding to the r.h.s.
of \eqref{ewallV}) to $-1$ on the left (corresponding to the inverse of the l.h.s. of \eqref{ewallV}).
The total translation along the fiber of $\lb_{\pcZ}\to \pcZ$ is now given by
\be
\label{dalphatot}
\Delta\talp =  \frac1{2\pi^2} \sum_{s} \epsilon_s \, \Omega(\gamma_{s} )\,
L_{\sigma(\gamma_{s} )} \big(  \cX_{\gamma_s}(s)  \big) ,
\ee
where $\cX_{\gamma_{s}}(s)$ denotes the Fourier
mode $\cX_{\gamma_s}$  successively acted upon by all preceding gauge transformations:
\be
\cX_{\gamma_s}(s)= U_{\gamma_{s-1}}\circ U_{\gamma_{s-2}}\circ
\cdots \circ U_{\gamma_2}\circ U_{\gamma_1}\cdot \cX_{\gamma_s}.
\ee
We thus need to show that the total shift $\Delta\talp$ in \eqref{dalphatot} vanishes
 modulo the natural periodicity of the variable $\talp$,
\be
\Delta\talp = 0 \ \, \text{mod}\ \, 2\, .
\label{deltaalphaidentity}
\ee
Fortunately, we shall now see that  the  motivic wall-crossing formula of Kontsevich and Soibelman
\cite{ks} ensures that the non-trivial functional identity \eqref{deltaalphaidentity}
for the Rogers dilogarithm indeed holds. Our strategy will be to consider
the semi-classical limit of the motivic wall-crossing formula,
using the techniques of \cite{Faddeev:1993rs,Kashaev:2011}.

Recall that the motivic wall-crossing formula pertains to the `refined index' \cite{Dimofte:2009bv}
(more accurately, the protected spin character  \cite{Gaiotto:2010be})
\be
\Omega(\gamma,y,z^a)=\Tr'(-y)^{2J_3} = \sum_{n\in\IZ} (-y)^n \Omega_n(\gamma,z^a)
\ee
(here $\Tr'$ denotes a trace on the space orthogonal to the bosonic
and fermionic translational zero-modes).
Although this quantity is not protected in string theory, it is nevertheless a useful
construct, since its behavior under wall-crossing can be computed using localization
methods which would break down at $y=1$. The motivic wall-crossing formula takes a
similar form as \eqref{ewall},
\be
\label{ewallq}
\prod_{\substack{m_1\geq 0,m_2\geq 0, \\ m_1/m_2\downarrow}}
\hat U_{m_1\gamma_1+m_2\gamma_2}(z_+)
=
\prod_{\substack{m_1\geq 0,m_2\geq 0, \\ m_1/m_2\uparrow}}
\hat U_{m_1\gamma_1+m_2 \gamma_2}(z_-)
\ee
but the operators $\hat U_\gamma$ are now given by  \cite{ks,Dimofte:2009tm}\footnote{We use
the conventions of \cite{Manschot:2010qz}. Note that $\hbar$ differs by a factor of $\pi$
from the one used in  \cite{Kashaev:2011}, and that $y$ differs by a sign from the
one used in \cite{Gaiotto:2010be}.}
\be
\hat U_\gamma = \prod_{n\in\IZ}\[ \qli2{q^{1/2}}
(y^n \hat\cX_{\gamma})\]^{-(-1)^{n}\Omega_n(\gamma,z^a)}\, ,
\quad
y=-q^{1/2}=e^{\I\pi\hbar},
\ee
where $\qli2{q^{1/2}}(x)$ is the quantum dilogarithm defined in \eqref{defqdilog}, and
$\hat\cX_\gamma$ are generators of the quantum torus
\be
\label{qtorus}
\hat\cX_\gamma \,\hat \cX_{\gamma'} = (-y)^{\langle \gamma,\gamma'\rangle} \,
\hat \cX_{\gamma+\gamma'}\, .
\ee
In particular, for a hypermultiplet BPS state with $\Omega(y)=1$, $\hat U_\gamma =
\qli2{q^{1/2}}^{-1}(\hat\cX_{\gamma})$. In the classical limit $y\to 1$, the adjoint action
\be
{\rm Ad}\, \hat U_\gamma\ : \  \hat \cX_{\gamma'} \mapsto
\hat U_\gamma  \hat \cX_{\gamma'}
(\hat U_\gamma)^{-1}
\ee
reduces to the usual twisted symplectomorphism \eqref{ugamma}.
Thus, the motivic wall-crossing formula  \eqref{ewallq} implies the numerical wall-crossing
formula  \eqref{ewall}. However, we shall see that it also implies a functional identity
for the Rogers dilogarithm, which yields the stronger contact wall-crossing formula
\eqref{ewallV}.

To see this, we proceed as in \cite{Faddeev:1993rs,Kashaev:2011}, and realize the
generators of the quantum
torus \eqref{qtorus} as unitary operators acting on $L^2(\IR^{2r})$:
\be
\hat \cX_{\gamma} = \sigma(\gamma) \exp\left( Q^i ( \hat p_i + \epsilon_{ij} \hat u^j ) \right),
\ee
where $\gamma=Q^i e_i$ (so $Q^i$ contains both the electric and magnetic charges),
$\langle \gamma,\gamma'\rangle=\eps_{ij} Q^i Q'^j$, and
\be
\[ \hat u^i, \hat u^j\] = \[ \hat p_i, \hat p_j\]=0\, ,
\qquad
\[ \hat p_i, \hat u^j\]=-\I\pi\hbar\delta_{ij} \, .
\ee
It will be convenient to use a complete basis of wave functions $\langle u |$ and $| p \rangle$
which diagonalize the action of $\hat u^i$ and $\hat p_i$, respectively:
\be
\langle u | \, \hat u^i = u^i\, \langle u | \, ,
\qquad
\langle u | \, \hat p_i = -\I \pi \hbar \, \pa_{u^i}\, \langle u |
\ee
and similarly
\be
\hat p_i \, |p\rangle  =  p_i \, |p\rangle \, ,
\qquad
\hat u^i \, |p\rangle  =  \I \pi \hbar \, \pa_{p_i}  \, |p\rangle\, .
\ee
The inner products and completeness relations are
\be
\langle u | p\rangle = \langle p | u\rangle^{-1} = e^{\frac{\I}{\pi\hbar} u^i p_i}\, ,
\quad
\int \de u  \, | u\rangle \langle u | =\frac{1}{(2\pi^2 \hbar)^n} \int \de p \, | p\rangle \langle p | =\unit\, .
\ee

We now assume that both sides of \eqref{ewallq} have a finite number of factors, and
rewrite it in a similar way as \eqref{ewallV2}:
\be
\label{ewallq1}
\hat U \equiv \prod_{\ind=1}^N \hat U_{\gamma_\ind }^{\eps_\ind} = \unit \, ,
\ee
where, as before, $\eps_\ind$ changes from $+1$ on the right of the product (corresponding to the r.h.s.
of \eqref{ewallq}) to $-1$ on the left (corresponding to the inverse of the l.h.s. of \eqref{ewallq}).
Thus, for any $u,p$, we have $\langle u | \hat U | p \rangle/\langle u | p \rangle=1$.
Inserting a complete basis of states between each of the factors in \eqref{ewallq1},
we arrive, as in Eq. 5.10 of  \cite{Kashaev:2011}, at
\be
\begin{split}
(2\pi^2\hbar)^{-n(N-1)} \int \de p(1) \de u(2)
\de p(2)  \dots  \de u(N) \\
\langle p(0)|u(1)\rangle \,
\frac{\langle u(1)|\hat U_{\gamma_{1} }^{\eps_1} |p(1)\rangle}{\langle u(1)|p(1)\rangle}
\langle u(1)|p(1)\rangle \times
\langle p(1)|u(2)\rangle \,
\frac{\langle u(2)| \hat U_{\gamma_{2} }^{\eps_\ind} |p(2)\rangle}{\langle u(2)|p(2)\rangle}
\langle u(2)|p(2)\rangle \\
 \dots
\times \langle p(N-1)|u(N)\rangle \,
\frac{\langle u(N)| \hat U_{\gamma_{N} }^{\eps_N}  |p(N)\rangle}{\langle u(N)|p(N)\rangle}
 \langle u(N)|p(N)\rangle = 1\, ,
\end{split}
\label{qintegral}
\ee
where  $u\equiv u(1)\equiv u(N+1), p\equiv p(N)\equiv p(0)$. Now, we use the fact that
\be
\frac{\langle u | \hat U_{\gamma}^\eps  | p\rangle}
{\langle u | p \rangle}
= \prod_{n\in\IZ}  \left( \qli2{q^{1/2}}\left[  y^n
\sigma(\gamma) \exp\left( Q^i ( p_i + \epsilon_{ij} u^j  \right) \right]\right)^{(-1)^{n+1} \Omega_n(\gamma)\, \eps}\, .
\ee
In the semi-classical limit $\hbar\to 0$, the integral \eqref{qintegral} can be evaluated in the saddle point
approximation. Using \eqref{qli2clas}, we arrive at
\be
(2\pi^2\hbar)^{-n(N-1)} \int \de p(1) \de u(2)
\de p(2)  \dots  \de u(N) \, \exp\left( \frac{1}{\pi\I\hbar}\, S \right)\sim 1\, ,
\ee
where
\be
S=\sum_{s=1}^{N} \left[ \frac{\epsilon_s}{2} \, \Omega(\gamma_{s} )
{\rm Li_2}\left[ \sigma(\gamma_{s} )\,  \cY_s \right] -u^i(s) ( p_i(s) - p_i(s-1)) \right]
\ee
and we defined $\cY_s\equiv e^{Q^i_s \[p_i(s) + \epsilon_{ij} u^j(s)\]}$.

We thus need to extremize $S$ with respect to $p_i(s), s=1\dots N-1$, and
$u^i(s), s=2\dots N$. Using \eqref{dLi2z}, we arrive at
\be
\label{sadpteq}
\begin{split}
p_i(s)-p_i(s-1) =&\, \frac12\, \epsilon_{ij} Q^j_s \, \eps_s \,\Omega(\gamma_{s} )\,
\log\left[1-\sigma(\gamma_{s} ) \,\cY_s \right],
\qquad
s=2,\dots, N
 \\
u^i(s)-u^i(s+1) =&\, -\frac12\, Q^i_s \, \eps_s \,  \Omega(\gamma_{s} )\,
\log\left[1-\sigma(\gamma_{s}) \,\cY_s \right],
\qquad
s=1,\dots, N-1 .
\end{split}
\ee
From this we conclude that for $s=2,...,N$ the quantity $\delta_i(s)\equiv p_i(s-1)-\epsilon_{ij} u^j(s)$ is independent of $s$.
 Furthermore, we can choose the initial and final states to satisfy the same relation, i.e.
$p_i(0)-\epsilon_{ij} u^j(1)=\delta_i$.
On the other hand, the same quantity can be evaluated using the saddle point equations \eqref{sadpteq}.
Equating the result to $\delta_i$, one arrives at the following requirement
\be
\sum_{s=1}^N\epsilon_{ij} Q^j(s) \, \eps_s \,\Omega(\gamma_{s} )\,
\log\left[1-\sigma(\gamma_{s} ) \,\cY_s \right]=0.
\label{delsum}
\ee
The left-hand side is recognized as the product of
KS factors corresponding to the product \eqref{ewallq1}
of quantum dilogarithms, provided that we can
identify $\sigma(\gamma_s)\cY_s$ with $\cX_{\gamma_s}$.
To establish this identification, note
that for all $s$ one has
\be
Q^i_s \epsilon_{ij} u^j(s) = Q^i_s (p_i(s)-\delta_i) .
\ee
Therefore, for arbitrary charge $\gamma$ we can define $\cY_{\gamma}(s)\equiv e^{Q^i(2p_i(s)-\delta_i) }$
such that $\cY_{\gamma_s}(s)=\cY_s$. The advantage of these new functions is that for all $s$
they satisfy the following recurrence relation
\be
\cY_{\gamma'}(s-1) = \cY_{\gamma'}(s)\, \left( 1-\sigma(\gamma_{s} ) \cY_s\right)^{\eps_s \langle
\gamma_{s} , \gamma'\rangle \, \Omega(\gamma_{s} )}.
\ee
This is precisely the symplectomorphism \eqref{ugamma} for $\eps_s=1$, or
its inverse for $\eps_s=-1$, which allows to identify $\cX_\gamma=\sigma(\gamma)\cY_\gamma$.

Given these results, in particular the constraint \eqref{delsum},
it is now easy to show that the action $S$ at the saddle point can be rewritten as
\be
S=\frac12\sum_{s=1}^{N} \epsilon_s\, \Omega(\gamma_{s} )\,  \[
 {\rm Li_2}\left( \sigma(\gamma_{s} ) \cY_s \right)
+ \frac12\,Q^i_s\(2 p_i(s)-\delta_i\)  \log \left( 1-\sigma(\gamma_{s})  \cY_s \right) \] .
\ee
Thus, the vanishing of $S$ at the saddle point leads to  the dilogarithm identity
\be
 \sum_{s=1}^{N} \epsilon_s \, \Omega(\gamma_{s} )\,
L_{\sigma(\gamma_{s} )} \big( \sigma(\gamma_{s} ) \cY_s  \big)  = 0 \, .
\label{generaldilogidentity}
\ee
 This formula generalizes the ``non-simply laced'' Rogers dilogarithm identities \eqref{naka},
proven and conjectured in \cite{Nakanishi:2010,Keller:2010,Keller:2010b} using techniques from
the theory of cluster categories. The general identity \eqref{generaldilogidentity} shows
that the constant shift \eqref{dalphatot} vanishes identically, at least
on the slice where all the Fourier modes $\cX_\gamma$ are real. By analytic
continuation, it will continue to vanish on the universal cover of the complex
torus. In the next subsection, we carry out this analytic continuation in detail
for some simple examples of wall-crossing, where \eqref{generaldilogidentity}
reduces to the known 5-term, 6-term and 8-term relations for the Rogers dilogarithm.

\subsection{Analytic continuation of the pentagon, hexagon and octagon identities}

\label{pentagonalcontact}

In this subsection, we show the consistency of the prescription \eqref{lga} with wall-crossing
in three simple examples which involve only a finite number of BPS states on either side of the
wall. Since wall-crossing involves only a two-dimensional sublattice of the total charge lattice,
we can restrict to the rank 2 case, and parametrize the complex torus by two $\IC^\times$-valued variables
\be
x=e^{2\pi\I\txi}=e^{u},
\qquad
y=e^{-2\pi\I\xi}=e^{\tilde u} .
\ee
The KS symplectomorphism \eqref{ugamma} acts as on $x,y$ as
\be
\label{Upq}
U^{(\Omega)}_{p,q}: \quad [x,y]\mapsto
[(1-\zz_{p,q})^{q\, \Omega} x, (1-\zz_{p,q})^{-p\, \Omega} y],
\quad \zz_{p,q}\equiv\sigma_{p,q} x^p y^q\, ,
\ee
preserving  the symplectic form
\be
\de \xi \wedge \de\txi  =  \frac{1}{4\pi^2} \frac{\de x}{x} \wedge \frac{\de y}{y}  Ê\, .
\ee
When $\Omega=1$, we omit the superscript and denote $U_{p,q}=U^{(1)}_{p,q}$.
The inverse of $U^{(\Omega)}_{p,q}$ is $U^{(-\Omega)}_{p,q}$. The contact transformation
$V_\gamma\equiv V_{p,q}^{(\Omega)}$ is obtained by supplementing the action
\eqref{Upq} by a translation of the contact variable $\talp\equiv z/(2\pi^2)$,
\be
\label{Valph}
\talp \mapsto \talp + \frac{\Omega}{2\pi^2} \,
L_{\sigma_{p,q}}\left( \zz_{p,q}\right).
\ee
More accurately, one should choose logarithms $u_{p,q}, v_{p,q}$, such that
\be
e^{u_{p,q}}= \zz_{p,q}\, ,
\qquad
e^{v_{p,q}}= 1-\zz_{p,q}\, ,
\qquad
u_{p,q} = p u + q \tilde u + 2\pi \I c_{p,q}\, ,
\ee
where $c_{p,q}$ is an element of $\IR/\IZ$ such that $\sigma_{p,q}=(-1)^{2c_{p,q}}$,
and express the variation of $\talp$ in terms
of the enhanced Rogers dilogarithm, whose definition and
basic properties are recalled in Appendix \ref{sec_enhrog}:
\be
\label{Valphuv}
\talp \mapsto \talp + \frac{\Omega}{2\pi^2} \left[
L(u_{p,q},v_{p,q})-\I\pi c_{p,q} v_{p,q} \right] .
\ee
By construction, $V_{p,q}^{(\Omega)}$ preserves the contact one-form
\be
-2\cX=\de\talp+\txi\de\xi-\xi \de\txi =
\frac{1}{2\pi^2}\left( \de z+ \frac12\left(  u \, \de \tilde u -  \tilde u \, \de u \right) \right) .
\ee
Moreover, the inverse of $V_{p,q}^{(\Omega)}$ is $V_{p,q}^{(-\Omega)}$.

The simplest example involves a single BPS state of charge $\gamma_1+\gamma_2$
with $\langle\gamma_1,\gamma_2\rangle=1$ decaying into its components
of charge $\gamma_1$ and $\gamma_2$. The corresponding product of KS
symplectomorphisms is the usual `pentagon identity'
\be
\label{pentU}
U_{0,1}^{-1}\, U_{1,0}^{-1}\, U_{0,1}\, U_{1,1}\, U_{1,0} = 1\, ,
\ee
which holds whenever $\sigma_{1,0}\sigma_{0,1}=-\sigma_{1,1}$, as required by
the quadratic refinement condition \eqref{qrifprop}. The successive
images  $(x_\ind,y_\ind)_{\ind=0,\dots 4}$ of $(x_0,y_0)\equiv(x,y)$ under the sequence of
symplectomorphisms \eqref{pentU} (from right to left), as well as the monomials
$\zz_\ind\equiv \zz_{\gamma_\ind}^{-\eps_\ind}$ and $1-\zz_\ind$
  are displayed in table \ref{pentagontable}.
\begin{table}[t]
\begin{center}
\begin{tabular}{|c||c|c|c|c|c|}
\hline
$\ind $ & $ \eps_\ind $ & $ x_\ind $ & $ y_\ind $ & $  \zz_\ind $ & $  1-\zz_\ind
$ \\ \hline
$0 $ & $ 1 $ & $ x $ & $ y $ & $ \frac{\sigma_{1,0}}{x} $ & $ \frac{x-\sigma_{1,0}}{x}
$ \\
$1 $ & $ 1 $ & $ x $ & $ \frac{y}{1-\sigma_{1,0}x} $ & $ \frac{\sigma_{1,1}(1-\sigma_{1,0}x)}{xy}$ & $
- \frac{\sigma_{1,1}(1-\sigma_{1,0}x-\sigma_{1,1}xy)}{xy}
$ \\
$2 $ & $ 1 $ & $ \frac{x(1-\sigma_{1,0} x -\sigma_{1,1} xy)}{1-\sigma_{1,0}x} $ & $
\frac{y}{1-\sigma_{1,0}x-\sigma_{1,1} xy} $ & $ \frac{\sigma_{0,1}(1-\sigma_{1,0}x-\sigma_{1,1}xy)}{y}
 $ & $ \frac{(\sigma_{1,0}x-1)(\sigma_{0,1}-y)}{y}
$ \\
$3 $ & $ -1 $ & $ x(1-\sigma_{0,1} y) $ & $  \frac{y}{1-\sigma_{1,0}x-\sigma_{1,1}xy}
$ & $ \sigma_{1,0} x(1-\sigma_{0,1} y) $ & $ 1-\sigma_{1,0}x-\sigma_{1,1}xy
$ \\
$4 $ & $ -1 $ & $ x(1-\sigma_{0,1}y)$ & $ y
$ & $\sigma_{0,1} y $ & $ 1-\sigma_{0,1}y
$ \\
$5 $ & $ 1 $ & $ x $ & $ y$ & $ \frac{\sigma_{1,0}}{x} $ & $ \frac{x-\sigma_{1,0}}{x} $
\\
\hline
\end{tabular}
\end{center}
\caption{The sequence of symplectomorphisms $U_\gamma$ corresponding to the pentagon identity.}
\label{pentagontable}
\end{table}
Upon extending the range of $\ind$ from $0,\dots 4$ to $\IZ$ by requiring
periodicity modulo 5, one easily checks that $\cX_s$ satisfies the recursion relation
\be
\label{modzz}
\zz_{\ind-1} \zz_{\ind+1}= 1-\zz_\ind\, ,
\ee
with periodicity 5. As we discuss in Appendix \ref{app_cluster}, this recursion relation
finds its origin in the periodicity of  mutations of the cluster algebra associated to the
Dynkin quiver $A_2$. The cluster algebras associated to $B_2$ and $G_2$ lead to
two other simple examples of wall-crossing described by the `hexagon formula'
\be
\label{hexagon}
U_{0,1}^{(-2)} U_{1,0}^{(-1)}  U_{0,1}^{(2)} U_{1,2}^{(1)} U_{1,1}^{(2)} U_{1,0}^{(1)} = 1
\ee
and the `octagon formula'
\be
\label{octogon}
U_{0,1}^{(-3)}  U_{1,0}^{(-1)}  U_{0,1}^{(3)}
U_{1,3}^{(1)} U_{1,2}^{(3)} U_{2,3}^{(1)} U_{1,1}^{(3)} U_{1,0}^{(1)} = 1 ,
\ee
respectively. Using \eqref{Upq}, it is straightforward to check that the products of symplectomorphisms
\eqref{hexagon} and \eqref{octogon} are indeed equal to the identity. In fact,
the monomials $\zz_\ind\equiv \zz_{\gamma_\ind}^{-\eps_\ind}$ in all
 three cases satisfy the recursion relation
 \be
\label{modzzgen}
\zz_{\ind-1} \zz_{\ind+1}= (1-\zz_\ind)^{\Omega_s} ,
\ee
with periodicity $N=5,6,8$ in the $A_2,B_2,G_2$ cases, respectively.
Here $\Omega_\ind=1$ if $\ind$ is even and $\Omega_\ind=1,2,3$ if $\ind$ is odd,
respectively. In the rest of this section, we shall show that the corresponding
product of contact transformations $V_\gamma$ is indeed the identity for these
three cases.

For this purpose, let us denote by $u_\ind, v_\ind, u'_\ind, v'_\ind$ the
logarithms of $(\zz_\ind)^{\pm 1}$ and $1-(\zz_\ind)^{\pm 1}$:
\be
\label{listlog}
e^{u_\ind}=\zz_\ind\, ,
\qquad
e^{v_\ind}=1-\zz_\ind\, ,
\qquad
e^{u'_\ind}=1/\zz_\ind\, ,
\qquad
e^{v'_\ind}=1-1/\zz_\ind\, .
\ee
The logarithms $(u_\ind, v_\ind)$ and $(u'_\ind, v'_\ind)$ are related by
\be
u'_\ind=-u_\ind+2\pi\I\eta_\ind,
\qquad
v'_\ind=v_\ind-u_\ind+\I\pi\eta_\ind,
\ee
where $\eta_\ind$ are odd integers. We choose the logarithms \eqref{listlog} such that
the recursion relation \eqref{modzzgen} lifts to
\be
\label{recurvu2}
\Omega_\ind\, v_\ind = u_{\ind-1} + u_{\ind+1}-\I\pi(\eta_{\ind-1} + \eta_{\ind+1})\, .
\ee
Now, according to \eqref{Valphuv}, the total variation of $z$ under the composition
of the contact transformations $V_\gamma$ is given by
\be
\label{delz4}
\Delta z = \sum_{s: \eps_s=1}
\Omega_s \left(L(u'_s,v'_s) - {\I\pi} c_s \, v'_s\right)
-  \sum_{s: \eps_s=-1}
\Omega_s\left(L(u_s,v_s)  - {\I\pi} c_s \, v_s\right) ,
\ee
where we recall that $c_s$ is a half integer chosen such that $\sigma(\gamma_s)=(-1)^{2 c_s}$.
Since the symplectomorphisms $U_\gamma$ compose to the identity, $\Delta z$
is constant. We shall now show that this constant vanishes modulo $4\pi^2$, provided
the  odd integers $\eta_\ind$ are suitably chosen. Since $z$ is related to the contact
coordinate $\talp$ by $z=2\pi^2\talp$, this $4\pi^2$-ambiguity is consistent with the
mod 2 periodicity of $\talp$.

To show that \eqref{delz4} vanishes modulo $4\pi^2$, we combine \eqref{LinvLuv} together with \eqref{covtr} to obtain
\be
\label{funclm3}
L(u_s,v_s)+L(u'_s,v'_s)=\frac{\I\pi\eta_s}{2}(2v_s-u_s)-{\pi^2} \eta_s^2 +2 L(1)\, .
\ee
Now, we use the key property
\be
 \sum_{s=0\dots N-1}\Omega_s L(u_s,v_s) =\frac{\I\pi}{2} \sum_{s=0\dots N-1}\Omega_s \eta_s v_s - N_+ \,L(1)\, ,
 \label{keyprop}
\ee
where $N_+$ is the total BPS index in the ``positive chamber'',
\be
N_+= \sum_{\substack{s=0\dots N-1\\ \eps_s=1}} \Omega_{s}\, .
\ee
In the $A_2$ case, the relation \eqref{keyprop} agrees with \eqref{Luv5term} upon using \eqref{covtr}.
More generally, \eqref{keyprop} can be justified as follows. First we note that the differential of
the same sum as on the l.h.s. of \eqref{keyprop} for the variant $L_{-1}(u,v)$ of Rogers dilogarithm
vanishes, i.e. $\de\big(\sum_{s} \Omega_s L_{-1}(u_s, v_s)\big)=0$, and hence this sum must be constant:
\be
\sum_{s=0\dots N-1}\Omega_s L_{-1}(u_s, v_s)=- N_+ L(1).
\label{keppropdef}
\ee
The precise value of the constant can be easily verified for the $A_2, B_2, G_2$-examples analyzed in \S\ref{A2example}, and in
fact for finite Dynkin quivers the formula \eqref{keppropdef} follows directly from the dilogarithm
identities proven in \cite{Nakanishi:2010,Keller:2010,Keller:2010b}. We do not know how to establish the identity \eqref{keppropdef}
in complete generality, but we note that it is consistent with the conjectural
eqs. 6.35 and 6.36 in \cite{Nakanishi:2010} (reproduced in \eqref{naka}).
The desired formula \eqref{keyprop} now follows from \eqref{keppropdef} by using eq. \eqref{Lm1vsL}.

Combining \eqref{delz4}, \eqref{funclm3} and \eqref{keyprop},
we can rewrite the total variation of $z$ as
\be
\label{delz5}
\Delta z = \frac{\I\pi}{2}  \sum_{s: \eps_s=1}  \Omega_s \, (\eta_s-2c_s) v'_s
- \frac{\I\pi}{2}  \sum_{s: \eps_s=-1} \Omega_s\, (\eta_s-2c_s) v_s
+ \frac{\pi^2}{2}  \sum_{s: \eps_s=1}  \Omega_s \, (1-\eta_s^2)\, .
\ee
Since the $\eta_s$ are odd, the last term vanishes modulo $4\pi^2$. Using
the recursion relation \eqref{recurvu2}, \eqref{delz5} can be rewritten as
\be
\Delta z
=\frac{\I\pi}{2}   \sum_{s=0}^{N-1} \eps_s
(\eta_s-2c_s) \big( u_{s-1}+u_{s+1} -\I\pi (\eta_{s-1}+\eta_{s+1}) \big)
-\frac{\I\pi}{2} \sum_{s: \eps_s=1} \Omega_s (\eta_s-2c_s) (u_s - \I\pi \eta_s)\, ,
\ee
where the equality holds modulo $4\pi^2$.
In the special case where $2c_s$ is odd for all $s$ (corresponding to $\sigma(\gamma)=-1$
for all BPS states), one may simply choose $\eta_s=2c_s$ for
all $s$, so that $\Delta z$ indeed vanishes. More generally, however,
we find that the condition that
$\Delta z$ should vanish (modulo $4\pi^2$)
for all $u_s, v_s$ subject to \eqref{recurvu2} selects
a two-dimensional linear subspace\footnote{These conditions are essentially
equivalent to the flattening conditions in \cite{MR2033484,MR2350461}.}
in the space of the $\eta_s$. E.g.
in the pentagonal case one has
\be
\eta_0-\eta_3=2(c_0-c_3) ,
\qquad
\eta_2-\eta_4=2(c_2-c_4) ,
\qquad
\eta_1-\eta_3-\eta_4=2(c_1-c_3-c_4)
\ee
while in the hexagonal case we find
\be
\begin{array}{rclcrcl}
\eta_0-\eta_4&=&2(c_0-c_4)\, ,
&\qquad &
\eta_1-\eta_4-\eta_5&=&2(c_1-c_4-c_5) ,
\\
\eta_2-\eta_4-2\eta_5&=&2(c_2-c_4-2c_5) ,
&\qquad &
\eta_3-\eta_5&=&2(c_3-c_5) ,
\end{array}
\ee
and, finally, in the octagonal case one has
\be
\begin{array}{rclcrcl}
\eta_0-\eta_6&=&2(c_0-c_6)\, ,
&\qquad &
\eta_1-\eta_6-\eta_7&=&2(c_1-c_6-c_7),
\\
\eta_3-\eta_6-2\eta_7&=&2(\eta_3-c_6-2c_7) ,
&\qquad &
\eta_4-\eta_6-3\eta_7&=&2(\eta_4-c_6-3c_7) ,
\\
\eta_2-2\eta_6-3\eta_7&=&2(\eta_2-2c_6-3c_7),
&\qquad &
\eta_5-\eta_7&=&2(c_5-c_7).
\end{array}
\ee
In all these cases, a solution with odd $\eta_s$'s can be shown to exist for any half integer $c_s$
obeying the  quadratic refinement condition \eqref{qrifprop}. Thus,  at least in these cases, we
have shown that the product of KS symplectomorphisms can be lifted to a product
of contact transformations consistent with wall-crossing.

\section{Discussion}

In the first part of this work we have presented
a general duality between quaternion-K\"ahler and hyperk\"ahler manifolds with
isometric circle actions.
More precisely, this QK/HK correspondence associates, to a real $4n$-dimensional
QK manifold $\cM$ with a quaternionic $S^{1}$-isometry, a HK manifold $\pcM$ of
the same dimension with a rotational $S^1$-isometry, equipped with a hyperholomorphic circle bundle
$\cP$ and a connection $\lambda$. The construction proceeds by lifting the $S^1$-isometry
of $\cM$ to a triholomorphic isometry of the associated Swann bundle
$\cS\to \cM$, and then performing the standard hyperk\"ahler quotient at non-zero level $\vec r$.
The circle bundle $\cP$ is the level set $\vec\mu=\vec r$, and $\lambda$
is induced from the Levi-Cevita connection on $\cS$. $\cP$
arises as the unit circle bundle in a holomorphic line bundle $\lb$ over $\pcM$
with unitary connection $\conhyper$ in complex structure determined by $\vec r$. By
the usual twistor correspondence, $\lb$ can be lifted to a holomorphic line bundle $\lb_{\pcZ}$
on the twistor space $\pcZ$ over $\pcM$. Unlike the twistor space $\cZ$ over $\cM$,
the former is a trivial product $\CP\times \pcM$. Thus, the QK/HK correspondence
gives a way to bypass the non-trivial topology of the twistor space $\cZ$,
at least for QK spaces with a quaternionic circle action.

In the second part of the paper, we have applied this correspondence to the hypermultiplet moduli space
$\cM$Ê in type II string theory on a Calabi-Yau threefold. In the absence of NS5-brane or
Kaluza-Klein monopole corrections (i.e. for weak coupling, or large radius), the latter
has a quaternionic circle action corresponding to shifts of the NS-axion (respectively, NUT
scalar). We have shown that the twistorial construction of the D-instanton corrected metric on $\cM$
given in \cite{Alexandrov:2008gh} can be reformulated as the construction
of a certain hyperholomorphic circle bundle
$\cP$ over the dual hyperk\"ahler manifold $\pcM$ (or equivalently,
a holomorphic line bundle $\lb_{\pcZ}$ on the twistor space $\pcZ$),
whose transition functions are expressed
in terms of the BPS degeneracies $\Omega(\gamma)$ by means of the Rogers dilogarithm function. The existence
of $\lb_{\pcZ}$  is ensured by the semi-classical limit of the motivic Kontsevich-Soibelman wall-crossing formula.
This reformulation clarifies the geometric origin of the similarity with the construction
of the HK metric on the Coulomb branch of $\cN=2$ gauge theories in 3
dimensions \cite{Gaiotto:2008cd}. In particular, it
provides a rigorous basis for the notion of `complex contact transformations
across BPS rays' used in
\cite{Alexandrov:2008gh}, which should be interpreted as transition functions for
the holomorphic line bundle $\lb_{\pcZ}$ over the twistor space $\pcZ$ of the HK
space $\pcM$ which is dual to the QK-space $\cM$.

Our work also reveals new aspects of the intriguing links between wall-crossing in $\cN~=~2$ theories,
dilogarithm identities and cluster algebras, which have emerged in recent years
(see~\cite{ks,Gaiotto:2010be,Cecotti:2010fi,Keller:2011}). The
generalized wall-crossing formula \eqref{ewallV} for gauge transformations $V_\gamma$
acting on $\lb_{\pcZ}$ generates a wealth of new functional identities \eqref{generaldilogidentity}
for the Rogers dilogarithm $L(x)$, which generalize the identities established (or conjectured)
in \cite{Nakanishi:2010,Keller:2010,Keller:2010b} using cluster algebra techniques.
Moreover, as mentioned in \S\ref{intro},
our construction of the line bundle $\lb_{\pcZ}$ is very reminiscent of recent work of Fock
and Goncharov \cite{FGgeometricquantization}, pertaining to the geometric quantization of
cluster $\cA$-varieties (see \S\ref{app_cluster}), where the Rogers dilogarithm also plays
the central role. This suggests that the complex torus $\pcM(\varpihk)\cong (\mathbb{C}^{\times})^{2n}$,
constructed from $\cM$ via the QK/HK correspondence, should be identified with a cluster seed torus
whose associated cluster variety $\cA$ is equipped with a hyperk\"ahler metric. In this picture
the holomorphic fibration $\lb_{\pcZ}\to \pcZ$ arises as the prequantum line bundle over the $\cA$-cluster variety.
Further support for this relation is found in the fact that the contact one-form $\cX$ in
\eqref{holconnection} defines a holomorphic connection on the line bundle $\lb_{\pcZ}$,
whose curvature $\de \cX$ is proportional to the holomorphic symplectic
form $\pomega(\varpihk)$
on the torus $\pcM(\varpihk)$, as is characteristic for geometric quantization.
The fact that $\lb_{\pcZ}$ is equipped with a connection goes beyond the standard relation
between hyperholomorphic connections on $\pcM$ and holomorphic line bundles
on $\pcZ$, which usually do not carry a natural connection \cite{ward-wells}.

It is natural to speculate that other
semi-classical limits of the motivic KS formula, where the quantization parameter
$q=e^{2\pi\I\hbar}$ approaches other roots of unity, may ensure the existence of higher rank
hyperholomorphic bundles on $\pcM$ (which would be
Morita-equivalent to the rank 1 bundle constructed
in this work). Indeed, on the cluster algebra side, the
limit $\hbar\to s/k\in\IQ$  produces a holomorphic vector bundle
$\mathscr{V}_{\hbar}\to\cA$ of rank $k^{\hf{\rm rk}\cB}$, where $\cB$ Êis the exchange matrix
of $\cA$ (see \S\ref{clustervarieties}), and first Chern class proportional
to $s$. Specializing for simplicity to $s=1$ and ${\rm rk}\cB=2$,
holomorphic sections of $\mathscr{V}_{\hbar}$  have a `non-Abelian' Fourier expansion with
respect to translations along the cluster seed torus and along the $\mathbb{C}^{\times}$-fiber,
which  is equivalent to the Fourier expansion
\be
H_k(\xi, \txi)\Upsilon^{k}=\sum_{\ell\in \mathbb{Z}/(|k|\mathbb{Z})}
\sum_{m\in \mathbb{Z}+\ell/|k|} \tilde{\Psi}_{k, \ell}(\xi-m) e^{2\pi \I km\txi}\Upsilon^{k}
\label{nonabelian}
\ee
of the holomorphic sections of $H^{1}(\cZ, \cO(2))$ which parametrize
deformations of $\cM$  consistent with
invariance under the large gauge transformations
\eqref{heisalgz} \cite{Pioline:2009qt,Bao:2009fg,Bao:2010cc,Alexandrov:2010ca} (see \cite{Persson:2011xi} for a recent survey).
Thus, the hyperholomorphic vector bundle on $\pcM$ arising from a variant of our construction
at $\hbar=1/k$ appears to be the right framework to discuss instanton corrections
from $k$ NS5-branes consistently with wall-crossing, at least  perturbatively away from the D-instanton corrected geometry.

Finally, we note that our construction of the hyperholomorphic line bundle $\lb$Ê
makes sense also in the context of the Hitchin moduli space of Higgs bundles,
and more generally in the context of
$\cN=2$ gauge theories in 3 dimensions.
 It would be very interesting to understand their
physical significance.

\acknowledgments

B. P. is especially thankful to A. Neitzke for many discussions since 2008
on $c$-map constructions and  hyperholomorphic line bundles.
It is also a pleasure to thank T.~Dimofte,
B.~Keller, J.~Manschot, S.~Monnier, S.~Vandoren, and D.~Zagier for helpful
discussions and correspondence.

\appendix

\section{Properties of the Rogers and quantum dilogarithms}
\label{sec_enhrog}

In this appendix we recall the definition and main properties of the Rogers dilogarithm
and its variants. More details can be found in \cite{MR2290758,Nahm:2004ch}
and references therein. We also include a brief summary of the most important properties of
the (non-compact) quantum dilogarithm

\subsection{The Rogers dilogarithm and its analytic continuations}

The Spence dilogarithm ${\rm Li}_2(z)$ is defined for $|z|<1$ by the absolutely convergent series
\be
{\rm Li}_2(z) = \sum_{n=1}^{\infty} \frac{z^n}{n^2}\, .
\ee
By analytic continuation, it defines a multi-valued function on $\IC$,
with a logarithmic branch cut  from $z=1$ to $z=+\infty$ (more precisely,
a univalued function on the universal cover of $\IC\backslash\{1\}$) . For a given contour
$\gamma$ extending from $0$ to $z$, $  {\rm Li}_2(z)$ is given by
\be
 {\rm Li}_2(z) = -\int_{\gamma} \frac{\log(1-y)}{y} \, \de y,
\ee
where $-\log(1-y)$ is the analytic continuation of the series $\sum_{n=1}^{\infty} y^n/n$
along the path $\gamma$. In particular,
\be
\label{dLi2z}
\de  {\rm Li}_2(z) = - \frac{\log(1-z)}{z} \de z\, ,
\qquad
{\rm Li_2}(0)=0\, ,
\qquad
{\rm Li_2}(1)=\frac{\pi^{2}}{6}\, .
\ee
The Spence dilogarithm satisfies many functional relations, which however take a more
pleasant form when expressed in terms of the Rogers dilogarithm.

For $|z|<1$ and $|1-z|<1$, the Rogers dilogarithm is defined by
\be
L(z) = {\rm Li}_2(z)+\frac12 \log z \log(1-z)\, .
\ee
In particular, $L(z)$ takes the special values
\be
L(0)=0\, ,
\qquad
L(1/2) = \frac{\pi^2}{12}\, ,
\qquad
L(1)=\frac{\pi^2}{6}\, .
\ee
By checking that the derivative of the left-hand side vanishes and evaluating $L(z)$
at one of these special values, one easily shows that for $x,y,z$ close to
the real interval $[0,1]$, such that all arguments of $L$ below satisfy $|z|,|1-z|<1$,
the following functional relations are obeyed:
\be
\label{Lz1mz}
L(z)+L(1-z)=L(1),
\ee
\be
\label{Lreal5term2}
L(x) -L\left(\frac{x(1-y)}{1-xy}\right)- L\left(\frac{y(1-x)}{1-xy}\right) + L(y) - L(xy)=0\, .
\ee
Using \eqref{Lz1mz}, one may rewrite this last relation as
\be
\label{Lreal5term}
L(x) +L\left(\frac{1-x}{1-xy}\right)+ L\left(\frac{1-y}{1-xy}\right) + L(y) + L(1-xy)=3L(1)\, ,
\ee
which has the advantage of making all terms appear with the same sign. Moreover,
the arguments $z_\ind$ of $L$ appearing from left to right  in \eqref{Lreal5term}
satisfy the period 5 recursion relation
\be
\label{recurz}
1-z_\ind=z_{\ind-1} z_{\ind+1}\, .
\ee
It is possible to extend $L(z)$ from the interval $[0,1]$ to the real axis such that the above
relations, together with the additional identity  $L(z)+L(1/z)=2L(1)$, are satisfied modulo
$3L(1)$, but we shall not make use of this extension in this work, as we instead
need an analytic extension of $L(z)$ modulo $24L(1)$ into the full complex plane.
Let us also record the nine-term relation, which follows by applying the five-term
relation three times \cite{Kirillov_1994}:
\be
\label{nineterm}
\begin{split}
L(a b c)+L\left(\frac{a (b-1) c}{1-a c}\right)+L\left(\frac{c (b-a)}{1-a
   c}\right)+L\left(\frac{\left(1-\frac{1}{a}\right) (1-a
   c)}{\left(1-\frac{1}{b}\right) (1-b c)}\right)
   \\
   +L\left(\frac{a (b
   c-1)}{1-a}\right)+L\left(\frac{b}{a}\right)+L\left(\frac{b-a}{1-a}\right)
   +L\left(-\frac{b}{1-b}\right) = 0\, .
\end{split}
\ee
This reduces to the five-term relation upon setting $a=b=x, c=y/x$.

By analytic continuation, the Rogers dilogarithm extends to a multi-valued function on $\IC$,
with two logarithmic branch cuts,  from $z=1$ to $z=+\infty$ and from
$z=0$ to $z=-\infty$ (more precisely,
a univalued function on the universal cover of $\IC\backslash\{0,1\}$).
For a given contour $\gamma$ extending from $1/2$ to $z$,
$L(z)$ is given by
\be
L(z) = \frac{\pi^2}{12} -\frac12 \int_{\gamma} \left[ \frac{\log(1-y)}{y} + \frac{\log(y)}{1-y} \right] \, \de y,
\ee
where $\log(1-y)$, $\log(y)$ are analytically continued away from $y=1/2$.
In particular, the derivative of $L(z)$ is given by
\be
\de  L(z) = - \frac12 \left[ \frac{\log(1-z)}{z} + \frac{\log(z)}{1-z} \right]\, \de z.
\ee

If one is interested only in the value of $L(z)$ modulo $\IZ(2)\equiv (2\pi\I)^2\IZ=24L(1)\IZ$ (which
suffices for the purpose of this work),
one may trade the universal cover of $\IC\backslash\{0,1,\infty\}$ for the
Abelian cover $\hat Y_+$ of $Y_+=\CP\backslash\{0,1,\infty\}$, defined as
\be
\label{reluv}
\hat Y_+ = \{ (u,v)\in\IC^2 \ \vert \ e^u + e^v=1\} ,
\ee
with covering map $\hat Y_+\to Y_+$ given by
\be
(u,v)\mapsto e^u=z=1-e^v\, .
\ee
In other words,  $u$ and $v$ run over all possible choices of logarithms of $z$ and $1-z$.
The analytically continued Rogers dilogarithm (sometimes referred to as `enhanced')
is then the univalued
function
\be
L\, :\, \hat Y_+\to \IC/\IZ(2),
\ee
 defined by
\be
L(u,v) = \Li_2(e^u) + \frac12\, u v\, .
\qquad
\ee
The enhanced Rogers dilogarithm is ambiguous modulo $\IZ(2)$,  due to the fact that the derivative
\be
\label{eqdL}
\de L = \hf\(u \de v - v \de u\)
\ee
(subject to the relation \eqref{reluv})
has simple poles at $v\in 2\pi \I \IZ$ with residues belonging to $2\pi \I \IZ$.
Under covering transformations $\hat Y_+\to Y_+$ (i.e. upon changing the choice
of logarithms of $z$ and $1-z$) one has
\be
\label{covtr}
L(u+2\pi\I r,v+2\pi\I s) = L(u,v) + \I\pi(rv-s u)+2\pi^2 r s\, .
\ee

This construction of the enhanced Rogers dilogarithm $L(u,v)$ is essentially identical
to the one presented in \cite{MR2033484,MR2350461} in the context of Chern-Simons
invariants of hyperbolic three-manifolds.
We recall that the Bloch-Wigner dilogarithm $D(u,v) $, defined by
\be
D(u,v) = \Im \left[L(u,v) +\frac12\, \bar u v\right],
\ee
is invariant under the covering transformations \eqref{covtr},
so descends to a univalued function $D\, :\, Y_+\to \IR$,
which computes the volume of an ideal hyperbolic tetrahedron with vertices at
$0,1,\infty,z$. In contrast, the real part of $L/(2\pi)$ is
inherently ambiguous modulo
$2\pi$, as it computes the Chern-Simons invariant of the same
tetrahedron (see e.g. \cite{MR2033484} and references therein).

At the special points $(0,\infty)$ and $(\infty,0)$, corresponding to $z=1$ and $z=0$,
the enhanced Rogers dilogarithm is continuous and takes the values
\be
L(0,\infty)=L(1)=\frac{\pi^2}{6}\, ,
\qquad
L(\infty,0)=L(0)=0\, .
\ee
The functional relations \eqref{Lz1mz}, \eqref{Lreal5term}, transcribed as
\be
\label{Luvvu}
L(u,v)+L(v,u)= L(1) \mod \IZ(2),
\ee
\be
\label{Luv5term}
v_\ind = u_{\ind-1} + u_{\ind+1}\quad \Rightarrow\quad
\sum_{\ind\, {\rm mod}\, 5} L(u_\ind,v_\ind) =
3L(1) \mod \IZ(2)\, ,
\ee
now hold throughout the Abelian cover $\hat Y_+$, as one can check by differentiation
and evaluating the l.h.s at
\be
(u_1,v_1)= (u_4,v_4)=(\infty,0),\quad (u_2,v_2)= (u_3,v_3)=
(u_5,v_5)=(0,\infty).
\ee
In addition, the analogue of the functional relation $L(z)+L(1/z)=2L(1)$ alluded
to below \eqref{recurz} becomes
\be
L(u,v)+L(-u,v-u+\I\pi\eta) = 2L(1) + \frac{\I\pi\eta}{2}\, u \mod \IZ(2)\, ,
\label{LinvLuv}
\ee
where $\eta$ is any {\it odd} integer (as is necessary for the argument to belong
to $\hat Y_+$).

As explained in \S \ref{sec_lifting}, it is also advantageous to introduce a variant of the
Rogers dilogarithm defined for $|z|,1-|z|<1$ by
\be
\label{defrogersvA}
L_{-1}(-z) = {\rm Li}_2(-z)+\frac12 \,\log (z) \log(1+z)\,  .
\ee
As before, one may analytically continue this to a function $L_{-1}\ :\ \hat{Y}_-\to \mathbb{C}/\mathbb{Z}(2)$,
defined by
\be
L_{-1}(\hu,\hv) = \Li_2\big( e^{\hu}\big) + \frac12\, \hu \hv
\ee
where
\be
\hat Y_- = \{ (u,v)\in\IC^2 \ \vert \ e^{\hu} - e^{\hv}=-1\},
\ee
is the Abelian cover  of $Y_-=\CP\backslash\{0,-1,\infty\}$, with covering map
\be
z=-e^{\hu}=e^{\hv}-1\, .
\ee
The function $L_{-1}(\hat u,\hat v)$ satisfies properties analogous to \eqref{eqdL},
\eqref{LinvLuv}, \eqref{Luv5term}:
\be
\label{dlm1uv}
\de L_{-1}(\hu,\hv) = -\frac12\, ( \hu \de \hv- \hv\de \hu) \ \mod \IZ(2)\, ,
\ee
\be
\label{funclm1}
L_{-1}(\hu,\hv) + L_{-1}(-\hu,\hv-\hu) = -L(1) \, \mod \IZ(2)\, ,
\ee
\be
\label{Luvm15term}
\hat v_\ind = \hat u_{\ind-1} + \hat u_{\ind+1}\quad \Rightarrow\quad
\sum_{\ind\, {\rm mod}\, 5} L(\hat u_\ind,\hat v_\ind) =
-3L(1) \mod \IZ(2)\, .
\ee
The relation between $L(u,v)$ and $L(\hu, \hv)$ is given by
\be
\label{Lm1vsL}
\hu = u- \I\pi \eta\, ,
\quad
\hv=v \quad \Rightarrow \quad L_{-1}(\hu,\hv) =
L(u,v) -\frac{\I\pi\eta}{2}\, v\ \mod \IZ(2)
\ee
whenever $\eta_s$ is an {\it odd} integer.

\subsection{The quantum dilogarithm}

We now turn to the quantum dilogarithm, defined by \cite{Faddeev:1993rs}
\be
\qli2{q^{1/2}}(x) =
\prod_{n=0}^{\infty} (1+q^{n+\frac12} x)^{-1}= \frac{1}{\left( -x q^{1/2};q\right)_\infty}
\label{defqdilog}
\ee
where $(x;q)_\infty\equiv \prod_{n=0}^{\infty}(1-q^n x)$.
Alternatively,
\be
\qli2{q^{1/2}}(x) = \sum_{n=0}^{\infty} \frac{(-q^{1/2} x)^n}{(1-q)\cdots (1-q^n)}
= \exp\left[ \sum_{n=1}^{\infty} \frac{(-q^{1/2} x)^n}{n(1-q^n)}\right].
\label{qdilog}
\ee
The main property of the quantum dilogarithm is the pentagon identity
\be
\label{penta}
\qli2{q^{1/2}}(x_1) \qli2{q^{1/2}}(x_2) = \qli2{q^{1/2}}(x_2) \qli2{q^{1/2}}(x_{12}) \qli2{q^{1/2}}(x_1),
\ee
where $x_1 x_2=q x_2 x_1$ and $x_{12}=q^{-1/2} x_1 x_2$.
In the classical limit $\hbar \to 0$, $q^{1/2}=-e^{\I\pi\hbar}$, the quantum dilogarithm
reduces to the ordinary dilogarithm,
\be
\label{qli2clas}
\qli2{q^{1/2}}(x) = \exp\left( -\frac{1}{2\I\pi\hbar} \Li_2(x) +  \frac{\I \pi  \hbar\, x}{12(1-x)}
+\cO(\hbar^3) \right).
\ee
In this limit,  the pentagon identity \eqref{penta}
reduces to the five-term relation \eqref{Lreal5term2}  \cite{Faddeev:1993rs}.

\section{Cluster varieties and dilogarithm identities\label{app_cluster}}

In this appendix we will introduce and apply some technology from the theory of cluster algebras and
cluster varieties, as developed by Fomin-Zelevinsky \cite{MR1887642,MR2295199}
and Fock-Goncharov \cite{FG,MR2470108}. This formalism gives powerful
algorithmic methods of finding dilogarithm identities, and consequently wall crossing formulas. We begin by
reviewing some basic properties of cluster varieties and
cluster mutations. In \S \ref{trop} we discuss cluster
transformations over a tropical semi-field, a point of
view which elucidates the periodicity
properties of sequences of cluster transformations. In \S \ref{nuperiod}
we introduce the notion of a framed quiver, which is useful for
extracting a particular class of (quasi-)periodic mutation
sequences, called $\nu$-periods. This allows us to give an explicit expression for
the Kontsevich-Soibelman symplectomorphisms $U_\gamma$ in terms of
certain birational cluster automorphisms, conjugated by products of
simple monomial transformations. In \S \ref{A2example}
we discuss some explicit examples corresponding to
the cluster algebras of type $A_2, B_2$ and $G_2$, which are associated with the pentagon,
hexagon and octagon dilogarithm identities studied in \S\ref{pentagonalcontact}.

\subsection{Cluster Varieties}
\label{clustervarieties}

The defining data for a cluster variety (or, more generally, a cluster ensemble) consists of
a finite set $I$ of cardinality $n$, a subset $I_0\subset I$ of cardinality $n_0$, a
$\IQ$-valued function $ \exm_{ij} $ on $I\times I$, such that
$ \exm_{ij} \in \IZ$ unless $(i,j)\in  I_0\times I_0$, and a set of coprime integers $d_i$
such that the function $\hat \exm_{ij} = \exm_{ij} /d_j$
is antisymmetric \cite{FG}. The function $ \exm_{ij} $ is often
called the exchange matrix. It is customary to represent
this data by a quiver diagram $Q$ with $n$ nodes, $| \exm_{ij} /d_j|$ arrows
going from node $i$ to node $j$ if $ \exm_{ij} >0$, or from $j$ to $i$ if $ \exm_{ij} <0$,
and with each node decorated by the integer $d_i$. The nodes associated to $I_0$
are called frozen nodes. In addition, we introduce a set of distinguished
$\IC^\times$-coordinates $\{x_i\}_{i\in I}$ and $\{a_i\}_{i\in I}$ on two complex
$n$-dimensional tori $\cX$ and $\cA$, which respectively carry a Poisson structure
\be
P=d_j^{-1} \exm_{ij}  x_i x_j \frac{\partial}{\partial x_i}\frac{\partial}{\partial x_j},
\label{Poissonstructure}
\ee
and a pre-symplectic structure (i.e. possibly degenerate) given by the closed 2-form
\be
\omega=d_i  \exm_{ij}  \frac{\de a_i}{a_i}\wedge\frac{\de a_j}{a_j} \, .
\label{symplecticform}
\ee
The data $(I,I_0,\exm_{ij}, d_i,\cX,\cA)$ together with the distinguished
coordinates $(x_i, a_i)$ are sometimes called the initial seed.

For any initial seed $(I,I_0,\exm_{ij}, d_i,\cX,\cA;x_i,a_i)$ and any choice of
$k\in I\backslash I_0$, one may construct a new seed $(I,I_0,\exm'_{ij},d_i,\cX',\cA';x'_i,a'_i)$
with the same set of nodes and multipliers, but with a new exchange matrix given by the explicit formula
\be
 \exm'_{ij} =  \left\{Ê\begin{array}{cc}
- \exm_{ij}  &\ i=k \,\,  \text{or}\,\,   j=k
\\
 \exm_{ij}  & \ \exm_{ik}  \exm_{kj} < 0
\\
 \exm_{ij} + \exm_{ik} | \exm_{kj} | &\  \exm_{ik}  \exm_{kj} \geq 0
\\
\end{array}\right. \, ,
\qquad
d'_i=d_i\, ,
\label{exchangeisomorphism}
\ee
and new distinguished $\IC^\times$-coordinates related by the birational
transformation
\beq
x'_i=
\left\{ \begin{array}{cc}
x_i\big(1+x_k^{-\text{sgn}( \exm_{ik} )}\big)^{- \exm_{ik} } &\ i\neq k
\\
x_k^{-1} &\ i=k
\\
\end{array} \right. ,
&\qquad &
a'_i=   \left\{ \begin{array}{cc}
a_i &\ i\neq k
\\
a_k^{-1}\big(\mathbb{A}_k^{+}+ \mathbb{A}_k^{-}\big) &\ i=k
\\
\end{array}\right. ,
\label{birational}
\eeq
where we defined
\be
\mathbb{A}_k^{+} \equiv \prod_{j:  \exm_{kj} > 0} a_j^{ \exm_{kj} },
\qquad \qquad
\mathbb{A}_k^{-}\equiv \prod_{j:  \exm_{kj} < 0} a_j^{- \exm_{kj} }.
\label{fatA}
\ee
The combined transformation \eqref{exchangeisomorphism},
\eqref{birational} is involutive and preserves the Poisson structure \eqref{Poissonstructure}
and closed 2-form \eqref{symplecticform}. Such transformations were first
introduced by Fomin and Zelevinsky \cite{MR1887642,MR2295199}\footnote{
In \cite{MR1887642,MR2295199},the coordinates $a_i$ and $x_i$ are called cluster
variables and principal coefficients and are denoted by $x_i$ and $y_i$,
respectively. The action on the variables $a_i$ given in \eqref{birational}
corresponds to the case where the coefficients are set to one;
moreover, the exchange matrix in \cite{MR1887642,MR2295199} is the transpose of the
one in \cite{FG}, whose conventions we follow.} and are
 called cluster transformations. We shall denote the combination
of  \eqref{exchangeisomorphism} and \eqref{birational} by $\mu_k$, and refer
to it as the mutation along the node $k$. Since $I,I_0,d_i$ are invariant under
mutation, we shall henceforth omit them.

Starting from an initial seed $(\exm_{ij}(0), \cX(0),\cA(0);x_i(0),a_i(0))$ and applying
sequences of mutations, we arrive at a collection of seeds $(\exm_{ij}(\ind), \cX(\ind),
\cA(\ind);x_i(\ind),a_i(\ind))$ attached to the vertices of an `exchange graph', whose edges
correspond to mutations. The seed tori $(\cX(\ind), \cA(\ind))$
together with the mutations $\mu_k$ between neighboring vertices of the exchange graph
form an atlas for the cluster varieties $(\cX,\cA)$ introduced by Fock and Goncharov. The former
carries a Poisson structure given locally by \eqref{Poissonstructure}, while the
latter carries a (possibly degenerate) symplectic structure  given locally
by \eqref{symplecticform}. In addition, there exists a homomorphism from
$\cA$ to $ \cX$, which maps the local coordinates
on seed tori as
\be
p\, :\, (a_i)_{i\in I}\, \mapsto \, (x_i)_{i\in I},
\qquad
x_i= \prod_{j\in I}a_j^{ \exm_{ij} }=\mathbb{A}_i^{+}/\mathbb{A}_i^{-}.
\ee
The fibers of the map $p$ are the leaves of the null-foliation of the 2-form $\omega$,
while the subtorus $p(\cA)$ is a symplectic leaf of the Poisson structure $P$.
The spaces $\cA$ and $\cX$ are in some sense ``Langlands-dual'' to each other \cite{FG}.
It is worth noting that the cluster variables $a_i(\ind)$
satisfy the Laurent phenomenon, in that
they always turn out to be finite Laurent polynomials in the initial cluster variables
$a_i(0)$ \cite{MR1888840}.

\subsection{Monomial transformations and birational automorphisms}

In order to relate the cluster transformations  \eqref{exchangeisomorphism}
with the KS symplectomorphisms $U_\gamma$ of \S \ref{GMNreview}, the first step is to decompose the birational
transformation  \eqref{exchangeisomorphism} into a `birational automorphism',
which preserves the symplectic form \eqref{symplecticform} for fixed exchange
matrix $\cB_{ij}$, and a `monomial map', which acts by a simple change of
basis on the seed tori. For this purpose, it is useful to
rewrite $x'_i$ ($i\neq k$) and $a'_k$ in \eqref{birational} as
\beq
\label{rewritemut}
x'_i &=& x_i\,x_k^{[\exm_{ik} ]_+}\, \big(1+x_k\big)^{- \exm_{ik} } =
x_i\, x_k^{[-\exm_{ik} ]_+}\, \big(1+1/x_k\big)^{- \exm_{ik} }
\\
a'_k &=& a_k^{-1} \, \mathbb{A}_k^{-} \big(1+ x_k \big)=
a_k^{-1} \, \mathbb{A}_k^{+} \big(1+ 1/x_k \big),
\eeq
where we defined $[z]_+=\text{max}(0,z)$ and identified $x_k$ with $p(a_k)=\mathbb{A}_k^{+}/
\mathbb{A}_k^{-}$ in the second line. It is then apparent that $\mu_k$ can be decomposed in
two different ways,
\be
\mu_k = \tau_+ \circ \mu_{k,+} =  \tau_- \circ \mu_{k,-}\, ,
\ee
where $\mu_{k,\epsilon}, \epsilon=\pm 1,$ acts via the birational map
\beq
\mu_{k,\epsilon} \, :\, x_i \mapsto
 \left\{ \begin{array}{cc}
x_i\, \big(1+x_k^{\eps}\big)^{- \exm_{ik} } &\ i\neq k
\\
x_k &\ i=k
\\
\end{array}\right. ,
&\qquad &
a_i\mapsto
 \left\{ \begin{array}{cc}
a_i &\ i\neq k
\\
a_k\big(1+x_k^{\eps}\big)^{-1} &\ i=k
\\
\end{array}\right.  ,
\label{birationalautomorphism2}
\eeq
keeping $\exm_{ij}$ unchanged, while $\tau_{\eps}$ acts by the monomial map
\be
\label{tautrop}
\tau_{k,\epsilon}\ : x_i\mapsto
\left\{\begin{array}{cc} x_i\, x_k^{[\epsilon\,  \exm_{ik} ]_+} &\ i\neq k \\ 1/x_k &\ i=k
\end{array} \right.  ,
\qquad
a_i\mapsto
 \left\{ \begin{array}{cc}
a_i &\ i\neq k
\\
\mathbb{A}_k^{-\epsilon}/a_k &\ i=k
\end{array} \right.
\ee
accompanied by the transformation $\exm_{ij}\mapsto\exm'_{ij}$ in
\eqref{exchangeisomorphism}. Importantly, $\mu_{k,\epsilon}$
preserves the symplectic form \eqref{symplecticform}, while $\tau_{k,\epsilon}$
is still involutive.

\subsection{Tropicalization and framed quivers}
\label{trop}
It is important that the action of $\mu_k$ on $x_i$ does not involve any subtraction,
implying that the variables $x_i(\ind)$  effectively lie in the universal semi-field spanned
by subtraction-free rational expressions in the initial variables $x_i(0)$.
Another natural semi-field is the tropical semi-field ${\rm Trop}(\{x_i(0)\})$, i.e. the free Abelian
multiplicative group generated by elements $x_i(0)$ with
the usual addition rule $+$ replaced by the tropical addition $\oplus$, defined by
\be
\label{troplus}
\prod_j x_j^{a_j} \oplus \prod_j x_j^{b_j} := \prod_j x_j^{\min(a_j,b_j)} .
\ee
There is a canonical
homomorphism  $x\mapsto x^T$ from the universal semi-field to the tropical
semi-field, obtained by replacing $x_i(\ind)$ by its leading Laurent monomial in the limit
where all the initial variables $x_i(0)$ are scaled to zero at the same rate.
This homomorphism, sometimes known as the tropicalization, commutes
with the mutations, and one may therefore ask how mutations act at the tropical
level. It turns out that at each step, the tropicalisation of the variable $x_i(\ind)$
is a monomial of either positive degree in all initial variables $x_i(0)$, or of
negative degree in all $x_i(0)$ (Conj. 5.4 and Prop. 5.6 in \cite{MR2295199})).
Equivalently, the `$c$-vectors' ${\bf c}(s)$ with components
\be
c_i(\ind)=\left(\frac{\partial\log x_i^T(\ind)}{\partial \log x^T_j(0)}\right)_{j\in I}
\ee
satisfy the `sign-coherence' property, i.e. all non-zero entries in the vector ${\bf c}(s)$ have
the same sign  \cite{Nakanishi:2011}. Following the terminology of \cite{Kashaev:2011},
we refer to this sign as the tropical sign of $x_i(s)$, and denote it by $\epsilon(x_i(s))$. The sign-coherence
property of the  $c$-vectors will play a crucial role below for establishing dilogarithm identities.

Upon replacing $+$ by $\oplus$ in the birational automorphism
\eqref{birationalautomorphism2} and choosing $\epsilon=\epsilon(x_k)$, it is then
clear that $\mu_{k,\epsilon(x_k)}$ acts trivially at the tropical level, and therefore
the tropicalization of $\mu_k$ reduces to the action of the monomial map
$\tau_{k,\epsilon(x_k)}$:
\be
\label{mutrop}
\mu_k^T \equiv \tau_{k,\epsilon(x_k)}\
:\ x_i\mapsto
\left\{\begin{array}{cc} x_i\, x_k^{[\epsilon(x_k)\,  \exm_{ik} ]_+} &\ i\neq k, \\ 1/x_k &\ i=k
\end{array} \right. ,\qquad
a_i\mapsto
 \left\{ \begin{array}{cc}
a_i &\ i\neq k \\ \mathbb{A}_k^{-\epsilon(x_k)}/a_k &\ i=k.
\end{array} \right.
\ee
Correspondingly, the action on the $c$-vectors is
obtained by
\be
\label{mutropt}
\log \mu_k^T\ :\  c_i \mapsto
\left\{\begin{array}{cc} c_i + [\epsilon(x_k)\,  \exm_{ik} ]_+ c_k & \ i\neq k \\ -c_k &\ i=k
\end{array} \right.  ,
\ee
which provides the gluing conditions for the tropical variety in \cite{FG}.
As we shall see momentarily, the birational automorphism $\mu_{k,\eps(x_k)}$, suitably conjugated
by a product of monomial transformations, is the one to be identified with the KS symplectomorphism $U_\gamma$.

A useful way to compute the tropical variables $x_i^T(\ind)$ is to extend the set of
nodes $I_u=I\backslash I_0$ by a copy $I'_u$, and extend the exchange matrix $\exm$
into $\tilde\exm$ such  that $\tilde \exm_{ij} = \exm_{ij} $ when $i,j\in I$,
$ \exm_{ii'} = \exm_{i'i} =1$ between a node in $i\in I_u$ and its copy in $i'\in I'_u$, and
$\exm_{i'i'}=0$ for $i',j'\in I'_u$. The full set of nodes is then $\tilde I=I_u\cup I'_u \cup I_0$,
where $\tilde I_0= I'_u \cup I_0$ are frozen (this construction is in fact the main reason for
introducing the notion of frozen node). We refer to $\tilde\exm$ as the framed exchange matrix,
and to the corresponding quiver $\tilde Q$ as the framed quiver. One can then show
that the $c$-vectors are given by the off-diagonal part of the framed exchange matrix,
namely $c_i(\ind)=\tilde \exm_{ii'}  e_{i'}$ where $e_{i'}$ is the unit vector in the $i'$-th
direction \cite{MR2295199}.  In particular, the sign $\epsilon(x_i(\ind))$ can be read off
straightforwardly from the sign of the $i$-th row in the upper-right hand
block of the framed exchange matrix (which is independent of the column by the
sign-coherence property).

\subsection{Wall crossing and dilogarithm identities from $\nu$-periods}
\label{nuperiod}

For a sequence of $N$ mutations
$\mu_{\bfk} \equiv \mu_{k_{N}}\circ \dots \circ \mu_{k_2} \circ \mu_{k_1}$ with
${\bf k}=(k_1,k_2, \dots k_n)\in (I\backslash I_0)^N$,
we shall denote by
\be
( \exm_{ij} (\ind),x_i(\ind),a_i(\ind))=\mu_{k_\ind}( \exm_{ij} (\ind-1),x_i(\ind-1),a_i(\ind-1))
\ee
the associated sequence of seeds, with $( \exm_{ij} (0),x_i(0),a_i(0))$ corresponding as before to the initial seed.
More generally, for $\ind\leq N$ we denote by ${\bfk}_\ind=(k_1,...,k_\ind)$ the
subsequence formed by the first $\ind$ mutations.
It is sometimes the case that a sequence of $N$ mutations
$\mu_{\bfk}$ composes to the identity transformation,
$\mu_{\bfk}( \exm_{ij} ,x_i,a_i)=( \exm_{ij} ,x_i,a_i)$.
More generally, it may happen that
\be
\mu_{\bfk}( \exm_{ij} ,x_i,a_i)=\nu( \exm_{ij} ,x_i,a_i),
\ee
where $\nu$ is an automorphism of the seed, i.e. a permutation of the nodes $i\in I$
(with the corresponding action on $ \exm_{ij} ,x_i,a_i$) which fixes the frozen nodes $i\in I_0$.
Such sequences of mutations are called $\nu$-periods of length $N$. An important theorem
asserts that $\mu_{\bfk}$ is $\nu$-periodic if and only if
its tropicalization $\mu_{\bfk}^T$  is
$\nu$-periodic \cite{Keller:2010,Plamondon:2010}.

The $\nu$-periods provide a powerful source of  wall-crossing identities,
as we will now demonstrate. As proposed in \cite{ks} and further elaborated upon
in \cite{Gaiotto:2009hg,Gaiotto:2010be}, we can identify the
$s$:th birational automorphism\footnote{We consider only the action of  $\mu_{k_s,\epsilon(x_{k_s})}$ on
$x_i$, the variables $a_i$ seem to play no role at this stage.}
 inside the sequence $\mu_{\bfk}$
with a KS symplectomorphism $U_{\gamma_\ind}$
for a suitable charge vector $\gamma_\ind$
and index $\Omega(\ind)$, up to conjugation
by a product of monomial maps.
More precisely, let $\{e_i\}$ be an integer basis of vectors  of the ``charge lattice''
$\Gamma$ equipped with the antisymmetric product\footnote{Although the integrality
of this antisymmetric product does not seem to be guaranteed by the axioms, it appears
to hold in all cases of interest.}
\be
\langle e_i, e_j \rangle =  \exm_{ij} /d_j ,
\ee
let $\mu^T_{{\bfk}_\ind}$ be the following product of monomial transformations
\be
\mu^T_{{\bfk}_\ind}= \mu^T_{k_{\ind-1}}\circ\dots\circ \mu^T_{k_1}\, ,
\ee
and let $\gamma_\ind$ be the $c$-vector ${\bf c}_{{\bfk}_\ind}(\ind)$.
Now denote an arbitrary monomial by $\cY_{\gamma} = \prod x_i^{c_i}$
where $c_i$ are the components of the c-vector ${\bf c}\equiv \gamma$.
One may then show that
the birational automorphism $\mu_{k_s,\epsilon(x_k(s))}$, conjugated
by $\mu^T_{{\bfk}_\ind}$, acts on any monomial $\cY_{\gamma'}$ according to
\be
(\mu^T_{{\bfk}_\ind})^{-1}\circ \mu_{k_\ind,\epsilon(x_k(\ind))}
\circ \mu^T_{{\bfk}_\ind}\quad :\quad
\cY_{\gamma'}\,  \longmapsto\,
\cY_{\gamma'} \left(1+ \cY_{\gamma_\ind}\right)^{ d_{k_\ind}
\epsilon_\ind\langle \gamma_\ind, \gamma' \rangle},
\ee
where $\epsilon_\ind=\epsilon(x_{k_\ind}(\ind))$ and
$\cY_{\gamma_\ind}=(x_{k_\ind}(\ind))^{\epsilon_\ind}$.
If we now identify $\cY_\gamma \equiv \sigma(\gamma)^{-1} \cX_\gamma$ Êas in \S\ref{secmot},
this agrees precisely with the KS symplectomorphism $U_{\gamma_\ind}$
defined in \eqref{ugamma} when $\epsilon_s=1$, or with its inverse $(U_{\gamma_\ind})^{-1}$
when $\epsilon_\ind=-1$, in the special case where the BPS index  and quadratic refinement
are taken to be
\be
\Omega(\gamma_\ind)=d_{k_\ind}\, ,
\qquad
\sigma(\gamma_\ind)=-1\, .
\ee
Under this identification, the periodicity $\nu$ of the mutation sequence
$\mu_{\bf k}$ and of its tropicalization $\mu^T_{\bf k}$ imply that
the following product is the identity:
\be
\label{cluspro}
\prod_{\ind=0}^{N-1} U^{\epsilon_\ind}_{\gamma_\ind} = 1 ,
\ee
where the product is ordered
from right to left. Moreover, it was shown\footnote{This statement
was proven for antisymmetric exchange matrix $ \exm_{ij} $, and conjectured to hold
in the antisymmetrizable case.}  in \cite{Nakanishi:2010} that
the following dilogarithm identity holds:
\be
\label{naka}
\sum_{\ind=0}^{N-1}\epsilon_\ind\, d_{k_\ind}\, L\left( \frac{(x_{k_\ind}(\ind))^{\epsilon_\ind}}
{1+(x_{k_\ind}(\ind))^{\epsilon_\ind}}\right) =0\ .
\ee
This is recognized as a special case of the general formula \eqref{generaldilogidentity}.

\subsection{Wall-crossing and dilogarithm identities for rank 2 Dynkin quivers}
\label{A2example}

To illustrate the general construction, we now consider the rank 2 cases with seed
\be
\label{epsA2}
I=\{1,2\}\, ,
\quad
I_0=\emptyset\, ,
\quad
 \exm_{ij}^\pm =\pm\left(\begin{array}{cc} 0 & c \\ -1 & 0 \\ \end{array}\right) ,
 \quad (d_1,d_2)=(1,c)\, ,
\ee
where $c=1,2$ or 3. The associated quiver then corresponds to the Dynkin diagram of $A_2$,
$B_2$ or $G_2$, respectively.  A mutation with respect to either of the nodes maps
$\exm_{ij}^\pm$ to $\exm_{ij}^\mp$. The  framed exchange matrix corresponding to the
choice of lower sign in \eqref{epsA2}, which we shall take as the initial seed, is then
\be
\tilde I=\{1,2,3,4\},
\quad
\tilde I_0=\{3,4\} ,
\quad
\tilde\exm_{ij}^- = \begin{pmatrix}
0 & -c & 1 & 0 \\
1 & 0 & 0 & 1 \\
-1 & 0 & 0 & 0 \\
0 & -1 & 0 & 0
\end{pmatrix},
\quad
(d_i)=(1,c,1,1).
\ee
The upper-right $2\times 2$ block of $\exm(\ind)$ then gives the tropicalization of the
variables $x_i(\ind)$ as row vectors. We shall denote by $\mu_i^\pm$ the mutations with respect to the node $i$
for the exchange matrix $\pm \exm_{ij}$. Moreover, for convenience we define
\be
\mu^{\pm\sharp}_i=\mu^{\pm}_{i,+},
\quad
\mu^{\pm\flat}_i=\mu^{\pm}_{i,-},Ê
\quad
\mu'^{\pm }_i=\tau_{i,+},
\quad
\mu''^{\pm }_i=\tau_{i,-}.
\ee
Clearly $\mu''^{\pm}_i=\mu'^{\mp}_i$
when acting on the $x$-variables, while $\mu^{+}_i=\mu^{-}_i$ when acting
on the $a$-variables. The action of these transformations
is summarized in table \ref{rank2table}.
\begin{table}[htb]
\begin{center}
\begin{tabular}{|c||c|c||c|c||}
\hline
& $a_1$ & $a_2$ & $x_1$ & $x_2$ \\
\hline
$\mu_1^+$ & $\frac{1+a_2^c}{a_1}$ & $a_2$ & $\frac{1}{x_1}$ & $x_2(1+x_1)$ \\
$\mu_1^{\sharp+}$&$\frac{a_1}{1+a_2^{c}} $&$ a_2$&$x_1$&$ x_2(1+x_1) $\\
$\mu'^+_1$&$1/a_1$&$ a_2 $&$1/x_1$&$ x_2 $ \\
$\mu_1^{\flat+} $&$ \frac{a_1} {1+a_2^{-c}} $&$ a_2 $&$ x_1 $&$ x_2 (1+1/x_1)$ \\
$\mu''^+_1 $&$ a_1/a_2^c $&$ a_2 $&$ 1/x_1 $&$ x_1 x_2 $ \\
\hline
$\mu_2^+ $&$ a_1$&$ \frac{1+a_1}{a_2} $&$\frac{x_1}{(1+1/x_2)^c}$&$ \frac{1}{x_2}  $ \\
$\mu_2^{\sharp+}$&$a_1$&$ \frac{a_1 a_2}{1+a_1}$&$x_1/(1+x_2)^c$&$ x_2 $ \\
$\mu'^+_2$&$a_1$&$a_1/a_2$&$x_1 x_2^c $&$1/x_2 $ \\
$\mu_2^{\flat+} $&$ a_1 $&$ \frac{a_2}{1+a_1} $&$ \frac{x_1}{(1+1/x_2)^c}$&$ x_2$ \\
$\mu''^+_2 $&$ a_1 $&$ a_1 a_2 $&$ x_1 $&$ 1/x_2$ \\
\hline
$\mu_1^-$&$\frac{1+a_2^c}{a_1}$&$ a_2 $&$\frac{1}{x_1}$&$\frac{x_1 x_2}{1+x_1}$ \\
$\mu_1^{\sharp-}$&$\frac{a_1}{1+a_2^{-c}}$&$ a_2 $&$x_1$&$x_2/(1+x_1)$ \\
$\mu'^-_1$&$a_2^c/a_1$&$a_2$&$1/x_1$&$x_1 x_2 $ \\
$\mu_1^{\flat-} $&$ \frac{a_1}{1+a_2^{c}} $&$ a_2 $&$ x_1 $&$ \frac{x_1 x_2}{1+x_1}$ \\
$\mu''^-_1 $&$ 1/a_1 $&$ a_2 $&$ 1/x_1 $&$ x_2 $ \\
\hline
$\mu_2^- $&$a_1$&$\frac{1+a_1}{a_2} $&$x_1(1+x_2)^c$&$\frac{1}{x_2}   $ \\
$\mu_2^{\sharp-}$&$a_1$&$ \frac{a_2}{1+a_1}$&$x_1(1+x_2)^c$&$  x_2$ \\
$\mu'^-_2$&$a_1 $&$ 1/a_2$&$x_1$&$1/x_2$ \\
$\mu_2^{\flat-} $&$ a_1 $&$ \frac{a_1 a_2}{1+a_1} $&$ x_1(1+1/x_2)^c$&$  x_2$ \\
$\mu''^-_2 $&$ a_1 $&$ a_2/a_1 $&$ x_1 x_2^c $&$ 1/x_2$ \\
\hline
\end{tabular}
\end{center}
\caption{Mutation sequences for rank 2 quivers of type $A_2, B_2, G_2$, corresponding to $c=1,2,3$.}
\label{rank2table}
\end{table}

\subsubsection{Example: $A_2$}
We now consider the $A_2$ case, corresponding to $c=1$ in \eqref{epsA2}.
The sequence of five mutations $\mu_{\bfk}=
\mu_1^- \mu_2^+ \mu_1^-  \mu_2^+ \mu_1^-$
is a $\nu$-period of length 5, where $\nu$ exchanges the nodes $1$ and $2$, as is evident from table \ref{A2table}.
\begin{table}[htb]
\begin{center}
\begin{tabular}{|c||c||c|c||c|c||c|c||}
\hline
$\ind $ &$ \hat B_{ij}(\ind) $&$ x_1(\ind) $&$ x_2(\ind) $& $x_1^T(\ind) $& $x_2^T(\ind)$ &$ a_1(\ind)$
&$ a_2(\ind)$\\
\hline
0 &{\scriptsize $\left[
\begin{array}{cccc}
 0 & -1 & 1 & 0 \\
 1 & 0 & 0 & 1 \\
 -1 & 0 & 0 & 0 \\
 0 & -1 & 0 & 0
\end{array}
\right]$}
 & $x_1$ & $x_2$&$ x_1 $&$ x_2$ &$   a_1$ &$ a_2 $ \\
1&
{\scriptsize
$\left[
\begin{array}{cccc}
 0 & 1 & -1 & 0 \\
 -1 & 0 & 1 & 1 \\
 1 & -1 & 0 & 0 \\
 0 & -1 & 0 & 0
\end{array}
\right]$}
&
$\frac{1}{x_1}  $& $\frac{x_1 x_2}{1+x_1} $&$ 1/x_1$ & $x_1 x_2 $&$  \frac{1+a_2}{a_1} $&$ a_2$\\
2&
{\scriptsize
$\left[
\begin{array}{cccc}
 0 & -1 & 0 & 1 \\
 1 & 0 & -1 & -1 \\
 0 & 1 & 0 & 0 \\
 -1 & 1 & 0 & 0
\end{array}
\right]$}
&$\frac{x_2}{1+x_1 x_2+x_1} $&$ \frac{1+x_1}{x_1 x_2} $&$ x_2$ &$ 1/x_1 x_2
$&$  \frac{1+a_2}{a_1} $& $\frac{1+a_1+a_2}{a_1a_2} $\\
3&
{\scriptsize
$\left[
\begin{array}{cccc}
 0 & 1 & 0 & -1 \\
 -1 & 0 & -1 & 0 \\
 0 & 1 & 0 & 0 \\
 1 & 0 & 0 & 0
\end{array}
\right]$}
&$\frac{1+x_1 x_2+x_1} {x_2}$ & $\frac{1}{x_1(1+x_2)} $&$ 1/x_2$ & $1/x_1 $&
$\frac{1+a_1}{a_2} $&$ \frac{1+a_1+a_2}{a_1 a_2}$ \\
4&
{\scriptsize
$\left[
\begin{array}{cccc}
 0 & -1 & 0 & -1 \\
 1 & 0 & 1 & 0 \\
 0 & -1 & 0 & 0 \\
 1 & 0 & 0 & 0
\end{array}
\right]$}
&$\frac{1}{x_2}$ & $x_1(1+x_2) $&$  1/x_2 $& $x_1$ &$ \frac{1+a_1}{a_2} $& $a_1$ \\
5&
{\scriptsize
$\left[
\begin{array}{cccc}
 0 & 1 & 0 & 1 \\
 -1 & 0 & 1 & 0 \\
 0 & -1 & 0 & 0 \\
 -1 & 0 & 0 & 0
\end{array}
\right]$}
&$x_2 $&$ x_1$ & $  x_2$ &$ x_1$ & $a_2$ &$ a_1$\\
\hline
\end{tabular}
\end{center}
\caption{The $\nu$-period $\mu_{\bfk}=
\mu_1^- \mu_2^+ \mu_1^-  \mu_2^+ \mu_1^-$ of length 5 for $A_2$.}
\label{A2table}
\end{table}
According to a general result of \cite{Keller:2010,Plamondon:2010}
quoted above, the periodicity of $\mu_{\bfk}$
follows from the periodicity of its tropicalization, i.e. from the identity
\be
(1+)(2-)(1-)(2+)(1-)=\nu  ,
\ee
where we abbreviated the monomial transformation $\mu'^{\pm}_i=\mu''^{\mp}_i$
acting on the $x$-variables by $(i\pm)$.
The tropical sequence is
\be
x^{T}_{{\bf k}_\ind}(\ind) = (x_1,x_1 x_2,x_2,1/x_1,1/x_2),
\ee
so the
tropical sign sequence is  $\epsilon_\ind=(+,+,+,-,-)$.
Thus, one may construct the sequence of symplectomorphisms
\beq
U_{1,0}&=& \mu_1^{\sharp -} ,
\nn\\
U_{1,1} &=& (1-)\, \mu_2^{\sharp +}\, (1-),
\nn \\
U_{0,1}&=&(1-)(2+)\, \mu_1^{\sharp -}  (2+)(1-),
\\
U_{1,0}^{-1} &=&  (1-)(2+)(1-)\, \mu_2^{\flat +} (1-)(2+)(1-),
\nn \\
U_{0,1}^{-1} &=& (1-)(2+)(1-)(2-) \, \mu_1^{\flat -}\, (2-)(1-)(2+)(1-) \nn
\eeq
so that the $\nu$-periodicity of $\mu_{\bfk}$
translates exactly into the pentagonal identity \eqref{pentU}.

Upon identifying the coordinates $x,y$ with the initial variables $x_1,x_2$,
one may check that the generators $U_\gamma$ take the standard
form given in \eqref{Upq} with quadratic refinement $\sigma_{p,q}=(-1)^{pq+p+q}$:
\be
\begin{split}
&
\qquad \qquad
U_{1,0}: [x,y]\mapsto [x,\frac{y}{1+x}] ,\quad
U_{1,1}: [x,y]\mapsto [(1+xy)x,\frac{y}{1+xy}] ,
\\
&
U_{0,1}: [x,y]\mapsto [(1+y)x,y] ,
\quad
U_{1,0}^{-1}: [x,y]\mapsto [x,y(1+x)] ,
\quad
U_{0,1}^{-1}: [x,y]\mapsto [\frac{x}{1+y},y] .
\end{split}
\ee
The dilogarithm identity \eqref{naka} specializes to
\be
L\left(\frac{x}{x+1}\right)+
L\left(\frac{x y}{x y+x+1}\right)
+L\left(\frac{y}{(x+1)(y+1)}\right)
-L\left(\frac{x (y+1)}{x y+x+1}\right)
   -L\left(\frac{y}{y+1}\right)=0 \, .
\ee
The validity of this formula can be most easily verified by taking the limit
$y\to 0$ and using the fact that $L(0)=0$. Setting $x'=x(1+y)/(1+x+xy), y'=y/(1+y)$, one recovers
\eqref{Lreal5term2}.

As a side remark, we
note that the tropical sequence $x^T_{{\bf k}_\ind}$, or equivalently the charge
vector $\gamma_\ind$, is in one-to-one correspondence with the positive
roots $\alpha_1, \alpha_1+\alpha_2, \alpha_2$ and the negative simple roots
$-\alpha_1, -\alpha_2$ of the finite Lie algebra $A_2$.

\subsubsection{Example: $B_2$}
\label{subsec-b2}

We now turn to the $B_2$ case, corresponding to $c=2$ in \eqref{epsA2}.
The sequence of mutations $(\mu_2^+ \mu_1^-)^3$
is a period of length $6$, as displayed in Table \ref{B2table}.
\begin{table}[t]
\begin{center}
\begin{tabular}{|c||c|c||c|c||c|c||}
\hline
$\ind $ & $ x_1(\ind) $ & $ x_2(\ind) $ & $ x_1^T(\ind) $ & $ x_2^T(\ind) $ & $ a_1(\ind)
$ & $ a_2(\ind)$ \\ \hline
$0 $ & $ x_1 $ & $ x_2 $ & $  x_1 $ & $ x_2 $ & $ a_1 $ & $ a_2
$ \\
$1$ & $ \frac{1}{x_1}  $ & $ \frac{x_1x_2}{1+x_1} $ & $  1/x_1 $ & $ x_1 x_2 $ & $ \frac{1+a_2^2}{a_1} $ & $ a_2
$ \\
$2$ & $\frac{x_1x_2^2}{(1+x_1+x_1 x_2)^2} $ & $
\frac{1+x_1}{x_1 x_2} $ & $  x_1x_2^2 $ & $
1/(x_1 x_2) $ & $ \frac{1+a_2^2}{a_1} $ & $ \frac{1+a_1+a_2^2}{a_1a_2}
$ \\
$3$ & $
\frac{(1+x_1+x_1 x_2)^2}{x_1 x_2^2} $ & $
\frac{x_2}{1+x_1(1+x_2)^2} $ & $
1/(x_1x_2^2) $ & $ x_2 $ & $
\frac{(1+a_1)^2+a_2^2}{a_1 a_2^2} $ & $ \frac{1+a_1+a_2^2}{a_1a_2}
$ \\
$4$ & $\frac{1}{x_1(1+x_2)^2} $ & $  \frac{1+x_1(1+x_2)^2} {x_2}$ & $ 1/x_1 $ & $ 1/x_2$ & $
\frac{(1+a_1)^2+a_2^2}{a_1 a_2^2} $ & $ \frac{1+a_1}{a_2}
$ \\
$5$ & $x_1(1+x_2)^2 $ & $ \frac{1}{x_2} $ & $ x_1 $ & $ 1/x_2 $ & $ a_1 $ & $ \frac{1+a_1}{a_2}
$ \\
$6$ & $x_1 $ & $ x_2 $ & $  x_1 $ & $ x_2 $ & $ a_1 $ & $ a_2$ \\ \hline
\end{tabular}
\end{center}
\caption{The length 6 mutation sequence $(\mu_2^+ \mu_1^-)^3$ for $B_2$.}
\label{B2table}
\end{table}
The tropical sequence is
\be
x^T_{k_\ind}(\ind)=(x_1,x_1 x_2, x_1 x_2^2,x_2,1/x_1,1/x_2)
\ee
so the tropical sign sequence is $\eps_\ind=(+,+,+,+,-,-)$.
Setting $x=x_1,y=x_2$, the dilogarithm identity \eqref{naka} reads
\be
\begin{split}
L\left(\frac{x}{x+1}\right)+2   L\left(\frac{x y}{x y+x+1}\right)
 +L\left(\frac{x y^2}{(x+1) \left(x (y+1)^2+1\right)}\right)
\\
+2 L\left(\frac{y}{(y+1) (x   y+x+1)}\right)
-L\left(\frac{x (y+1)^2}{x   (y+1)^2+1}\right)
  -2   L\left(\frac{y}{y+1}\right)
=0\, .
\end{split}
\ee
This is in fact a consequence of the nine-term relation \eqref{nineterm} with $a=-x-y-xy$, $b=-x$, $c=1$.
The associated wall-crossing identity (consistent with Eq.\, C.1 in \cite{Manschot:2010qz}) reads
\be
\label{UprodB2}
U_{0,1}^{(-2)} U_{1,0}^{(-1)}  U_{0,1}^{(2)} U_{1,2}^{(1)} U_{1,1}^{(2)} U_{1,0}^{(1)} = 1
\ee
where we denote the BPS index in superscript.
This arises from the mutations by identifying
\be
\begin{split}
U_{1,0}^{(1)} &= \mu_1^{\sharp -} , \\
U_{1,1}^{(2)} &= (1-) \, \mu_2^{\sharp +} \, (1-), \\
U_{1,2}^{(1)} &= (1-)(2+) \,  \mu_1^{\sharp -} \, (2+)(1-), \\
U_{0,1}^{(2)} &=  (1-)(2+)(1-) \, \mu_2^{\sharp +} \, (1-)(2+)(1-), \\
U_{1,0}^{(-1)} &=  (1-)(2+)(1-)(2+) \, \mu_1^{\flat -} \, (2+)(1-)(2+)(1-), \\
U_{0,1}^{(-2)} &=  (1-)(2+)(1-)(2+)(1+) \,  \mu_2^{\flat +} \, (1+)(2+)(1-)(2+)(1-),
\end{split}
\ee
and the identity \eqref{UprodB2} then follows from the tropical identity
\be
(2-)(1+)(2+)(1-)(2+)(1-)=1\, .
\ee

Finally, we note that the factors in \eqref{UprodB2} are in one-to-one correspondence with
the positive roots $\alpha_1,\alpha_1+\alpha_2,
\alpha_1+2\alpha_2,\alpha_2$ and negative simple roots $-\alpha_1,-\alpha_2$ of the Lie algebra $B_2$.
We also note that the cluster algebra for the quiver $B_2$ can be obtained by folding
the cluster algebra for the quiver $A_3$, i.e. by specializing to the locus $x_1=x_3$,
$a_1=a_3$.

\subsubsection{Example: $G_2$}
\label{subsec-g2}

Finally, we  turn to the $G_2$ case, corresponding to $c=3$ in \eqref{epsA2}.
The sequence of mutations $(\mu_2^+\mu_1^-)^4$
is now a period of length $8$, as displayed in Table \ref{G2table}.
\begin{table}[t]
\begin{center}
\begin{tabular}{|c||c|c||c|c||c|c||}
\hline
$\ind $ & $ x_1(\ind) $ & $ x_2(\ind) $ & $ x_1^T(\ind) $ & $ x_2^T(\ind) $ & $ a_1(\ind)
$ & $ a_2(\ind)$ \\ \hline
$0 $ & $ x_1 $ & $ x_2 $ & $  x_1 $ & $ x_2 $ & $ a_1 $ & $ a_2  $ \\
$1$ & $ \frac{1}{x_1}  $ & $ \frac{x_1 x_2}{1+x_1} $ & $  1/x_1 $ & $ x_1 x_2 $ & $ \frac{1+a_2^3}{a_1} $ & $ a_2$ \\
$2$ & $\frac{x_1^2 x_2^3}{(1+x_1 + x_1 x_2)^3} $ & $ \frac{1+x_1}{x_1 x_2} $ & $
x_1^2 x_2^3 $ & $ 1/x_1 x_2 $ & $ \frac{1+a_2^3}{a_1} $ & $ \frac{1+a_1+a_2^3}{a_1a_2} $ \\
$3$ & $
\frac{(1+x_1 + x_1 x_2)^3}{x_1^2 x_2^3} $ & $ \frac{x_1 x_2^2}{1+x_1^2(1+x_2)^3+x_1(2+3x_2)} $ & $
1/(x_1^2 x_2^3) $ & $ x_1 x_2^2 $ & $ \frac{(1+a_1)^3+(2+3a_1)a_2^3+a_2^6}{a_1^2 a_2^3} $ & $ \frac{1+a_1+a_2^3}{a_1a_2}$ \\
$4$ & $\frac{x_1 x_2^3}{(1+x_1(1+x_2)^2)^3} $ & $ \frac{1+x_1^2(1+x_2)^3+x_1(2+3x_2)}{x_1 x_2^2}  $ & $
x_1 x_2^3 $ & $ 1/(x_1 x_2^2) $ & $
\frac{(1+a_1)^3+(2+3a_1)a_2^3+a_2^6}{a_1^2 a_2^3}
 $ & $ \frac{(1+a_1)^2+a_2^3}{a_1 a_2^2} $ \\
$ 5 $ & $ \frac{(1+x_1(1+x_2)^2)^3}{x_1 x_2^3}  $ & $ \frac{ x_2}{(1+x_1(1+x_2)^3} $ & $
 1/(x_1 x_2^3) $ & $ x_2 $ & $
 \frac{(1+a_1)^3+a_2^2}{a_1 a_2^3} $ & $ \frac{(1+a_1)^2+a_2^3}{a_1 a_2^2} $ \\
$ 6 $ & $\frac{1}{x_1(1+x_2)^3} $ & $\frac{(1+x_1(1+x_2)^3}{ x_2} $ & $
 1/x_1 $ & $ 1/x_2 $ & $
 \frac{(1+a_1)^3+a_2^2}{a_1 a_2^3} $ & $  \frac{1+a_1}{a_2} $ \\
$ 7 $ & $ x_1(1+x_2)^3 $ & $ \frac{1}{x_2} $ & $x_1 $ & $ 1/x_2 $ & $  a_1 $ & $ \frac{1+a_1}{a_2} $ \\
$8 $ & $x_1 $ & $ x_2 $ & $x_1 $ & $ x_2 $ & $  a_1 $ & $ a_2$ \\
\hline
\end{tabular}
\end{center}
\caption{The mutation sequence $(\mu_2^+\mu_1^-)^4$ for $G_2$.}
\label{G2table}
\end{table}
The tropical sequence is now
\be
\label{topseqg2}
x^T_{k_\ind}(\ind)=(x_1, x_1 x_2,x_1^2 x_2^3,
x_1 x_2^2,x_1 x_2^3,x_2,1/x_1,1/x_2)
\ee
so the tropical sign sequence is $\eps_\ind=(+,+,+,+,+,+,-,-)$.
Setting $x=x_1,y=x_1$, the dilogarithm identity \eqref{naka} reads
\be
\begin{split}
L\left(\frac{x}{x+1}\right)
+3 L\left(\frac{x y}{x   y+x+1}\right)
+ L\left(\frac{x^2 y^3}{1+x+xy)^3+x^2 y^3}\right)\\
+3 L\left(\frac{x y^2}{(x y+x+1)  \left(x (y+1)^2+1\right)}\right)
+L\left(\frac{x y^3}{\left(x  (y+1)^2+1\right)^3 + x y^3}\right)\\
+3 L\left(\frac{y}{(y+1) \left(x  (y+1)^2+1\right)}\right)
-L\left(\frac{x (y+1)^3}{x  (y+1)^3+1}\right)-3  L\left(\frac{y}{y+1}\right)
 = 0\, .
 \end{split}
   \ee
As in the previous case, we expect that this identity can be obtained by
specializing a 16-term identity in 4 variables arising from periods of
mutations of the $D_4$ quiver, and presumably accessible
by repeated use of the five-term relation. The associated wall-crossing identity is
\be
\label{UprodG2}
U_{0,1}^{(-3)}  U_{1,0}^{(-1)}  U_{0,1}^{(3)}
U_{1,3}^{(1)} U_{1,2}^{(3)} U_{2,3}^{(1)} U_{1,1}^{(3)} U_{1,0}^{(1)} = 1,
\ee
which arises from the mutations by identifying
\be
\begin{split}
U_{1,0}^{(1)} &= \mu_1^{\sharp -} ,\\
U_{1,1}^{(3)} &= (1-) \, \mu_2^{\sharp +} \, (1-), \\
U_{2,3}^{(1)} &= (1-)(2+) \,  \mu_1^{\sharp -} \, (2+)(1-), \\
U_{1,2}^{(3)} &=  (1-)(2+)(1-) \, \mu_2^{\sharp +} \, (1-)(2+)(1-),  \\
U_{1,3}^{(1)} &=  (1-)(2+)(1-)(2+) \, \mu_1^{\sharp -} \, (2+)(1-)(2+)(1-), \\
U_{0,1}^{(3)} &=  (1-)(2+)(1-)(2+)(1-) \,  \mu_2^{\sharp +} \, (1-)(2+)(1-)(2+)(1-), \\
U_{1,0}^{(-1)} &=  (1-)(2+)(1-)(2+)(1-)(2+) \,  \mu_1^{\flat -} \, (2+)(1-)(2+)(1-)(2+)(1-), \\
U_{0,1}^{(-3)} &=  (1-)(2+)(1-)(2+)(1-)(2+)(1+) \,  \mu_2^{\flat +} \, (1+)(2+)(1-)(2+)(1-)(2+)(1-).
\end{split}
\ee
The identity \eqref{UprodG2} then follows from the simpler tropical identity
\be
   (2-)(1+)(2+)(1-)(2+)(1-)(2+)(1-) =1.
\ee
As before, the factors in \eqref{UprodG2} are in one-to-one correspondence with
the positive roots $\alpha_1,\alpha_1+\alpha_2, \alpha_1+2\alpha_2, \alpha_1+3\alpha_2, 2\alpha_1+3\alpha_2$
and negative simple roots $-\alpha_1,-\alpha_2$ of the Lie algebra $G_2$.


\providecommand{\href}[2]{#2}\begingroup\raggedright\endgroup

\end{document}